\documentclass{article}
\usepackage[top=1in, bottom=1in, left=1in, right=1in]{geometry} 
\usepackage{soul, color}
\usepackage{float}
\usepackage[utf8]{inputenc}
\usepackage{authblk}
\usepackage{placeins}
\usepackage{booktabs}
\usepackage{makeidx}  
\usepackage[utf8]{inputenc}
\usepackage[misc]{ifsym}
\usepackage{soul}

\usepackage{graphicx}
\usepackage{amsmath}
\usepackage{comment}
\usepackage{multirow}
\usepackage{color,soul}
\usepackage{amssymb}
\usepackage{appendix}
\usepackage{adjustbox}

\usepackage{bm}
\usepackage{epsfig}
\usepackage{float} 
\usepackage{makecell}
\usepackage{colortbl}
\usepackage{mathtools}
\usepackage{extarrows}
\usepackage{algorithmic}
\usepackage[linesnumbered,ruled]{algorithm2e}
\SetKwComment{Comment}{$\triangleright$\ }{}
\usepackage{balance}
\usepackage{booktabs}
\usepackage{wrapfig}
\usepackage{subfigure}
\usepackage{lipsum}

\newcommand{\argmin}{\arg\!\min}

\usepackage{multirow}
\usepackage{comment}
\usepackage[pagebackref,breaklinks,colorlinks]{hyperref}
\usepackage{xcolor}
\hypersetup{
    colorlinks=true,       
    linkcolor=red,          
}
\graphicspath{{figures/}}

\definecolor{cYellow}{HTML}{FFCC33}
\definecolor{cPurple}{HTML}{9966FF}
\definecolor{cRed}{HTML}{CC3300} 
\definecolor{lightRed}{HTML}{FF9999}
\definecolor{lightBlue}{HTML}{89c3eb}

\title{This is title}
\author{Jiamian Wang\thanks{Santa Clara University. jwang16@scu.edu}}
\author{Yulun Zhang\thanks{ETH Z\"{u}rich. yulun100@gmail.com}}
\author{Xin Yuan\thanks{Westlake University. xyuan@westlake.edu.cn}}
\author{Ziyi Meng\thanks{Kuaishou Technology. mengziyi@163.com}}
\author{Zhiqiang Tao\thanks{Santa Clara University. ztao@scu.edu}}
\affil{}

\date{}

\begin{document}

\title{Modeling Mask Uncertainty in Hyperspectral Image Reconstruction}

\maketitle

\begin{abstract}
Recently, hyperspectral imaging (HSI) has attracted increasing research attention, especially for the ones based on a coded aperture snapshot spectral imaging (CASSI) system. Existing deep HSI reconstruction models are generally trained on paired data to retrieve original signals upon 2D compressed measurements given by a particular optical hardware mask in CASSI, during which the mask largely impacts the reconstruction performance and could work as a ``model hyperparameter'' governing on data augmentations. This mask-specific training style will lead to a hardware miscalibration issue, which sets up barriers to deploying deep HSI models among different hardware and noisy environments. To address this challenge, we introduce mask uncertainty for HSI with a complete variational Bayesian learning treatment and explicitly model it through a mask decomposition inspired by real hardware. Specifically, we propose a novel Graph-based Self-Tuning (GST) network to reason uncertainties adapting to varying spatial structures of masks among different hardware.  Moreover, we develop a bilevel optimization framework to balance HSI reconstruction and uncertainty estimation, accounting for the hyperparameter property of masks. Extensive experimental results and model discussions validate the effectiveness (over $33$/$30$ dB) of the proposed GST method under two miscalibration scenarios and demonstrate a highly competitive performance compared with the state-of-the-art well-calibrated methods. Our code and pre-trained model are available at \url{https://github.com/Jiamian-Wang/mask_uncertainty_spectral_SCI} 
\end{abstract}

\section{Introduction}\label{sec:intro}

\begin{figure}[t] 
	\centering 
	\subfigtopskip=2pt 
	\subfigbottomskip=2pt 
	\subfigcapskip=-2pt 
	\subfigure[Real mask decomposition]{
		\includegraphics[width=0.57\linewidth]{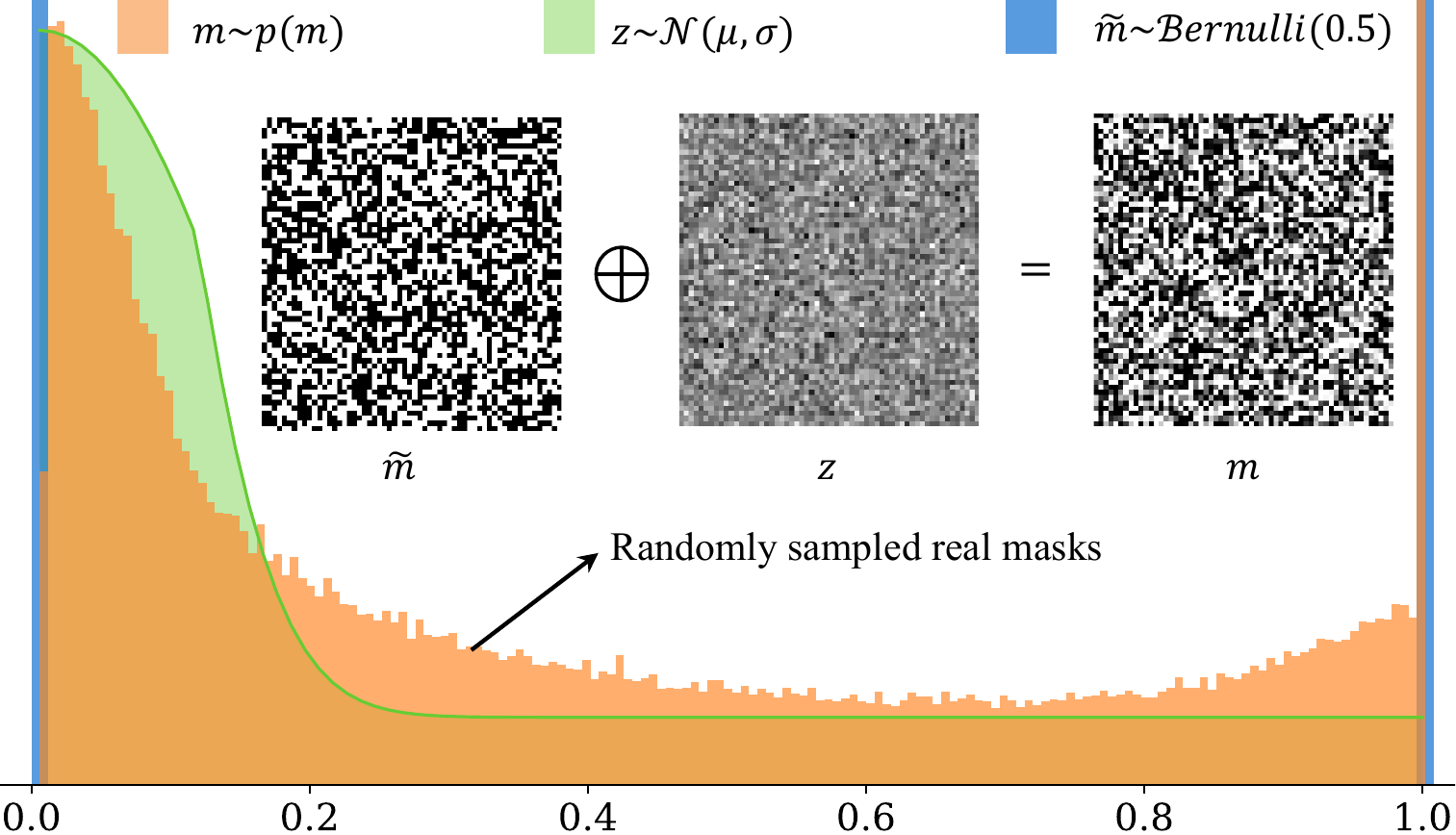}}
	\subfigure[Miscalibration performance]{
		\includegraphics[width=0.38\linewidth]{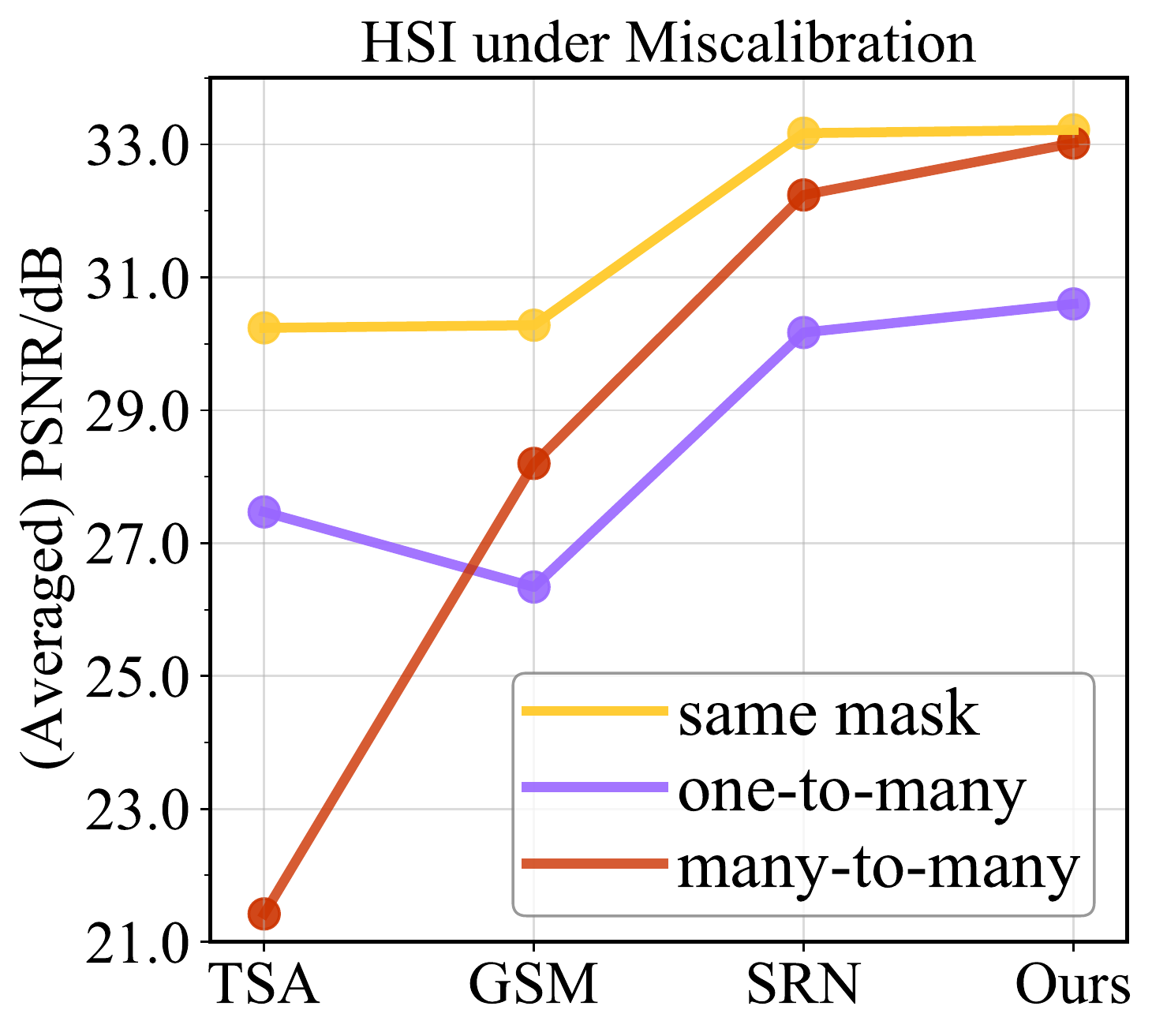}}
\vspace{-3mm}
\caption{(a) A real mask $m\sim p(m)$ can be decomposed into an unknown clean mask $\tilde{m}$ plus random noise $z$. The mask distribution is plotted by realistic hardware mask values. Note that the distributions are demonstrated in a \texttt{symlog} scale. 
(b) Performance comparison under three different settings, including 1) the {\color{cYellow}same mask} for training/testing, 2) training on one mask and testing on multiple masks ({\color{cPurple}one-to-many}), and 3) training with random masks and testing on a held-out mask set ({\color{cRed}many-to-many}).}
\vspace{-2mm} 
\label{fig: coverfig}
\end{figure}

Hyperspectral imaging (HSI) provides richer signals than the traditional RGB vision and has broad applications across agriculture~\cite{gowen2007hyperspectral,lu1999hyperspectral,lorente2012recent}, medical imaging~\cite{lu2014medical,johnson2007snapshot}, remote sensing~\cite{borengasser2007hyperspectral,yuan2017hyperspectral,fuying-2020-remotesensing}, etc. Various HSI systems have been built and studied in recent years, among which, the coded aperture snapshot spectral imaging (CASSI) system~\cite{gehm2007single,Yuan_review} stands out due to its passive modulation property and has attracted increasing research attentions~\cite{Wang17PAMI,liu2018rank,9495194,meng2020end,huang2021deep,meng2021self,zhang2019computational} in the computer vision community. The CASSI system adopts a hardware encoding \& software decoding schema. It first utilizes an optical hardware mask to compress hyperspectral signals into a 2D measurement and then develops software algorithms to retrieve original signals upon the coded measurement conditioning on one particular mask used in the system. Therefore, the hardware mask generally plays a key role in reconstructing hyperspectral images and may exhibit a strongly-coupled (\emph{one-to-one}) relationship with its reconstruction model.

While deep HSI networks~\cite{meng2020end,huang2021deep,wang2021new,wang2020dnu,zhang2019hyperspectral} have recently shown a promising performance on high-fidelity reconstruction and real-time inference, they mainly treat the hardware mask as a fixed ``model hyperparameter'' (governing data augmentations on the compressed measurements) and train the reconstruction network on the paired hyperspectral images and measurements given the same mask. Empirically, this will cause a \textbf{hardware miscalibration} issue -- the mask used in the pre-trained model is unpaired with the real captured measurement -- when deploying a single deep reconstruction network among multiple hardware systems (usually with different masks, or different responses even using the same mask due to fabrication errors). As shown in Fig.~\ref{fig: coverfig}, the performance of deep reconstruction networks pre-trained with one specific mask will badly degrade when applying in different unseen masks (\emph{one-to-many}). Rather than re-training models on each new mask, which is inflexible for practical usage, we are more interested in training a single model that could adapt to different hardware by exploring and exploiting uncertainties among masks\footnote{Note that in CASSI, the masks corresponding to different wavelengths are obtained by calibration with a single fixed wavelength to approximate a small range of band-limited signal. In~\cite{arguello2013higher}, a high-order model was proposed to address this issue, which is different from the uncertainty problem in the mask considered in the present work.}.

One possible solution is to train a deep network over multiple CASSI systems, \emph{i.e.}, using multiple masks and their corresponding encoded measurements following the deep ensemble~\cite{kendall2017uncertainties} strategy. However, due to the distinct spatial patterns of each mask and its hyperparameter property, directly training the network with randomly sampled masks still cannot achieve a well-calibrated performance and sometimes performs even worse, \emph{e.g.}, \emph{many-to-many} of TSA-Net~\cite{meng2020end} in Fig.~\ref{fig: coverfig}. Hence, we delve into one possible mask decomposition observed from the real hardware, which treats a mask as the unknown clean one plus random noise like Gaussian (see Fig.~\ref{fig: coverfig}). We consider the noise stemming from two practical sources: 1) the hardware fabrication in real CASSI systems and 2) the functional mask values caused by different lighting environments. Notably, rather than modeling the entire mask distribution, which is challenging due to the high-dimensionality of a 2D map, we explicitly model the mask uncertainty as Gaussian noise centering around a given mask through its decomposition and resort to learn \emph{self-tuning variances} adapting to different mask spatial patterns.

\begin{figure}[t]
	\begin{center}
		\includegraphics[width=.9\textwidth]{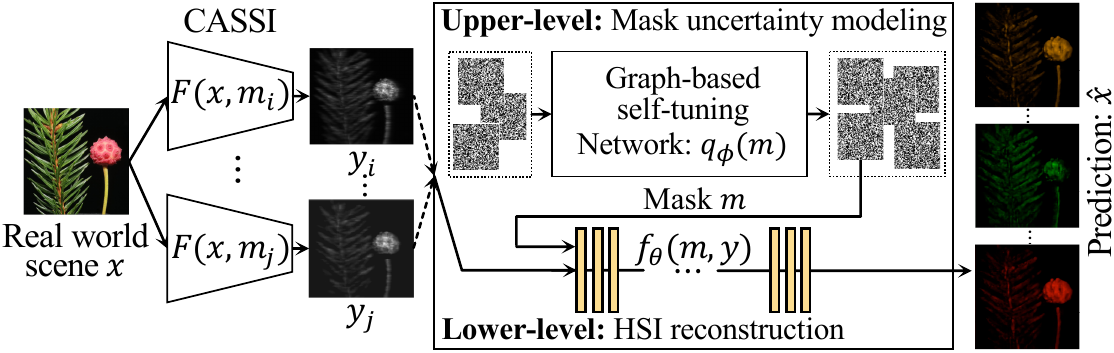}\vspace{-4mm}
	\end{center}
\caption{Illustration of the proposed method with a bilevel optimization framework. The upper-level employs the proposed Graph-based Self-Tuning (GST) network to model mask uncertainty through approximating the mask posterior distribution, while the lower-level adopts a reconstruction network $f_{\theta}(m,y)$ to retrieve the original hyperspectral images $\widehat{x}$ by taking masks as hyperparameters. Our model could be applied in multiple CASSI systems using different unseen masks.}
\vspace{-2mm}
\label{fig: framework}
\end{figure}

In this study, we propose a novel Graph-based Self-Tuning (GST) network to model mask uncertainty upon variational Bayesian learning and hyperparameter optimization techniques. On the one hand, we approximate the mask posterior distribution with variational inference under the given prior from real mask values, leading to a smoother mask distribution with smaller variance supported by empirical evidence. On the other hand, we leverage graph convolution neural networks to instantiate a stochastic encoder to reason uncertainties varying to different spatial structures of masks. Moreover, we develop a bilevel optimization framework (Fig.~\ref{fig: framework}) to balance the HSI reconstruction performance and the mask uncertainty estimation, accounting for the high sensitive network responses to the mask changes. We summarize the contributions of this work as follows. 
\begin{itemize}
    \setlength\itemsep{0em}
    \item We introduce \emph{mask uncertainty} for CASSI to calibrate a single deep reconstruction network applying in multiple hardware, which brings a promising research direction to improve the robustness and flexibility of deploying CASSI systems to retrieve hyperspectral signals in real-world applications. To our best knowledge, this is the first work to explicitly explore and model mask uncertainty in the HSI reconstruction problem. 

    \item A complete variational Bayesian learning framework has been provided to approximate the mask posterior distribution based on a mask decomposition inspired by real hardware mask observations. Moreover, we design and develop a bilevel optimization framework (see Fig.~\ref{fig: framework}) to jointly achieve high-fidelity HSI reconstruction and mask uncertainty estimation.
    
    \item We propose a novel Graph-based Self-Tuning (GST) network to automatically capture uncertainties varying to different spatial structures of 2D masks, leading to a smoother mask distribution over real samples and working as an effective data augmentation method. 
    
    \item Extensive experimental results on both simulation data and real data demonstrate the effectiveness (over $33$/$30$ dB) of our approach under two miscalibration cases. Besides, the proposed method also shows a highly competitive performance compared with state-of-the-art methods under the traditional well-calibrated setting. 
\end{itemize}

\section{Related Work}
\label{sec:related work}
Recently, plenty of advanced algorithms have been designed from diverse perspectives to reconstruct the HSI data from measurements encoded by CASSI system. Among them, optimization-based methods solve the problem by introducing different priors, eg, GPSR~\cite{figueiredo2007gradient}, TwIST~\cite{bioucas2007new} and GAP-TV~\cite{yuan2016generalized} etc., in which DeSCI~\cite{liu2018rank} leads to the best results. Another novel stream is to empower optimization-based method by deep learning. For example, deep unfolding methods~\cite{hershey2014deep,ma2019deep,wang2019hyperspectral} and Plug-and-Play (PnP) structures~\cite{qiao2020snapshot,yuan2020plug,qiao2020deep,meng2021self} have been raised. Although their well interpretability and robustness to masks to a  certain degree, they suffer from low efficiency and unstable convergence.

Besides, a number of deep reconstruction networks~\cite{meng2020snapshot,miao2019net,meng2020end,wang2021new,huang2021deep} have been proposed, yielding SOTA performance with high reconstructive efficiency. Specifically, $\lambda$-net~\cite{miao2019net} is introduced under a generative adversarial framework. TSA-Net~\cite{meng2020end} reconstructs the HSI by considering spatial/spectral self-attentions, outperforming counterparts of the day. SRN~\cite{wang2021new} provides a lightweight reconstruction backbone based on residual learning. Recently, a Gaussian Scale Mixture (GSM) based method~\cite{huang2021deep} is proposed, showing robustness on masks by enabling an approximation on different sensing matrices. However, all the above pre-trained networks perform unsatisfactorily on distinct unseen masks, raising a question that how to deploy a single reconstruction network among different hardware systems.

In this work, we enable a single well-trained deep reconstruction network to adapt to distinct real masks by proposing a variational approach for estimating the mask uncertainty. Popular uncertainty estimation methods include 1) Bayesian neural networks (BNN)~\cite{m-bmam-92,blundell2015weight,pmlr-v48-gal16} and 2) deep ensemble~\cite{kendall2017uncertainties,fort2019deep,liu2019accurate}. The former usually approximates the weight posterior distribution by using variational inference~\cite{blundell2015weight} or MC-dropout~\cite{pmlr-v48-gal16}, while the latter generally trains a group of networks from random weight initialization. 
However, it is challenging to directly quantify mask uncertainty via BNNs or deep ensemble, since treating masks as model weights contraries to their hyperparameter properties.

\begin{figure*}[t] 
\centering 
\includegraphics[width=1.0\textwidth]{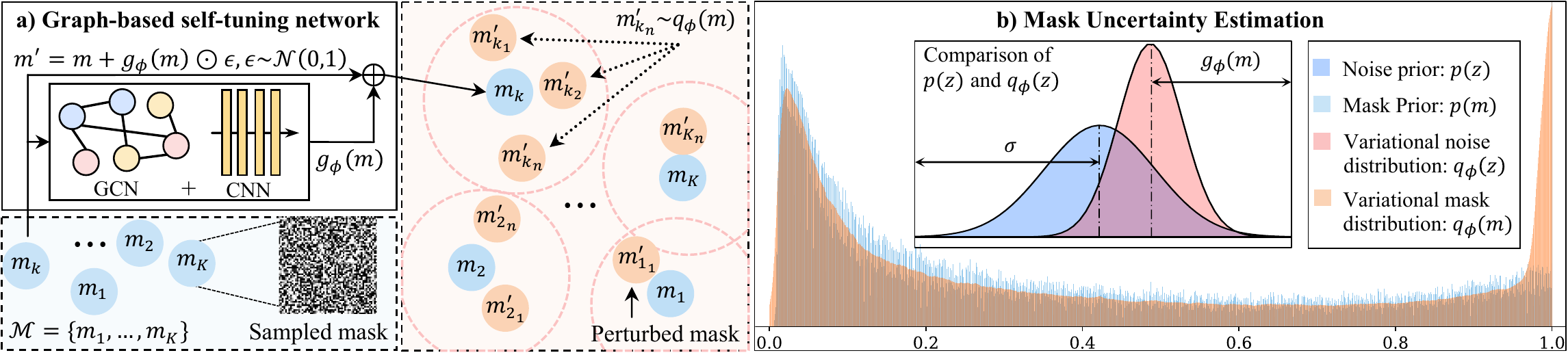} \vspace{-6mm}
\caption{
    Illustration of modeling mask uncertainty with the proposed Graph-based Self-Tuning (GST) network. a) GST takes as input a real mask $m_k$ randomly sampled from different hardware masks $\mathcal{M}$ and obtains perturbed masks $m'_{k_n}$ by learning self-tuning variance centering on $m_k$. b) GST estimates mask uncertainty by approximating the mask posterior with a variational distribution $q_{\phi}(m)$, leading to a smoother mask distribution over the mask prior $p(m)$. More discussions are given in Section~\ref{subsec: model analysis}.
}
\vspace{-2mm}
\label{fig: bayesian_learning} 
\end{figure*}

\section{Methodology}\label{sec:method}

\subsection{Preliminaries \label{subsec: preliminaries}}
\textbf{HSI reconstruction.} The reconstruction based on the CASSI system~\cite{Yuan_review,meng2020end} generally includes a hardware-encoding forward process and a software-decoding inverse process. Let $x$ be a 3D hyperspectral image with the size of $H \times W\times \Lambda$, where $H$, $W$, and $\Lambda$ represent the height, width, and the number of spectral channels. The optical hardware encoder will compress the datacube $x$ into a 2D measurement $y$ upon a fixed physical mask $m$. The forward model of CASSI is
\begin{equation}\label{eq: CASSI}
    y=F(x;m)=\sum\nolimits_{\lambda}^{\Lambda} \texttt{shift}(x)_{\lambda}\odot \texttt{shift}(m)_{\lambda}+ \zeta,
\end{equation}
where $\lambda$ refers to a spectral channel, $\odot$ represents the element-wise product, and $\zeta$ denotes the measurement noise. The shift operation is implemented by a single disperser as $\texttt{shift}(x)(u, v, i)=x(h,w+d(i-\lambda),i)$. In essence, the measurement $y$ is captured by spectral modulation\footnote{We used a two-pixel shift for neighbored spectral channels following~\cite{meng2020end,wang2021new}. More details about spectral modulation could be found in~\cite{Yuan_review}.} conditioning on the hardware mask $m$.

In the inverse process, we adopt a deep reconstruction network as the decoder: $\widehat{x}=f_{\theta}(m, y)$ where $\widehat{x}$ is the retrieved hyperspectral image, and $\theta$ represents all the learnable parameters. Let $\mathcal{D} = \{(x_i,y_i)\}_{i=1}^{N}$ be the dataset. The reconstruction network $f_{\theta}$ is generally trained to minimize an $\ell_1$ or $\ell_2$ loss as the following:
\begin{equation}\label{eq: original loss}
\min\limits_{\theta} \sum_{x,y\in \mathcal{D}}\ell(f_{\theta}(m,y)-x)\ ~\textup{where} ~\ y=F(x; m).
\end{equation}
We instantiate $f_{\theta}$ as a recent HSI backbone model provided in~\cite{wang2021new}, which benefits from nested residual learning and spatial/spectral-invariant learning. We employ this backbone for its lightweight structure to simplify the training\footnote{We provide more details toward SRN in Append~\ref{sec: original backbone}. More discussion on alternating reconstruction backbones could be found in Append~\ref{sec: backbones}.}.

\textbf{Hardware miscalibration.} As shown in Eq.~\eqref{eq: original loss}, there is a {\em paired relationship} between the parameter $\theta$ and mask $m$ in deep HSI models. Thus, for different CASSI systems (i.e., distinct masks), multiple pairs $\{m_1; \theta_1\},...,\{m_K; \theta_K\}$ are expected for previous works. Empirically, the miscalibration between $m$ and $\theta$ will lead to obvious performance degradation. This miscalibration issue inevitably impairs the flexibility and robustness of deploying deep HSI models across real systems, considering the expensive training time and various noises existing in hardware. To alleviate such a problem, one straight-forward solution is to train the model $f_{\theta}$ with multiple masks, i.e., $\mathcal{M}=\{m_1,...,m_K\}$, falling in a similar strategy to deep ensemble~\cite{kendall2017uncertainties}. 
However, directly training a single network with random masks cannot provide satisfactory performance to unseen masks (see Section~\ref{sec:exp}), since the lack of explicitly exploring the relationship between uncertainties and different mask structures.

\subsection{Mask Uncertainty \label{subsec: mask uncertainty}}

Modeling mask uncertainty is challenging due to 1) the high dimensionality of a 2D mask and the limited mask set size (\emph{i.e.}, $K$ for $\mathcal{M}$) and 2) the varying spatial structures (patterns) among masks. Hence, we propose to first estimate uncertainties around each mask through one possible mask decomposition in this section and then reason how the uncertainty will adapt to the change of mask structures with a self-tuning network in Section~\ref{subsec: self tuning network}.

Inspired by the distribution of real mask values (Fig.~\ref{fig: coverfig} and Fig.~\ref{fig: bayesian_learning}), which renders two peaks at $0$ and $1$ and appears a Gaussian shape spreading over the middle, we decompose a mask as two components
\begin{equation}\label{eq: real mask divide}
        m=\tilde{m}+z,
\end{equation}
where we assume each pixel in $z$ follows a Gaussian distribution. For simplicity, we slightly abuse the notations by denoting the noise prior as $p(z) = \mathcal{N}(\mu, \sigma)$. The $\tilde{m}$ denotes the underlying clean binary mask with a specific spatial structure.

We estimate the mask uncertainty by approximating the mask posterior $p(m|X,Y)$ following~\cite{blundell2015weight,Gal2016Uncertainty,wilson2020bayesian}, where $X = \{x_1,\dots,x_{N}\}$ and $Y = \{y_1,\dots,y_{N}\}$ indicate hyperspectral images and their corresponding measurements. To this end, we aim to learn a variational distribution $q_{\phi}(m)$ parameterized by $\phi$ to minimize the KL-divergence between $q_{\phi}(m)$ and $p(m|X,Y)$, \emph{i.e.}, $\min_{\phi} KL[q_{\phi}(m)||p(m|X,Y)]$, equivalent to maximizing the evidence lower bound (ELBO)~\cite{kingma2013auto,hoffman2016elbo} as
\begin{equation}\label{eq: global obj divide}
        \max_{\phi}\underbrace{\mathbb{E}_{q_{\phi}(m)}[\log p(X|Y,m)]}_{\rm reconstruction}\!-\!\underbrace{KL[q_{\phi}(m)||p(m)]}_{\rm regularization},
\end{equation}
where the first term measures the reconstruction (i.e., reconstructing the observations $X$ based on the measurements $Y$ and mask $m$ via $f_{\theta}(m,y)$), and the second term regularizes $q_{\phi}(m)$ given the mask prior $p(m)$.
Inspired by Eq.~\eqref{eq: real mask divide}, we treat the clean mask $\tilde{m}$ as a 2D constant with a particular structure and focus on  mask uncertainties arising from the noise $z$. Thus, the variational distribution $q_{\phi}(m)$ is defined as a Gaussian distribution centering on a given $m \in \mathcal{M}$ by
\begin{equation}\label{eq: self-tuning variance}
    q_{\phi}(m) = \mathcal{N}(m, g_{\phi}(m)),
\end{equation}
where $g_{\phi}(m)$ learns \emph{self-tuning variance} to model the uncertainty adapting to real masks sampled from $\mathcal{M}$. Correspondingly, the underlying variational noise distribution $q_{\phi}(z)$ follows Gaussian distribution with variance $g_{\phi}(m)$. We adopt the reparameterization trick~\cite{kingma2013auto} for computing stochastic gradients for the expectation w.r.t $q_{\phi}(m)$. Specifically, let $m' \sim q_{\phi}(m)$ be a random variable sampled from the variational distribution, we have 
\begin{equation}
\begin{aligned}\label{eq: mask define}
m' = t(\phi,\epsilon) = m + g_{\phi}(m)\odot\epsilon, ~~\epsilon\sim\mathcal{N}(0,1).
\end{aligned}
\end{equation}
Notably, we clamp all the pixel values of $m'$ in range $[0, 1]$.

The first term in Eq.~\eqref{eq: global obj divide} reconstructs $x$ with $p(x|y,m)\propto p(x|\widehat{x}=f_{\theta}(m,y))$, yielding a squared error when $x|\widehat{x}$ follows a Gaussian distribution~\cite{vincent2010stacked}. Similar to AutoEncoders, we implement the negative log-likelihood $\mathbb{E}_{q_{\phi}(m)}[-\log p(X|Y,m)]$ as a $\ell_2$ loss and compute its Monte Carlo estimates with Eq.~\eqref{eq: mask define} as
\begin{equation}
\label{eq: final reconstruction}
\ell(\phi, \theta; \mathcal{D}) = \textstyle \frac{N}{B}\sum_{i=1}^{B} \| f_\theta(y_i, t(\phi,\epsilon_i)) - x_i\|^2,
\end{equation}
where $(x_i,y_i) \in \mathcal{D}$, $B$ denotes the mini-batch size, and $t(\phi,\epsilon_i)$ represents the $i$-th sample from $q_{\phi}(m)$. We leverage $t(\phi, \epsilon_i)$ to sample $B$ perturbed masks from $q_{\phi}(m)$ centering on one randomly sampled mask $m \in \mathcal{M}$ per batch.

Since $p(m)$ is unknown due to various spatial structures of masks, we resort to approximating the KL term in Eq.~\eqref{eq: global obj divide} with the entropy of $q_{\phi}(m)$. Eventually, we implement the $\textup{ELBO}(q(m))$ with the following loss:
\begin{equation}
\begin{aligned}\label{eq: final loss}
\mathcal{L}(\phi, \theta; \mathcal{D}) = \ell(\phi, \theta; \mathcal{D}) + \beta \mathbb{H}[\log q_{\phi}(m)], 
\end{aligned}
\end{equation}
where $\mathbb{H}[\log q_{\phi}(m)]$ is computed by $\ln(g_{\phi}(m)\sqrt{2\pi e})$ and $\beta >0$ interprets the objective function between variational inference and variational optimization~\cite{mackay2019self,khan2018fast}.

\begin{figure}[t]
	\begin{center}
	\includegraphics[width=.8\textwidth]{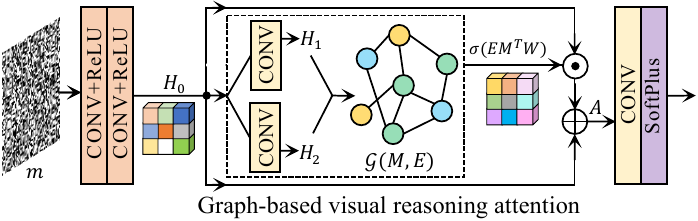}\vspace{-5mm}
	\end{center}
\caption{Structure of Graph-based Self-Tuning (GST) network. The model takes mask $m$ as input and outputs a 2D variance map, globally handling mask in a graph domain.}
\label{fig: graph_reasoning_attention}
\vspace{-2mm}
\end{figure} 

\subsection{Graph-based Self-Tuning Network} \label{subsec: self tuning network}

We propose a graph-based self-tuning (GST) network to instantiate the variance model $g_{\phi}(m)$ in Eq.~\eqref{eq: self-tuning variance}, which captures uncertainties around each mask and leads to a smoother mask distribution over real masks (see Fig.~\ref{fig: bayesian_learning}). The key of handling unseen masks (new hardware) is to learn how the distribution will change along with the varying spatial structures of masks. To this end, we implement the GST as a visual reasoning attention network~\cite{chen2018iterative,li2019visual,zhang2021mr}. It firstly computes pixel-wise correlations (visual reasoning) based on neural embeddings and then generates attention scores based on graph convolutional networks (GCN)~\cite{scarselli2008graph,GCN}. Unlike previous works~\cite{chen2018iterative,li2019visual,zhang2021mr}, the proposed GST model is tailored for building a stochastic probabilistic encoder to capture the mask distribution.

We show the network structure of GST in Fig.~\ref{fig: graph_reasoning_attention}. Given a real mask $m$, GST produces neural embedding $H_0$ by using two concatenated \texttt{CONV-ReLU} blocks. Then, we employ two \texttt{CONV} layers to convert $H_0$ into two different embeddings $H_1$ and $H_2$, and generate a graph representation by matrix multiplication $H_1^{T}H_2$, resulting in $\mathcal{G}(M, E)$, where the node matrix $M$ represents mask pixels and the edge matrix $E$ denotes the pixel-wise correlations. Let $W$ be the weight matrix of GCN. We obtain an enhanced attention cube by pixel-wise multiplication
\begin{equation}
    A=H_0 \odot (\sigma(EM^{T}W)+\mathbf{1}),
\end{equation}
where $\sigma$ is the sigmoid function. Finally, the self-tuning variance is obtained by
\begin{equation}
    g_{\phi}(m)=\delta(\texttt{CONV}(A)),
\end{equation}
where $\delta$ denotes the softplus function and $\phi$ denotes all the learnable parameters. Consequently, GST enables adaptive variance modeling to arbitrary real masks.

\begin{wrapfigure}[15]{r}{0.59\textwidth}
\begin{minipage}{0.59\textwidth}
\begin{algorithm}[H]\footnotesize 
\caption{\scriptsize GST Training Algorithm} \label{algo: training}
\KwIn{$\mathcal{D}^{trn}, \mathcal{D}^{val}$,$\mathcal{M}$; initialized $\theta$, $\phi$;}
\KwOut{$\theta^{*}, \phi^{*}$}
Pre-train $f_{\theta}(\cdot)$ on $\mathcal{D}^{trn}$ with $\alpha_0$ for $T^{init}$ epochs\;
\While{not converge}
{
    \For{$t=1,...,T^{trn}$}{
    $\{(x_i, y_i)\}_{i=1}^B \sim \mathcal{D}^{trn}$\;
    $\theta\leftarrow\theta-\alpha_{1}\frac{\partial}{\partial \theta}\ell(\phi, \theta; \mathcal{D}^{trn})$\;
    }
    \For{$t=1,...,T^{val}$}{
 $\{(x_i, y_i)\}_{i=1}^B \!\sim\! \mathcal{D}^{val}, m \!\sim\! \mathcal{M}, \epsilon \!\sim\! \mathcal{N}(0,1)$\;
    $\phi\leftarrow\phi-\alpha_{2}\frac{\partial}{\partial \phi}\mathcal{L}(\phi, \theta; \mathcal{D}^{val})$\;
    }
}
\end{algorithm}
\end{minipage}
\end{wrapfigure}

\subsection{Bilevel Optimization} \label{subsec: bilevel optimization}
While it is possible to jointly train the HSI reconstruction network $f_{\theta}$ and the self-tuning network $g_{\phi}$ using the loss in Eq.~\eqref{eq: final loss}, it is more proper to formulate the training of these two networks as a bilevel optimization framework accounting for two hyperparameter properties of masks. First, the deep reconstruction network is generally high-sensitive to the change/perturbation of masks. Thus, the model weight $\theta$ is largely subject to a mask $m$. Second, deep HSI methods~\cite{meng2020end,wang2021new} usually employ a single mask and a group of shifting operations to lift the 2D measurement as a multi-channel input, where the mask works as a hyperparameter of data augmentation for training deep networks. 

To be specific, we define the lower-level problem as HSI reconstruction and the upper-level problem as mask uncertainty estimation, and propose the final objective function of our GST model as the following:
\begin{equation}
    \min_{\phi} \mathcal{L}(\phi, \theta^{*}; \mathcal{D}^{val}) ~\textup{~s.t.~}~
    \theta^{*} = \argmin_{\theta} \ell(\phi, \theta; \mathcal{D}^{trn}), \label{eq: bilevel relationship}
\end{equation}
where $\ell(\phi, \theta; \mathcal{D}^{trn})$ is provided in Eq.~\eqref{eq: final reconstruction} with a training set and $\mathcal{L}(\phi, \theta^{*}; \mathcal{D}^{val})$ is given by Eq.~\eqref{eq: final loss} in a validation set. Upon Eq.~\eqref{eq: bilevel relationship}, $f_{\theta}$ and $g_{\phi}$ are alternatively updated by computing gradients $\frac{\partial l}{\partial\theta}$ and $\frac{\partial\mathcal{L}}{\partial\phi}$. To better initializing the parameter $\theta$, we pre-train the reconstruction network $f_{\theta}(m,y)$ for several epochs. The entire training procedure of the proposed method is summarized in Algorithm~\ref{algo: training}. Notably, introducing Eq.~\eqref{eq: bilevel relationship} brings two benefits. 1) It could balance the solutions of HSI reconstruction and mask uncertainty estimation. 2) It enables the proposed GST as a hyperparameter optimization method, which could provide high-fidelity reconstruction even working on a single mask (see Table~\ref{Tab: psnr_ssim_single_trn}).

\section{Experiments}\label{sec:exp}

\textbf{Simulation data.} We adapt the training set provided in~\cite{meng2020end}.
Simulated measurements are computed by mimicking the compressing process of SD-CASSI optical system~\cite{meng2020end}. For both metric and perceptual comparison, we employ a benchmark test set that contains ten 256$\times$256$\times$28 HSIs following~\cite{huang2021deep,wang2021new,meng2021self}. We build a validation set by splitting 40 HSIs from the training set. 

\textbf{Real data.} We adopt five real 660$\times$714 measurements provided in~\cite{meng2020end} for qualitative evaluation. We train the model on the expanded simulation training set by augmenting 37 HSIs originating from the KAIST dataset~\cite{choi2017high}. Also, the Gaussian noise ($\mathcal{N}(0, \varphi), \varphi \sim U[0,0.05]$) is added on the simulated measurements during training, for the sake of mimicking practical measurement noise $\zeta$. All the other settings are kept the same as compared deep reconstruction methods.

\textbf{Mask set. \label{mask set}} We adopt two 660$\times$660 hardware masks in our experiment. Both are produced by the same fabrication process. For training, the mask set $\mathcal{M}$ is created by randomly cropping (256$\times$256) from the mask provided in~\cite{meng2020end}. For testing, both masks are applied. In simulation, testing masks are differentiated from the training ones. For real HSI reconstruction, the second mask~\cite{meng2020snapshot} is applied, indicating hardware miscalibration scenario. 

\textbf{Implementation details.} The training procedures (Algorithm~\ref{algo: training}) for simulation and real case follow the same schedule: We apply the \texttt{xavier uniform}~\cite{glorot2010understanding} initializer with \texttt{gain=1}. Before alternating, the reconstruction network is trained for $T^{init}$=$20$ epochs (learning rate $\alpha_0$=4$\times$ $10^{-4}$). Then, the reconstruction network $f_{\theta}(\cdot)$ is updated on training phase for  $T^{trn}$=$5$ epochs ($\alpha_1$=4$\times$ $10^{-4}$) and the GST network is updated on validation phase for $T^{val}$=$3$ epochs ($\alpha_2$=1$\times$ $10^{-5}$). The learning rates are halved per 50 epochs and we adopt Adam optimizer~\cite{kingma2014adam} with default setting. In this work, we adopt SRN (\texttt{v1})~\cite{wang2021new} as the reconstructive backbone, i.e., the full network without rescaling pairs. All the experiments were conducted on 4 NVIDIA GeForce RTX 3090 GPUs. 

\textbf{Compared methods.}
For hardware miscalibration, masks for data pair setup (i.e., CASSI compressing procedure) and network training should be different from those for testing. We specifically consider two scenarios: 1) many-to-many, i.e., training the model on mask set $\mathcal{M}$ and testing it by unseen masks; 2) One-to-many, i.e., training the model on single mask and testing it by diverse unseen masks, which brings more challenges. For quantitative performance comparison, in this work all the testing results are computed upon 100 testing trials (100 random unseen masks). We compare with four SOTA methods: TSA-Net~\cite{meng2020end}, GSM-based method~\cite{huang2021deep}, SSI-ResU-Net (SRN)~\cite{wang2021new} and PnP-DIP~\cite{meng2021self}, among which the first three are deep reconstruction networks and the last one is an iterative optimization-based method. Note that 1) PnP-DIP is a self-supervised method. We test it by feeding the data encoded by different masks in the testing mask set and compute the performance over all obtained results. 2) For real-world HSI reconstruction, all models are trained on the same mask while tested on the other. Specifically, the network inputs are initialized by testing mask for TSA-Net and SRN. For GSM, as demonstrated by the authors, we directly compute the sensing matrix of testing mask and replace the corresponding approximation in the network. We use PSNR and SSIM~\cite{wang2004image} as metrics for quantitative comparison.

\begin{figure*}[t] 
\centering 
\includegraphics[width=0.96\textwidth]{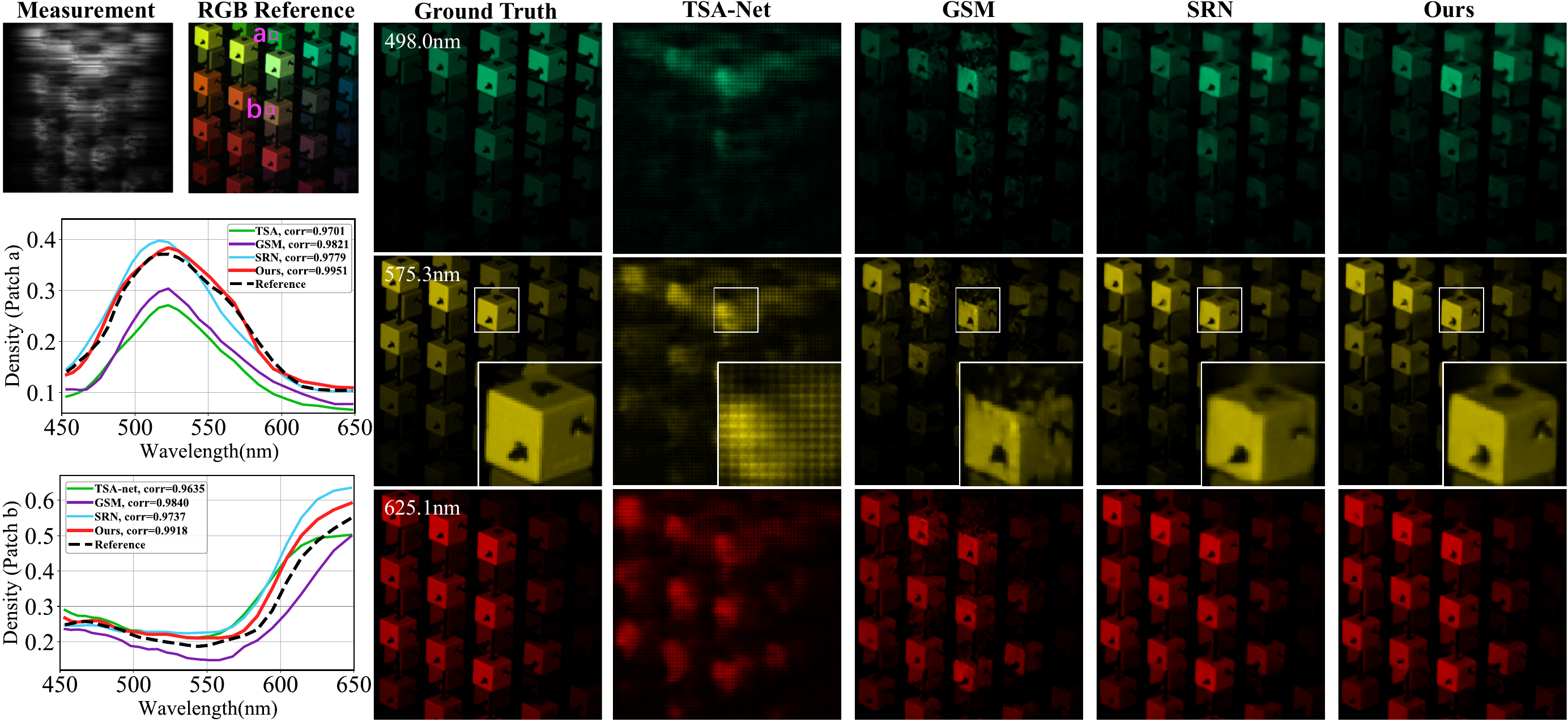}\vspace{-0.2cm}
\caption{Reconstruction results on one simulation scene under hardware miscalibration (many-to-many). All methods are trained on the mask set $\mathcal{M}$ and tested by one unseen mask. Density curves computed on chosen patches are compared to analysis the spectra.}\vspace{-0.4cm}
\label{fig: random_mask_simu_2}
\end{figure*}

\subsection{HSI Reconstruction Performance}
\label{subsec: performance}
We evaluate our method under different settings on both simulation and real data. More visualizations are provided in the supplementary material.

\textbf{Miscalibration (many-to-many).} Training the deep reconstruction networks with a mask ensemble strategy could improve the generalization ability and might be a potential solution to deploy a single network across different hardware, such as training TSA-Net~\cite{meng2020end}, GSM~\cite{huang2021deep}, and SRN~\cite{wang2021new} on a mask set. However, as shown in Table~\ref{Tab: nton_miscalibration} and Table~\ref{Tab: psnr_ssim_single_trn}, these methods generally suffer from a clear performance degradation under miscalibration compared with their well-calibrated performance. For example, TSA-Net~\cite{meng2020end} drops around 8db on PSNR. Benefiting from modeling mask uncertainty, our approach achieves high-fidelity results (over 33dB) on both cases, with only a 0.2db drop.

Fig.~\ref{fig: random_mask_simu_2} compares the reconstructive results of different methods perceptually. Our method retrieves more details at different spectral channels. We randomly choose two regions, corresponding to two colors separately (patch \texttt{a} and \texttt{b} in RGB reference), to analyze results regarding spectra. Density curves of spectral correlations to ground truth are compared in the bottom left of Fig.~\ref{fig: random_mask_simu_2}. 

\begin{table*}[t]
\caption{PSNR(dB)/SSIM by different methods on 10 simulation scenes under the \textbf{many-to-many} hardware miscalibration. All the methods are trained with a mask set $\mathcal{M}$ and tested by random unseen masks. TSA-Net~\cite{meng2020end}, GSM~\cite{huang2021deep}, and SRN~\cite{wang2021new} are obtained with a mask ensemble strategy. We report \texttt{mean}$_{\pm\texttt{std}}$ among 100 testing trials.
}\label{Tab: nton_miscalibration}
\vspace{1mm}
\centering
\resizebox{\textwidth}{!}{
\centering
\begin{tabular}{ccccccccccc} 
	\toprule
	\multirow{2}{*}{Scene} & \multicolumn{2}{c}{TSA-Net~\cite{meng2020end}} & \multicolumn{2}{c}{GSM~\cite{huang2021deep}} & \multicolumn{2}{c}{PnP-DIP$^\dagger$~\cite{meng2021self}} & \multicolumn{2}{c}{SRN~\cite{wang2021new}} & \multicolumn{2}{c}{Ours}\\
 	\cmidrule(lr){2-3} \cmidrule(lr){4-5} \cmidrule(lr){6-7} \cmidrule(lr){8-9} \cmidrule(lr){10-11} 
	& PSNR & SSIM & PSNR & SSIM & PSNR & SSIM & PSNR & SSIM & PSNR & SSIM\\
	\midrule
1&23.45$_{\pm0.29}$&0.6569$_{\pm0.0051}$&31.38$_{\pm0.20}$&0.8826$_{\pm0.0032}$&29.24$_{\pm0.98}$&0.7964$_{\pm0.0532}$&33.26$_{\pm0.16}$&0.9104$_{\pm0.0018}$&\bf{33.99$_{\pm0.14}$}&\bf{0.9258$_{\pm0.0013}$} \\
2&18.52$_{\pm0.12}$&0.5511$_{\pm0.0049}$&25.94$_{\pm0.22}$&0.8570$_{\pm0.0041}$&25.73$_{\pm0.54}$&0.7558$_{\pm0.0117}$&29.86$_{\pm0.23}$&0.8809$_{\pm0.0029}$&\bf{30.49$_{\pm0.17}$}&\bf{0.9002$_{\pm0.0022}$}  \\
3&18.42$_{\pm0.30}$&0.5929$_{\pm0.0127}$&26.11$_{\pm0.20}$&0.8874$_{\pm0.0034}$&29.61$_{\pm0.45}$&0.8541$_{\pm0.0125}$&31.69$_{\pm0.20}$&0.9093$_{\pm0.0020}$&\bf{32.63$_{\pm0.16}$}&\bf{0.9212$_{\pm0.0013}$} \\
4&30.44$_{\pm0.15}$&0.8940$_{\pm0.0043}$&34.72$_{\pm0.35}$&0.9473$_{\pm0.0023}$&38.21$_{\pm0.66}$&0.9280$_{\pm0.0078}$&39.90$_{\pm0.22}$&0.9469$_{\pm0.0012}$&\bf{41.04$_{\pm0.23}$}&\bf{0.9667$_{\pm0.0014}$} \\
5&20.89$_{\pm0.23}$&0.5648$_{\pm0.0077}$&26.15$_{\pm0.24}$&0.8256$_{\pm0.0061}$&28.59$_{\pm0.79}$&0.8481$_{\pm0.0183}$&30.86$_{\pm0.16}$&0.9232$_{\pm0.0019}$&\bf{31.49$_{\pm0.17}$}&\bf{0.9379$_{\pm0.0017}$}\\
6&23.04$_{\pm0.19}$&0.6099$_{\pm0.0060}$&30.97$_{\pm0.29}$&0.9224$_{\pm0.0025}$&29.70$_{\pm0.51}$&0.8484$_{\pm0.0186}$&34.20$_{\pm0.23}$&0.9405$_{\pm0.0014}$&\bf{34.89$_{\pm0.29}$}&\bf{0.9545$_{\pm0.0009}$} \\
7&15.97$_{\pm0.14}$&0.6260$_{\pm0.0042}$&22.58$_{\pm0.24}$&0.8459$_{\pm0.0054}$&27.13$_{\pm0.31}$&0.8666$_{\pm0.0079}$&27.27$_{\pm0.16}$&0.8515$_{\pm0.0026}$&\bf{27.63$_{\pm0.16}$}&\bf{0.8658$_{\pm0.0024}$} \\
8&22.64$_{\pm0.18}$&0.6366$_{\pm0.0066}$&29.76$_{\pm0.22}$&0.9059$_{\pm0.0021}$&28.38$_{\pm0.35}$&0.8325$_{\pm0.0203}$&32.35$_{\pm0.22}$&0.9320$_{\pm0.0015}$&\bf{33.02$_{\pm0.26}$}&\bf{0.9471$_{\pm0.0013}$} \\
9&18.91$_{\pm0.11}$&0.5946$_{\pm0.0083}$&27.23$_{\pm0.11}$&0.8899$_{\pm0.0021}$&33.63$_{\pm0.26}$&0.8779$_{\pm0.0073}$&32.83$_{\pm0.13}$&0.9205$_{\pm0.0016}$&\bf{33.45$_{\pm0.13}$}&\bf{0.9317$_{\pm0.0013}$} \\
10&21.90$_{\pm0.18}$&0.5249$_{\pm0.0110}$&28.05$_{\pm0.21}$&0.8877$_{\pm0.0055}$&27.24$_{\pm0.43}$&0.7957$_{\pm0.0226}$&30.25$_{\pm0.14}$&0.9053$_{\pm0.0019}$&\bf{31.49$_{\pm0.15}$}&\bf{0.9345$_{\pm0.0015}$} \\
    \midrule
\emph{Avg.}&21.42$_{\pm0.07}$&0.6162$_{\pm0.0030}$&28.20$_{\pm0.01}$&0.8852$_{\pm0.0001}$&29.66$_{\pm0.38}$&0.8375$_{\pm0.0093}$&32.24$_{\pm0.10}$&0.9121$_{\pm0.0010}$&\bf{33.02$_{\pm0.01}$}&\bf{0.9285$_{\pm0.0001}$} \\
	\bottomrule
\multicolumn{11}{l}{$^\dagger$PnP-DIP is a mask-free method which reconstructs from measurements encoded by random masks.}
\end{tabular}}\vspace{-2mm}
\end{table*}

\begin{table*}[t]
\caption{PSNR(dB)/SSIM by different methods on 10 simulation scenes under the \textbf{one-to-many} hardware miscalibration. All the methods are trained by a single mask and tested by random unseen masks. We report  \texttt{mean}$_{\pm\texttt{std}}$ among 100 testing trials.}\label{Tab: 1ton_miscalibration}
\vspace{1mm}
\centering
\resizebox{\textwidth}{!}{
\centering
\begin{tabular}{ccccccccccc} 
	\toprule
	\multirow{2}{*}{Scene} & \multicolumn{2}{c}{TSA-Net~\cite{meng2020end}} & \multicolumn{2}{c}{GSM~\cite{huang2021deep}} & \multicolumn{2}{c}{PnP-DIP$^\dagger$~\cite{meng2021self}} & \multicolumn{2}{c}{SRN~\cite{wang2021new}} & \multicolumn{2}{c}{Ours}\\
 	\cmidrule(lr){2-3} \cmidrule(lr){4-5} \cmidrule(lr){6-7} \cmidrule(lr){8-9} \cmidrule(lr){10-11} 
	& PSNR & SSIM & PSNR & SSIM & PSNR & SSIM & PSNR & SSIM & PSNR & SSIM\\
	\midrule
    1&28.49$_{\pm0.58}$&0.8520$_{\pm0.0081}$&28.20$_{\pm0.95}$&0.8553$_{\pm0.0185}$&29.24$_{\pm0.98}$&0.7964$_{\pm0.0532}$&31.24$_{\pm0.77}$&0.8878$_{\pm0.0117}$&\bf{31.72$_{\pm0.76}$}&\bf{0.8939$_{\pm0.0119}$} \\
    2&24.96$_{\pm0.51}$&0.8332$_{\pm0.0064}$&24.46$_{\pm0.96}$&0.8330$_{\pm0.0189}$&25.73$_{\pm0.54}$&0.7558$_{\pm0.0117}$&27.87$_{\pm0.82}$&0.8535$_{\pm0.0131}$&\bf{28.22$_{\pm0.85}$}&\bf{0.8552$_{\pm0.0144}$} \\
    3&26.14$_{\pm0.76}$&0.8829$_{\pm0.0108}$&23.71$_{\pm1.18}$&0.8077$_{\pm0.0221}$&29.61$_{\pm0.45}$&0.8541$_{\pm0.0125}$&28.31$_{\pm0.88}$&\bf{0.8415$_{\pm0.0213}$}&28.77$_{\pm1.13}$&0.8405$_{\pm0.0257}$ \\
    4&35.67$_{\pm0.47}$&0.9427$_{\pm0.0028}$&31.55$_{\pm0.75}$&0.9385$_{\pm0.0074}$&38.21$_{\pm0.66}$&0.9280$_{\pm0.0078}$&\bf{37.93$_{\pm0.72}$}&\bf{0.9476$_{\pm0.0057}$}&37.60$_{\pm0.81}$&0.9447$_{\pm0.0071}$\\
    5&25.40$_{\pm0.59}$&0.8280$_{\pm0.0108}$&24.44$_{\pm0.96}$&0.7744$_{\pm0.0291}$&28.59$_{\pm0.79}$&0.8481$_{\pm0.0183}$&27.99$_{\pm0.79}$&0.8680$_{\pm0.0194}$&\bf{28.58$_{\pm0.79}$}&\bf{0.8746$_{\pm0.0208}$} \\
    6&29.32$_{\pm0.60}$&0.8796$_{\pm0.0047}$&28.28$_{\pm0.92}$&0.9026$_{\pm0.0094}$&29.70$_{\pm0.51}$&0.8484$_{\pm0.0186}$&32.13$_{\pm0.87}$&\bf{0.9344$_{\pm0.0061}$}&\bf{32.72$_{\pm0.79}$}&0.9339$_{\pm0.0061}$ \\
    7&22.80$_{\pm0.65}$&0.8461$_{\pm0.0101}$&21.45$_{\pm0.79}$&0.8147$_{\pm0.0162}$&27.13$_{\pm0.31}$&0.8666$_{\pm0.0079}$&24.84$_{\pm0.73}$&0.7973$_{\pm0.0150}$&\bf{25.15$_{\pm0.76}$}&\bf{0.7935$_{\pm0.0173}$}\\
    8&28.09$_{\pm0.43}$&0.8738$_{\pm0.0043}$&28.08$_{\pm0.76}$&0.9024$_{\pm0.0089}$&28.38$_{\pm0.35}$&0.8325$_{\pm0.0203}$&31.32$_{\pm0.59}$&0.9324$_{\pm0.0043}$&\bf{31.84$_{\pm0.56}$}&\bf{0.9323$_{\pm0.0042}$} \\
    9&27.75$_{\pm0.55}$&0.8865$_{\pm0.0054}$&26.80$_{\pm0.78}$&0.8773$_{\pm0.0144}$&33.63$_{\pm0.26}$&0.8779$_{\pm0.0073}$&31.06$_{\pm0.66}$&0.8997$_{\pm0.0091}$&\bf{31.11$_{\pm0.72}$}&\bf{0.8988$_{\pm0.0104}$}\\
    10&26.05$_{\pm0.48}$&0.8114$_{\pm0.0072}$&26.40$_{\pm0.77}$&0.8771$_{\pm0.0124}$&27.24$_{\pm0.43}$&0.7957$_{\pm0.0226}$&29.01$_{\pm0.61}$&0.9028$_{\pm0.0092}$&\bf{29.50$_{\pm0.68}$}&\bf{0.9030$_{\pm0.0098}$} \\
    \midrule
    \emph{Avg.}&27.47$_{\pm0.46}$&0.8636$_{\pm0.0060}$&26.34$_{\pm0.06}$&0.8582$_{\pm0.0012}$&29.66$_{\pm0.38}$&0.8375$_{\pm0.0093}$&30.17$_{\pm0.63}$&0.8865$_{\pm0.0108}$&\bf{30.60$_{\pm0.08}$}&\bf{0.8881$_{\pm0.0013}$}\\
	\bottomrule
	\multicolumn{11}{l}{$^\dagger$PnP-DIP is a mask-free method which reconstructs from measurements encoded by random masks.}
\end{tabular}}
\vspace{-2mm}
\end{table*}


\textbf{Miscalibration (one-to-many).} 
In Table~\ref{Tab: 1ton_miscalibration}, all the methods are trained on a single mask and tested on multiple unseen masks. We pose this setting to further demonstrate the hardware miscalibration challenge. As can be seen, except for the mask-free method PnP-DIP, the methods usually experience large performance difference compared with using mask ensemble in Table~\ref{Tab: nton_miscalibration} and the well-calibrated case in Table~\ref{Tab: psnr_ssim_single_trn}. This observation supports the motivation of modeling mask uncertainty -- 1) simply using mask ensemble may aggravate the miscalibration (TSA-Net using ensemble performs even worse) and 2) the model trained with a single mask cannot be effectively deployed in different hardware.

\textbf{Same mask (one-to-one).} Table~\ref{Tab: psnr_ssim_single_trn} reports the well-calibrated performance for all the methods, \emph{i.e.}, training/testing models on the same real mask. While our approach is specially designed for training with multiple masks, it still consistently outperforms all the competitors by leveraging bilevel optimization.

\begin{figure*}[t] 
\centering 
\includegraphics[width=0.975\textwidth]{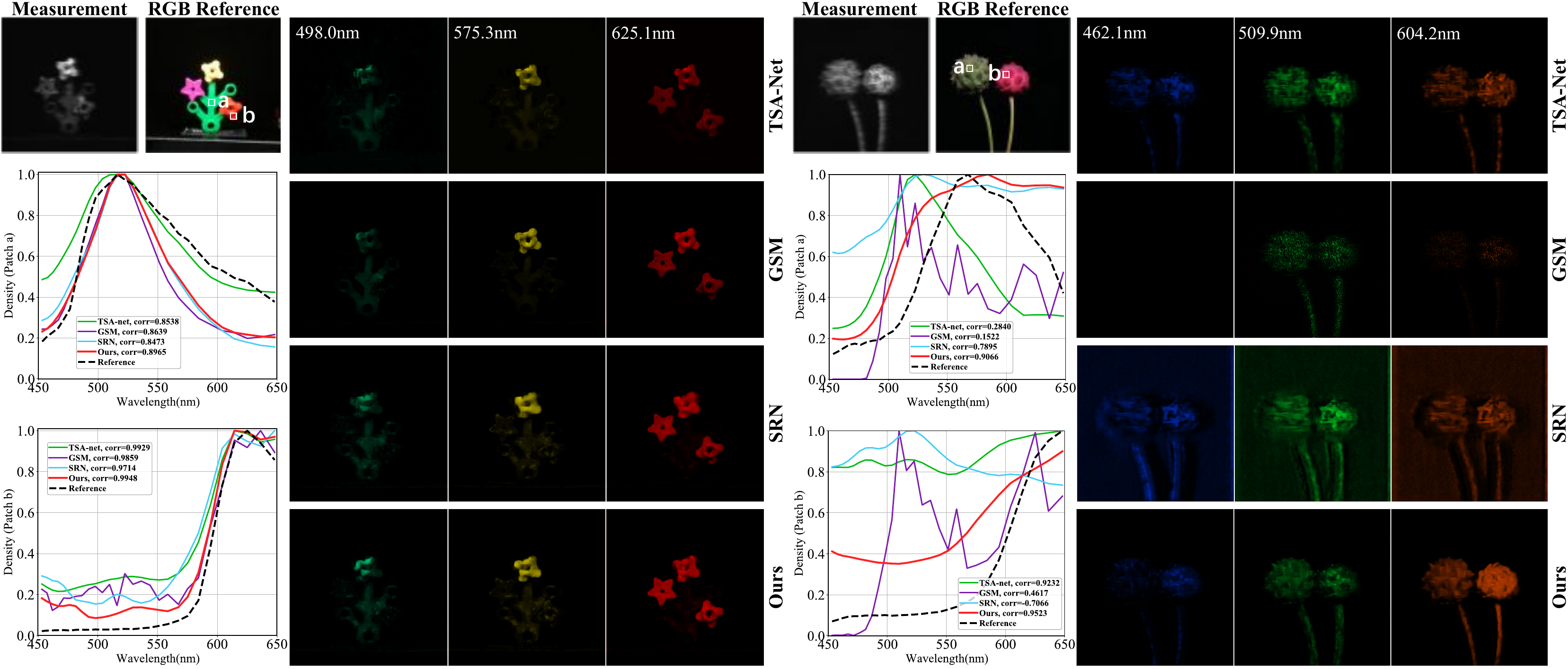}\vspace{-0.2cm}
\caption{Real HSI reconstruction. \textbf{\textit{Left}}: same mask (one-to-one) reconstruction, i.e., all methods are trained and tested on the same 660$\times$660 real mask. \textbf{\textit{Right}}: miscalibration (one-to-many) setting, i.e., all methods are trained on a single mask and tested by unseen masks (Here we adopt another 660$\times$660 real mask). Density curves computed on chosen patches are compared to analysis the spectra.} 
\label{fig: random_mask_real_1}
\vspace{-5mm}
\end{figure*}

\textbf{Results on real data.}
Fig.~\ref{fig: random_mask_real_1} visualizes reconstruction results on the real dataset, where the \emph{left} provides results using the same mask and the \emph{right} is under the one-to-many miscalibration setting. For the same mask, the proposed method is supposed to perform comparably. For the miscalibration, we train all the models on a single real mask provided in~\cite{meng2020end} and test them on the other one~\cite{meng2020snapshot}. As shown, the proposed method produces plausible results and improves over other methods visually. Besides, spectral fidelity is also demonstrated.

\begin{table*}[t]
\caption{PSNR (dB) and SSIM values by different algorithms on the simulation dataset under the well-calibrated setting (training/test on the \textbf{\textit{same mask}}). We adopt the same 256$\times$256 real mask provided in previuous works~\cite{meng2020end,huang2021deep} for a fair comparison.}\label{Tab: psnr_ssim_single_trn}
\vspace{1mm}
\centering
\resizebox{0.98\textwidth}{!}{
\setlength{\tabcolsep}{0.7mm}
\centering
\begin{tabular}{ccccccccccccccc} 
	\toprule
	\multirow{2}{*}{Scene} & \multicolumn{2}{c}{$\lambda$-net~\cite{miao2019net}} & \multicolumn{2}{c}{HSSP~\cite{wang2019hyperspectral}} & \multicolumn{2}{c}{TSA-Net~\cite{meng2020end}} 
	& \multicolumn{2}{c}{GSM~\cite{huang2021deep}} & \multicolumn{2}{c}{PnP-DIP~\cite{meng2021self}} & \multicolumn{2}{c}{SRN~\cite{wang2021new}} & \multicolumn{2}{c}{Ours} \\
 	\cmidrule(lr){2-3} \cmidrule(lr){4-5} \cmidrule(lr){6-7} \cmidrule(lr){8-9}
 	\cmidrule(lr){10-11} \cmidrule(lr){12-13} \cmidrule(lr){14-15} 
	& PSNR & SSIM & PSNR & SSIM & PSNR & SSIM & PSNR & SSIM & PSNR & SSIM & PSNR & SSIM & PSNR & SSIM \\
	\midrule
		1  & 30.82& 0.8492& 31.07& 0.8577& 31.26& 0.8920& 32.38& 0.9152& 31.99& 0.8633& 34.13& 0.9260& \bf{34.19}  &\bf{0.9292}\\
		2  & 26.30& 0.8054& 26.30& 0.8422& 26.88& 0.8583& 27.56& 0.8977& 26.56& 0.7603& 30.60& 0.8985& \bf{31.04}& \bf{0.9014}\\
		3  &29.42& 0.8696& 29.00& 0.8231& 30.03& 0.9145& 29.02& 0.9251& 30.06& 0.8596& 32.87& 0.9221& \bf{32.93}& \bf{0.9224}\\
		4  &37.37& 0.9338& 38.24& 0.9018& 39.90& 0.9528& 36.37& 0.9636& 38.99& 0.9303& \bf{41.27}& \bf{0.9687}& 40.71& 0.9672\\
	    5  &27.84& 0.8166& 27.98& 0.8084& 28.89& 0.8835& 28.56& 0.8820& 29.09& 0.8490& 31.66& 0.9376& \bf{31.83}& \bf{0.9415}\\
		6  &30.69& 0.8527& 29.16& 0.8766& 31.30& 0.9076& 32.49& 0.9372& 29.68& 0.8481& 35.14 &\bf{0.9561}& \bf{35.14}& 0.9543\\
		7  &24.20& 0.8062& 24.11& 0.8236& 25.16& 0.8782& 25.19& 0.8860& 27.68& 0.8639& 27.93& \bf{0.8638}& \bf{28.08}& 0.8628\\
		8  &28.86 &0.8307& 27.94& 0.8811& 29.69& 0.8884& 31.06& 0.9234& 29.01& 0.8412& 33.14& \bf{0.9488}& \bf{33.18}& 0.9486\\
		9  &29.32& 0.8258& 29.14& 0.8676& 30.03& 0.8901& 29.40& 0.9110& 33.35& 0.8802& 33.49& 0.9326& \bf{33.50}& \bf{0.9332}\\
		10  &27.66& 0.8163& 26.44& 0.8416& 28.32& 0.8740& 30.74& 0.9247& 27.98& 0.8327& 31.43& \bf{0.9338}& \bf{31.59}& 0.9311\\
		\midrule
		\emph{Avg.} &29.25& 0.8406& 28.93& 0.8524& 30.24& 0.8939& 30.28& 0.9166& 30.44& 0.8529& 33.17& 0.9288& \bf{33.22}& \bf{0.9292}\\
	\bottomrule
\end{tabular}}
\vspace{-0.2cm}
\end{table*}

\subsection{Model Discussion}
\label{subsec: model analysis}

\textbf{Ablation study}. Table~\ref{Tab: ablation studies} compares the performance and complexity of the proposed \texttt{full model} with three ablated models as follows. 
1) The model \texttt{w/o GST} is equivalent to training the reconstruction backbone SRN~\cite{wang2021new} with a mask ensemble strategy. 
2) The model \texttt{w/o Bi-Opt} is implemented by training the proposed method without using the bilevel optimization framework. 
3) In the model \texttt{w/o GCN}, we replace the GCN module in GST with convolutional layers carrying a similar size of parameters. 
The bilevel optimization achieves 0.59dB improvement without overburdening the complexity. The GCN contributes 0.2dB with 0.09G FLOPs increase. Overall, the proposed GST yields 0.8dB improvement with negligible cost (i.e., $+$0.02M $\#$params, $+$1.03G FLOPs, and $+$1.14 days training), and could be used in multiple unseen masks without re-training.

\begin{table}[t]
\footnotesize
\caption{Ablation study and complexity analysis. All the methods are tested on simulation test set under the many-to-many setting with one NVIDIA RTX 3090 GPU. We report the PSNR (dB)/SSIM among 100 testing trials, the total training time, and the test time per sample. PnP-DIP is self-supervised, thus no training is required.}\label{Tab: ablation studies}
\vspace{-1mm}
\begin{center}
\scalebox{0.82}{
\begin{tabular}{lcccccc}
\toprule
    Settings~~ & PSNR~~ & SSIM~~ &    $\#$params (M) & FLOPs (G) & Training (day) & ~~Test (sec.) \\
\midrule
    TSA-Net~\cite{meng2020end}  & 21.42$_{\pm0.07}$~~ & 0.6162$_{\pm0.0030}$~~ & 44.25 & 110.06 & 1.23 & 0.068 \\
    GSM~\cite{huang2021deep}    & 28.20$_{\pm0.01}$~~ & 0.8852$_{\pm0.0001}$~~ & 3.76  & 646.35 & 6.05 & 0.084 \\
    PnP-DIP~\cite{meng2021self} & 29.66$_{\pm0.38}$~~ & 0.8375$_{\pm0.0093}$~~ & 33.85 & 64.26  & --   &  482.78\\
\midrule
    \texttt{w/o GST}         & 32.24$_{\pm0.10}$~~    & 0.9121$_{\pm0.0010}$~~ & 1.25 & 81.84 & 1.14  &  0.061\\
    \texttt{w/o Bi-Opt}      & 32.43$_{\pm0.02}$~~    & 0.9206$_{\pm0.0001}$~~ & 1.27 & 82.87 & 1.83  &  0.061\\
    \texttt{w/o GCN}         & 32.82$_{\pm0.01}$~~    & 0.9262$_{\pm0.0001}$~~ & 1.27 & 82.78 & 1.63  & 0.062\\
\midrule
    Ours (full model)~~ & 33.02$_{\pm0.01}$~~     & 0.9285$_{\pm0.0001}$~~ & 1.27 & 82.87 & 2.56 & 0.062\\
\bottomrule
\end{tabular}}\vspace{-6mm}
\end{center}
\end{table}

\begin{figure}[t] 
\centering 
\includegraphics[width=\textwidth]{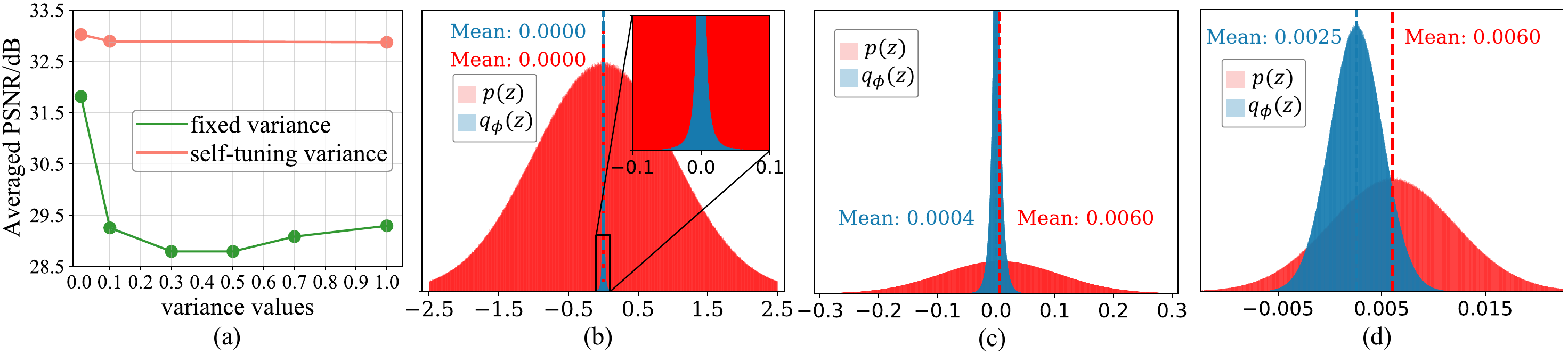}\vspace{-5mm}
\caption{Discussion on self-tuning variance. (a) Performance comparison between self-tuning variance and fixed ones. (b) The standard normal prior $\mathcal{N}(0,1)$. (c) Set the prior as $\mathcal{N}(0.006,0.1)$ by observing real masks. (d) Set the prior as $\mathcal{N}(0.006,0.005)$ by observing real masks and the performance curve in (a).} 
\label{fig: hist_compare}
\vspace{-2mm}
\end{figure}

\begin{figure}[t] 
\centering 
\includegraphics[width=\textwidth]{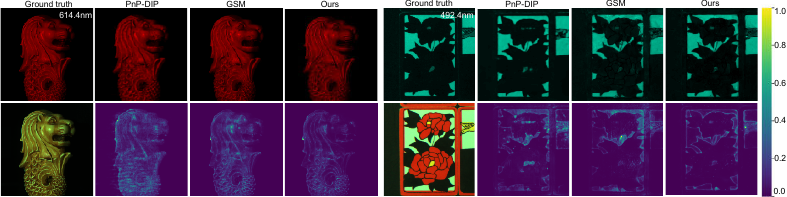}\vspace{-3mm}
\caption{Illustration of epistemic uncertainty induced by multiple masks. For each block, the first row shows the averaged reconstruction results of selected channels given by different methods and the second demonstrates the corresponding epistemic uncertainty.}
\label{fig: pred_var}
\vspace{-4mm}
\end{figure}

\textbf{Complexity comparison}.
In Table~\ref{Tab: ablation studies}, we further compare the complexity of the proposed method with several recent HSI methods. The proposed method possess one of the smallest model size. Besides, our method shows a comparable FLOPs and training time with others. Notably, given $M$ distinct masks, TSA-Net, GSM, and SRN require $M\times$ training time as reported to achieve well-calibrated performance. Instead, the proposed method only needs to be trained one time to provide calibrated reconstructions over multiple unseen masks. 

\textbf{Self-tuning variance under different priors.}
We first validate the effectiveness of the self-tuning variance by comparing with the fix-valued variance, i.e., scalars from 0 to 1. As shown by the green curve in Fig.~\ref{fig: hist_compare} (a), fixed variance only achieves no more than 32dB performance. And the best performance by 0.005 indicates a strong approximation nature to the mask noise.  By comparison, we explore the behaviour of the self-tuning variance upon different noise priors and achieve no less than 32.5dB performance (red curve in Fig.~\ref{fig: hist_compare} (a)). Specifically, we implement the noise prior $p(z)$ by exchanging the standard normal distribution of auxiliary variable $\epsilon$ in Eq.~\eqref{eq: mask define}. We start from $\mathcal{N}(0,1)$, which is so broad that the GST network tries to centralize variational noise and  restricting the randomness as Fig.~\ref{fig: hist_compare} (b) shown. Then, we constraint the variance and approximate the mean value by the minimum of the real mask histogram to emphasize the near-zero noise, proposing $\mathcal{N}(0.006,0.1)$. The corresponding variational noise distribution deviates from the prior as shown in Fig.~\ref{fig: hist_compare} (c), indicating the underlying impact of GST network. Finally, we further combine the previous fixed-variance observation and propose $\mathcal{N}(0.006,0.005)$. As the red curve in Fig.~\ref{fig: hist_compare} (a) indicated, the best reconstruction performance is also obtained. In summary, the proposed method effectively restricts the posited noise prior, leading to the variational noise distribution with a reduced range.

\textbf{From mask uncertainty to epistemic uncertainty}. The hardware mask plays a similar role to model hyperparameter and largely impacts the weights of reconstruction networks. Thus, marginalizing over the mask posterior distribution will induce the epistemic uncertainty (also known as model uncertainty~\cite{Gal2016Uncertainty,NIPS2017_2650d608}) and reflect as pixel-wise variances (the second row in Fig.~\ref{fig: pred_var}) of the reconstruction results over multiple unseen masks. As can be seen, the mask-free method PnP-DIP~\cite{meng2021self} still produces high uncertainties given measurements of the same scene coded by different hardware masks. While employing a deep ensemble strategy could alleviate this issue, such as training GSM~\cite{huang2021deep} with mask ensemble, it lacks an explicit way to quantify mask uncertainty and may lead to unsatisfactory performance (see Table~\ref{Tab: nton_miscalibration}). Differently, the proposed GST method models mask uncertainty by approximating the mask posterior through a variational Bayesian treatment, exhibiting high-fidelity reconstruction result with low epistemic uncertainties across different masks as shown in~Fig.~\ref{fig: pred_var}.

\section{Conclusions}\label{sec:conclusion}
In this work, we have explored a practical hardware miscalibration issue when deploying deep HSI models in real CASSI systems. Our solution is to calibrate a single reconstruction network via modeling mask uncertainty. We proposed a complete variational Bayesian learning treatment upon one possible mask decomposition inspired by observations on real masks. Bearing the objectives of variational mask distribution modeling and HSI retrieval, we introduced and implemented a novel Graph-based Self-Tuning (GST) network that proceeds HSI reconstruction and uncertainty reasoning under a bilevel optimization framework. The proposed method enabled a smoothed distribution and achieved promising performance under two different miscalibration scenarios. We hope the proposed insight will benefit future work in this novel research direction.

%
%
\section*{Appendix}
\appendix

\section{Overview}
\label{sec:overview}
In this appendix, we present additional results and analyses about the proposed method as follows.
\begin{itemize}
    \setlength\itemsep{0em}
    \item \textbf{Reconstruction Backbone SRN}: A detailed introduction to the reconstruction backbone network used in the manuscript (Section~\ref{sec: original backbone}).
    \item \textbf{Alternating Reconstruction Backbones}: More analyses on alternating reconstruction backbones employed in the proposed method (Section~\ref{sec: backbones}).
    \item \textbf{Spectral Fidelity Analysis}: Evaluation on the spectral fidelity of the reconstruction results by the proposed method (Section~\ref{sec: spectral fidelity}).
    \item \textbf{Epistemic Uncertainty Analysis}: More visualization and discussion on epistemic uncertainty of the proposed method (Section~\ref{sec: epistemic uncertianty}).
    \item \textbf{Complementary Ablation Studies}: Ablation studies under one-to-many miscalibration and the same mask setting (one-to-one) (Section~\ref{sec: ablation}).
    \item \textbf{Self-tuning variance Analysis}: More discussions about the proposed self-tuning variance. Specifically, more results for fixed variance are provided. Also, we demonstrate the convergence of $g_{\phi}(m)$, variational noise distributions given distinct noise priors (Section~\ref{sec: model discussion}).
    \item \textbf{Datasets}: More illustrations on the dataset, includes training data, validation data, testing data, and mask set (Section~\ref{sec: data}).
   
\end{itemize}

\section{Reconstruction Backbone: SRN}\label{sec: original backbone}

In the manuscript, we adopt a recent deep reconstruction network, SRN~\cite{wang2021new} as the backbone $f_{\theta}(\cdot)$. Specifically, the network input $x_{in}\in\mathbb{R}^{H\times W\times \Lambda}$ is initialized by the measurement $y\in\mathbb{R}^{H\times (W+\Lambda-1)\times\Lambda}$ and the mask $m\in \mathbb{R}^{H\times W}$ 
\begin{equation}\label{eq: initialization}
    x_{in}[:,:,\lambda]=\texttt{shift}(y)_{\lambda}\odot m,
\end{equation}
where $\odot$ is a Hadamard product and the \texttt{shift} is the reverse operation applied in the forward process (see Eq. ({\color{red}1}) in the manuscript for more details). 

The model is composed of a 1) \textit{main body}, which is simultaneously bridged by a global skip connection, 2) a \textit{head} operation and 3) a \textit{tail} operation, both of which are conducted by a \texttt{CONV-ReLU} structure. Let $x_{head}$ and $x_{body}$ be the output of the \textit{head} operation and the \textit{main body}, respectively. We have
\begin{equation}
    x_{body}= f_{res}^{J}(f_{res}^{J-1}(...(f_{res}^{1}(x_{head}))...)),
\end{equation}
where $J=$ 16 concatenated residual blocks share the same structure, i.e., $f_{res}(x)$ $=x+ (\texttt{CONV}(\texttt{ReLU}(\texttt{CONV}(x))))$. 

\vspace{1mm}

\section{Alternating Reconstruction Backbones}\label{sec: backbones}

The performance and the robustness toward masks of the deep reconstruction networks largely depend on their constructions.
Thus, we validate the effectiveness of the proposed method upon backbones with different architectures. 

\noindent\textbf{SwinIR Backbone.} 
In this supplementary material, we consider transformer architectures as the backbone. Specifically, transformer acquires modeling ability from attention mechanism~\cite{vaswani2017attention}, which has been proved to behave quite differently from the traditional ConvNets~\cite{raghu2021vision}.

\begin{table}[tp]
\scriptsize
\caption{Averaged PSNR(dB)/SSIM of the different models. We consider the miscalibration many-to-many scenario for a fair comparison. For three types of backbones, this is implemented by training upon a mask ensemble and testing on random masks. The \textit{mean} and \textit{std} are obtained upon 100 testing trials.}\label{Tab: different backbones}
\begin{center}
\resizebox{0.6\textwidth}{!}{
\begin{tabular}{lcc}
\toprule
    Models & PSNR (dB) & SSIM \\
\midrule
    SRN~\cite{wang2021new}              & 32.24$_{\pm0.10}$ & 0.9121$_{\pm0.0010}$ \\
    Spectral ViT~\cite{ali2021xcit}     & 31.62$_{\pm0.09}$ & 0.9282$_{\pm0.0010}$ \\
    SwinIR~\cite{liang2021swinir}       & 33.49$_{\pm0.10}$ & 0.9501$_{\pm0.0010}$ \\
\midrule
    SRN$+$GST (Ours)                    & \textbf{33.02$_{\pm0.01}$} & \textbf{0.9285$_{\pm0.0001}$} \\
    Spectral ViT$+$GST (Ours)~~~~~~     & \textbf{32.15$_{\pm0.01}$} & \textbf{0.9330$_{\pm0.0001}$} \\
    SwinIR$+$GST (Ours)                 & \textbf{34.15$_{\pm0.01}$} & \textbf{0.9548$_{\pm0.0001}$} \\
\bottomrule
\end{tabular}}
\end{center}
\vspace{-5mm}
\end{table}

Given the initialized input $x_{in}$ by Eq.~\eqref{eq: initialization}, we firstly implement the backbone by Swin transformer structure~\cite{liu2021swin}, which computes the spatial self-attention. It is composed of three modules: (1) shallow feature extraction by a \texttt{CONV3$\times$3} layer, i.e., $x_{\texttt{SF}} = \texttt{CONV}(x_{in})$, (2) deep feature extraction module consisting of $K$ concatenated residual Swin transformer blocks, i.e, $x_{\texttt{DF}}=f_{\texttt{DF}}(x_{\texttt{SF}})$ where $f_{\texttt{DF}}(\cdot)= f_{\texttt{RSTB}}^{K}(f_{\texttt{RSTB}}^{K-1}(...(f_{\texttt{RSTB}}^{1}(\cdot))...))$, and (3) a reconstruction module by a \texttt{CONV3$\times$3} layer, i.e., $\widehat{x} = \texttt{CONV}(x_{\texttt{DF}})$. 

For each residual Swin transformer block $f_{\texttt{RSTB}}(\cdot)$, we have $L$ Swin transformer layers, which conducts window-based MSA and MLP
\begin{equation}
   x = f_{\texttt{W-MSA}}(f_{\texttt{LN}}(x)) + x, ~~~
   x = f_{\texttt{MLP}}(f_{\texttt{LN}}(x)) + x,
\end{equation}
where the details of the $f_{\texttt{W-MSA}}(\cdot)$, $f_{\texttt{MLP}}(\cdot)$, and $f_{\texttt{LN}}(\cdot)$ could be found in~\cite{liu2021swin}. In the experiment, we set the $K=4$, $L=6$. For all the blocks, we let the embedding dimension to be 60 and number of heads to be 6.\footnote{For implementation please refer to \url{https://github.com/JingyunLiang/SwinIR}.}

\vspace{1mm}
\noindent \textbf{Spectral ViT.} We also provide another type of vision transformer, which exchanges the previous spatial self-attention with the spectral self-attention. Specifically, it treats the feature map of each embedding channel as a token. Given the query $\textbf{Q}$, key $\textbf{K}$ and value $\textbf{V}$, we have output $\textbf{X}$
\begin{equation}
    \textbf{X}=\textbf{V}\texttt{Attn}(\textbf{K},\textbf{Q}), ~~where~~ \texttt{Attn}(\textbf{K},\textbf{Q})=\texttt{softmax}(\textbf{K}^{T}\textbf{Q}/\delta),
\end{equation}
where $\delta$ denotes a learnable scalar. For more details please refer to~\cite{ali2021xcit}. 

\vspace{1mm}
\noindent\textbf{Comparison.} 
We summarize the performance of different backbones under  miscalibration scenario many-to-many in Table~\ref{Tab: different backbones}. 
For detailed illustration of the miscalibration setting, please refer to manuscript Section {\color{red}4.1}. Notably, the integration of the backbone into our method is implemented by  training the full model upon a mask dataset $\mathcal{M}$ in a bilevel optimization framework. 

By comparison, one can draw the following conclusions.
(1) For the metric comparison, our method brings performance gain $\Delta\texttt{PSNR}=$ 0.78dB, 0.53dB, and 0.66dB, respectively, for different backbones.
(2) The proposed method enables high-fidelity reconstruction with the highest confidence. Specifically,  in \texttt{many-to-many} case, $\texttt{PSNR}_{\textit{std}}>0.1$dB. Our method achieves 10$\times$ the randomness control, indicating a better epistemic modeling capacity (Please see Section~\ref{sec: epistemic uncertianty}).

\section{Spectral Fidelity.}\label{sec: spectral fidelity} 
In this work, we adopt two methods to demonstrate the spectral fidelity of the reconstruction results.

\begin{figure*}[t]
\scriptsize
\centering
\begin{tabular}{cc}
\begin{adjustbox}{valign=t}
\begin{tabular}{c}
\begin{adjustbox}{valign=t}
\begin{tabular}{cccccccccc}
\includegraphics[width=0.08\textwidth]{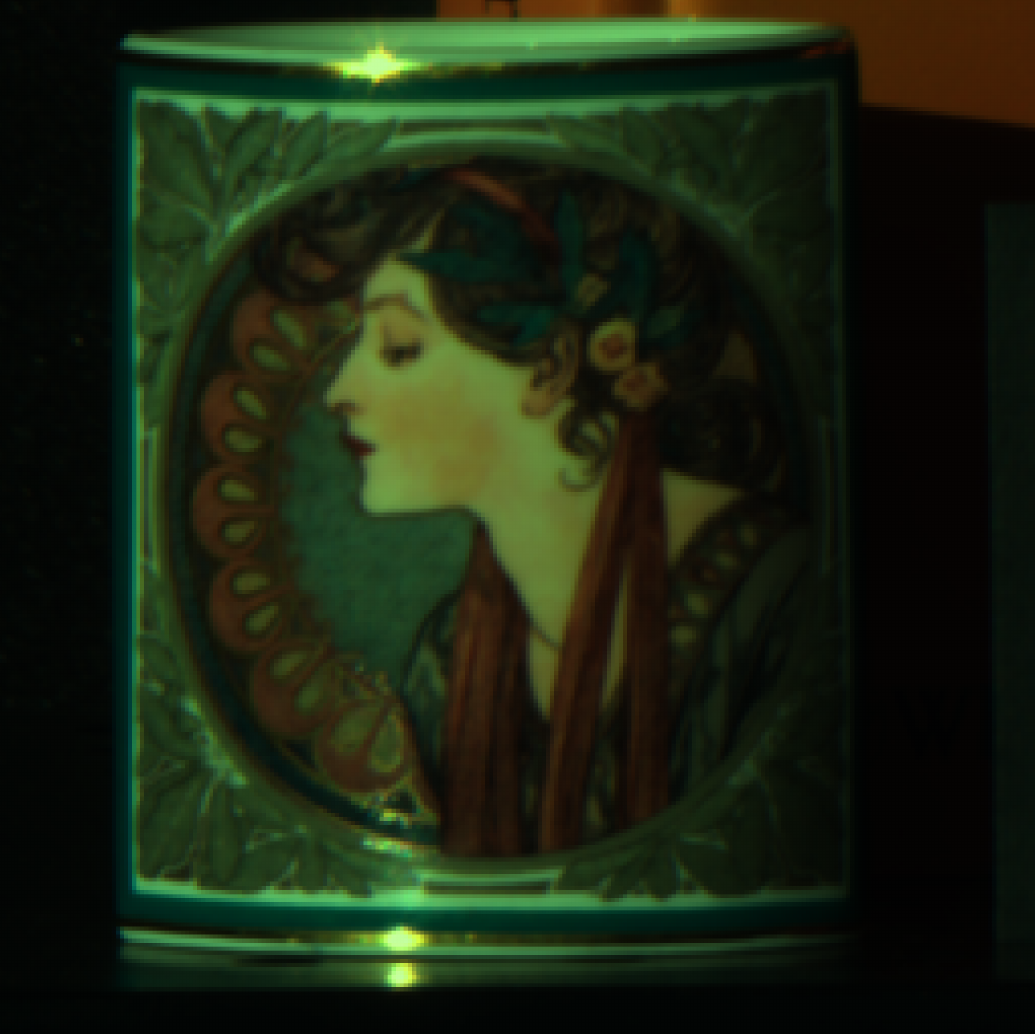}\hspace{-2.3mm}  &
\includegraphics[width=0.08\textwidth]{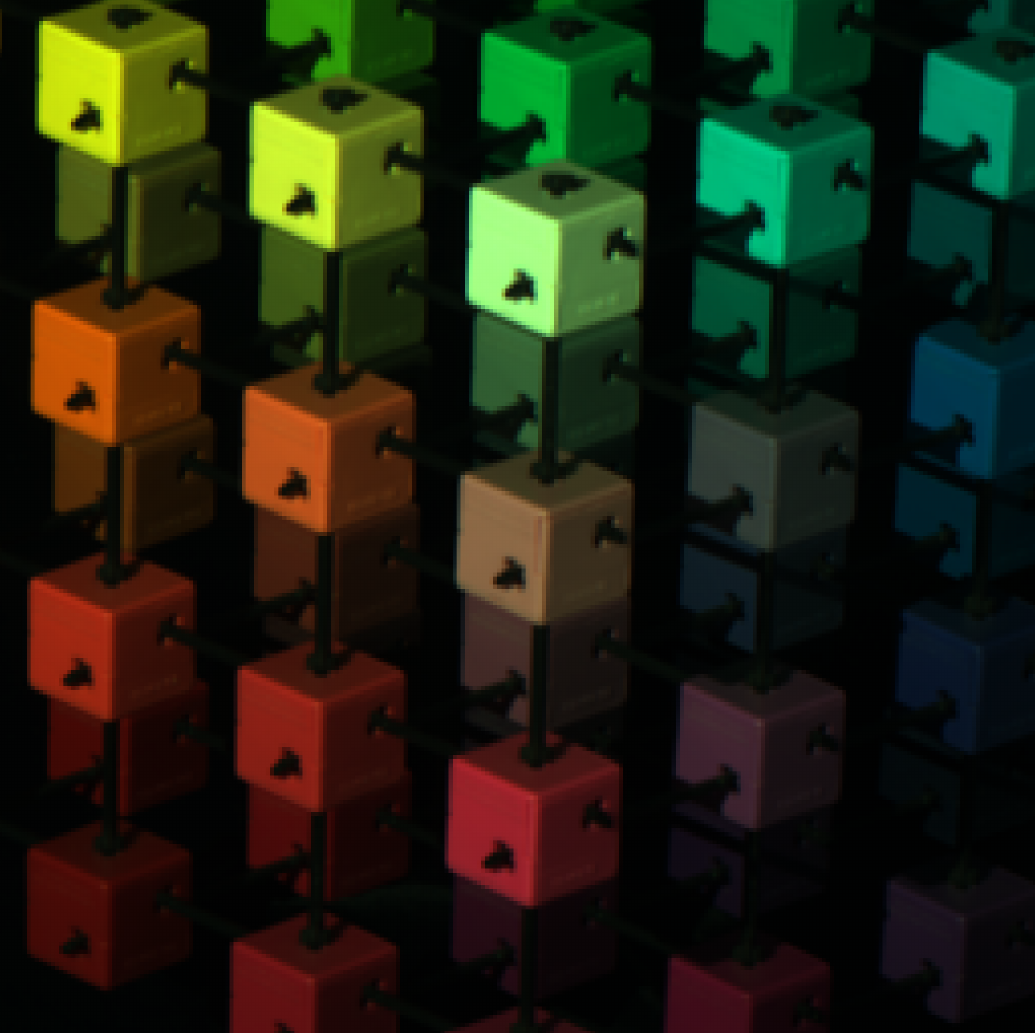}\hspace{-2.3mm}  &
\includegraphics[width=0.08\textwidth]{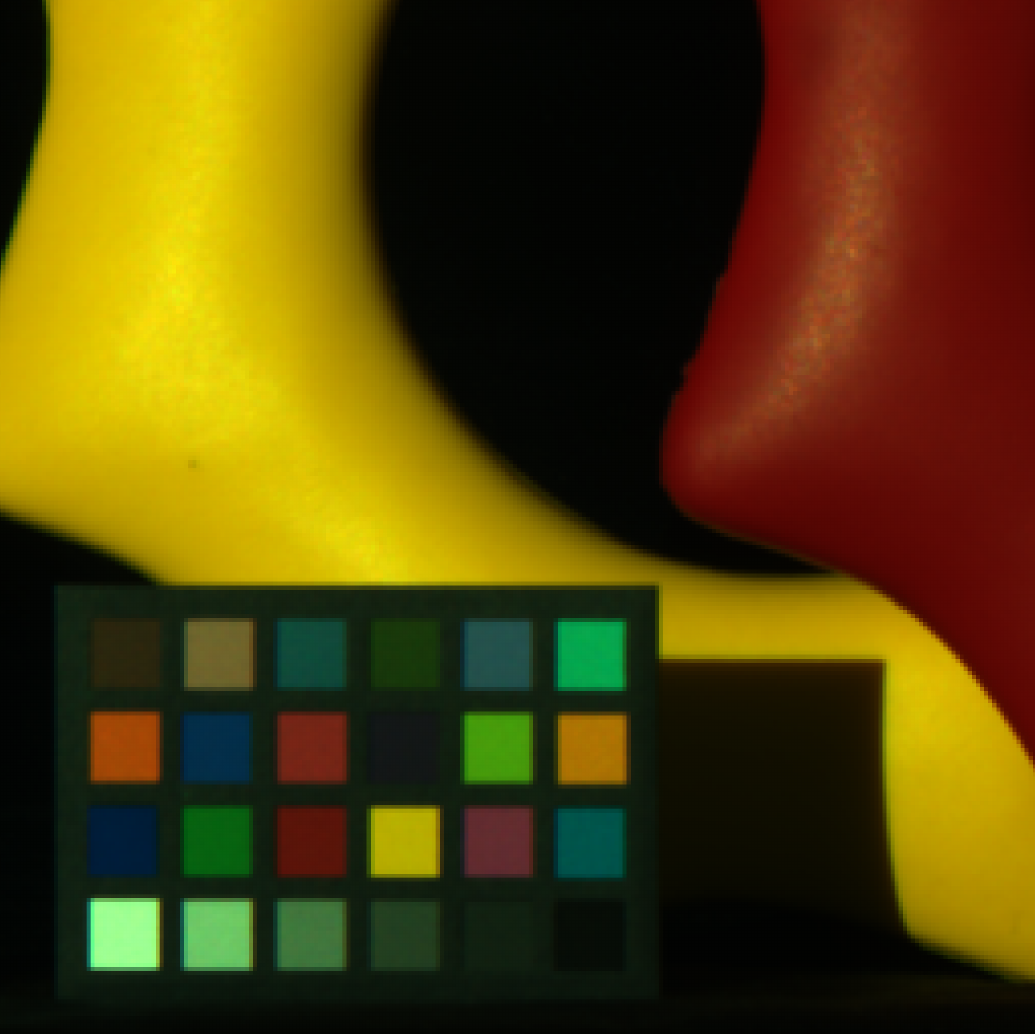}\hspace{-2.3mm}  &
\includegraphics[width=0.08\textwidth]{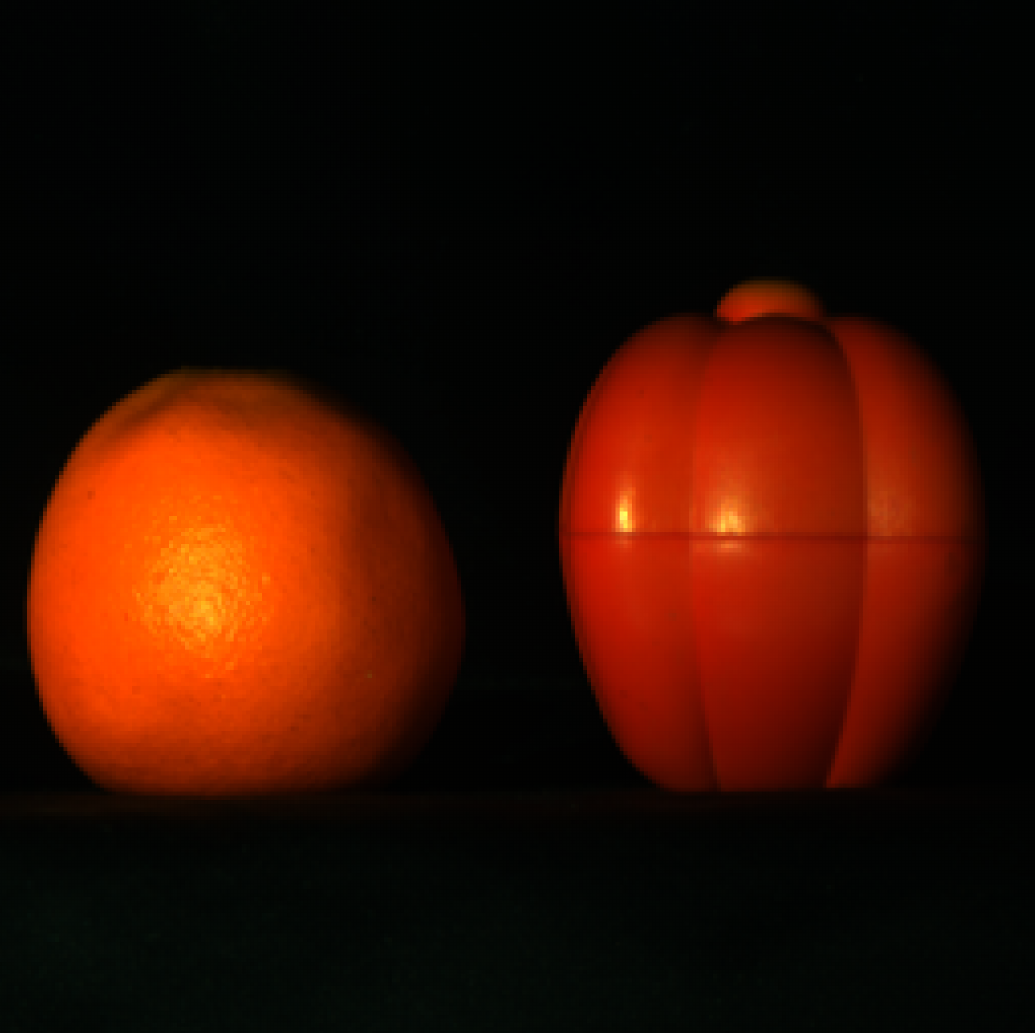}\hspace{-2.3mm}  &
\includegraphics[width=0.08\textwidth]{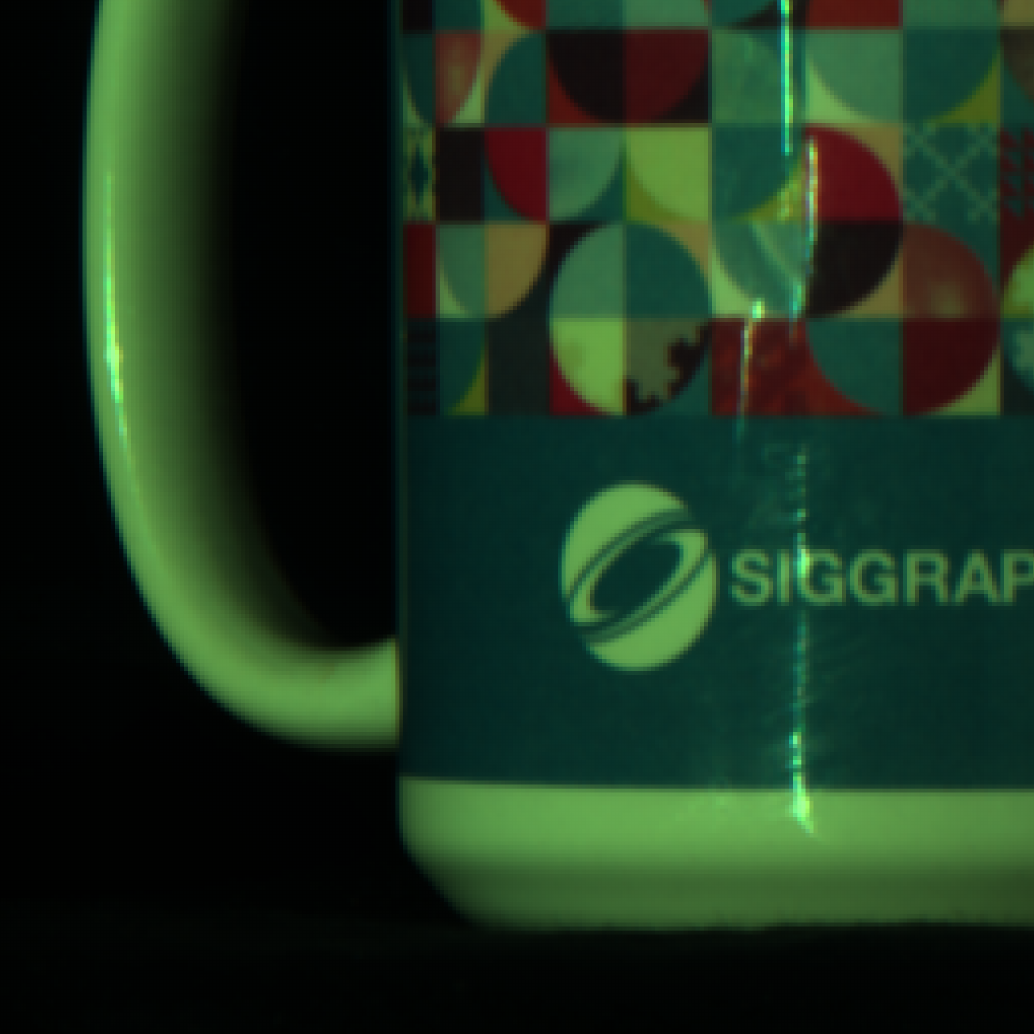}\hspace{-2.3mm}  &
\includegraphics[width=0.08\textwidth]{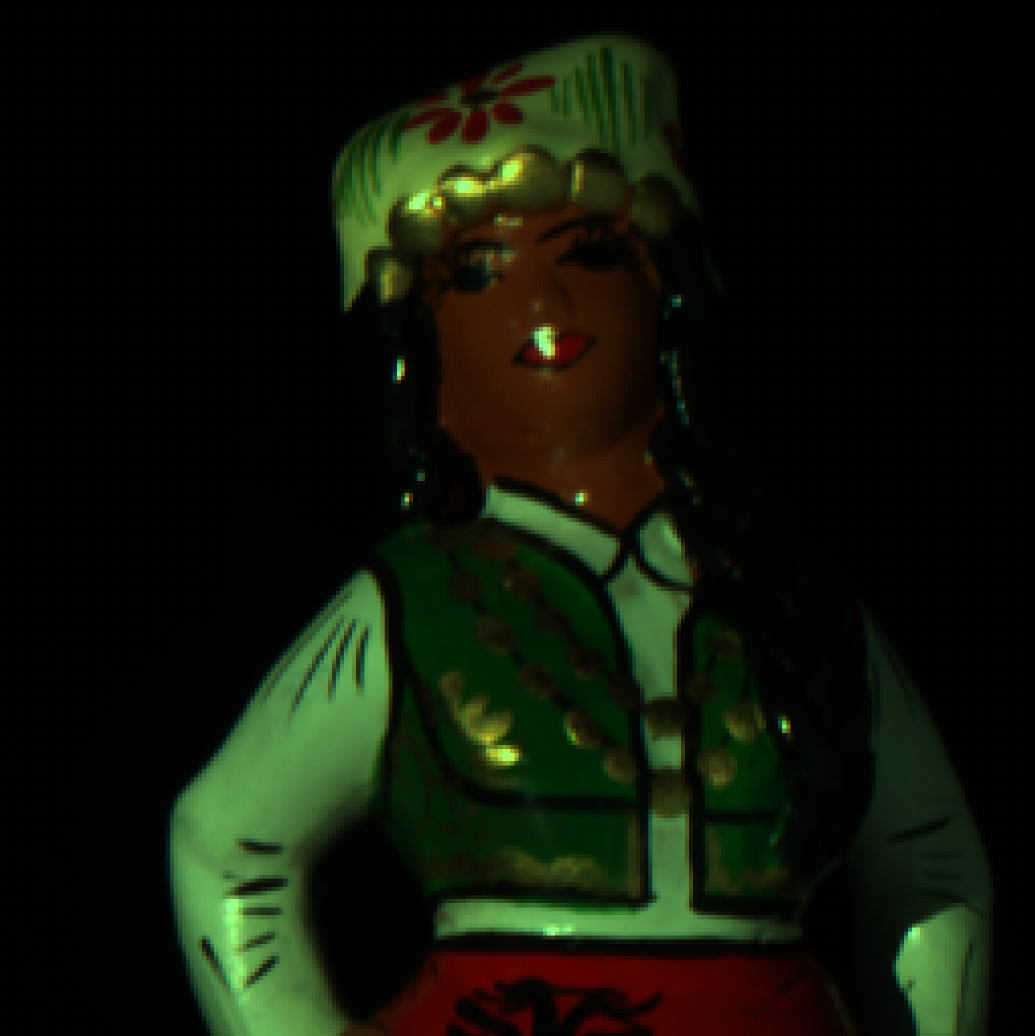}\hspace{-2.3mm}  &
\includegraphics[width=0.08\textwidth]{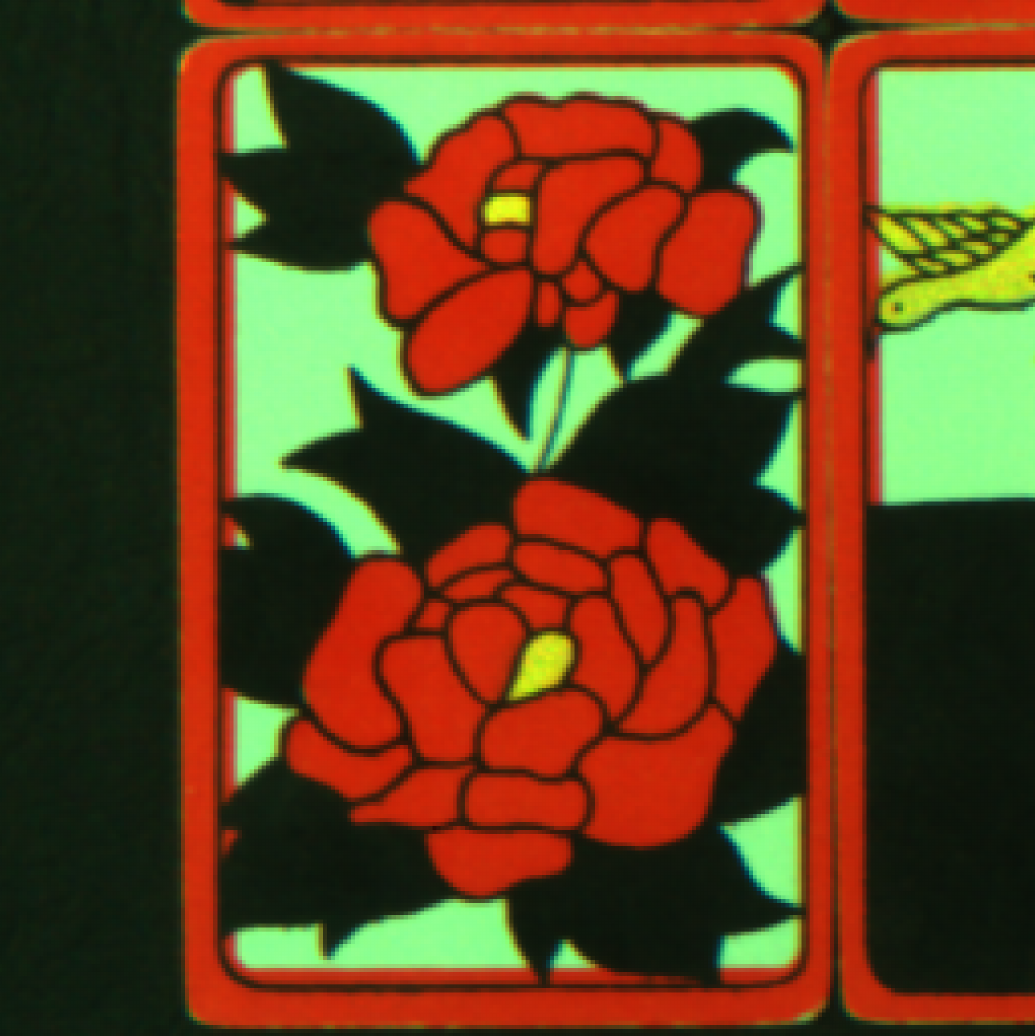}\hspace{-2.3mm}  &
\includegraphics[width=0.08\textwidth]{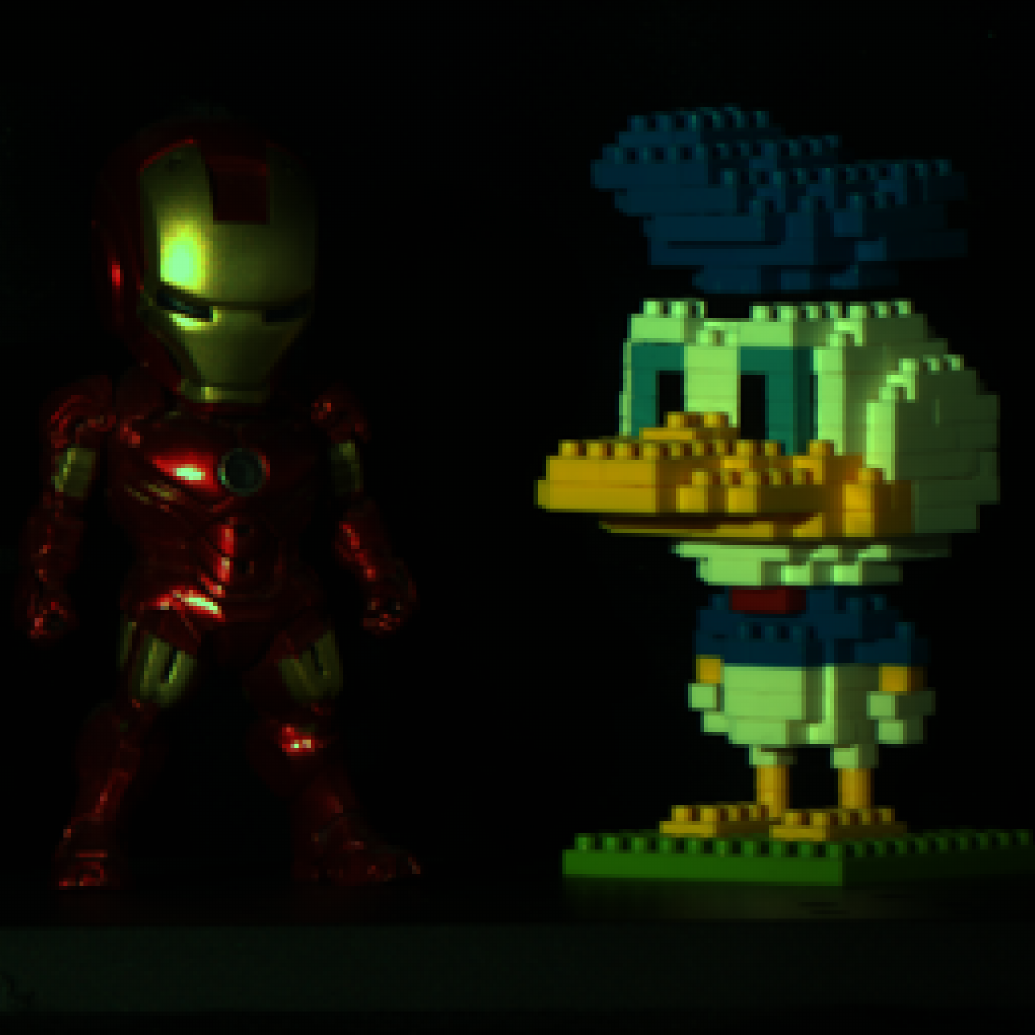}\hspace{-2.3mm}  &
\includegraphics[width=0.08\textwidth]{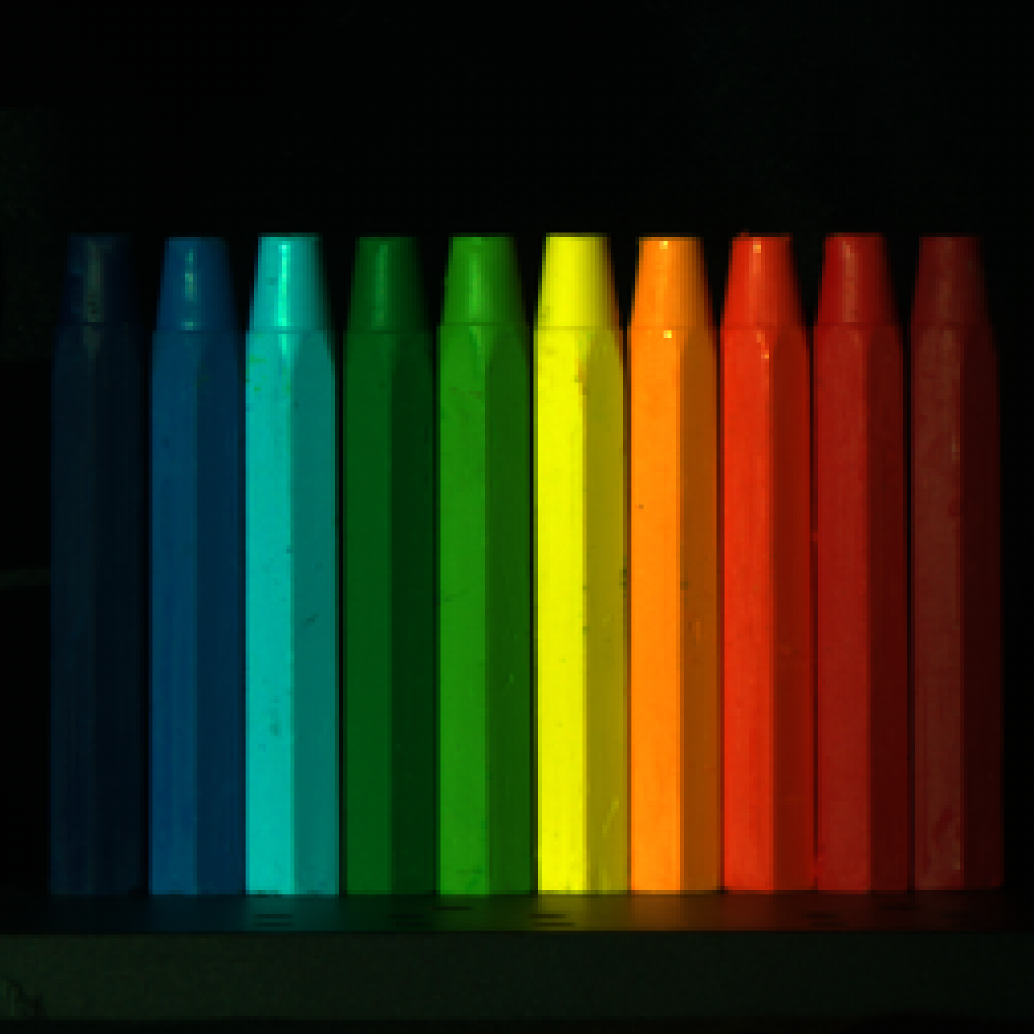}\hspace{-2.3mm}  &
\includegraphics[width=0.08\textwidth]{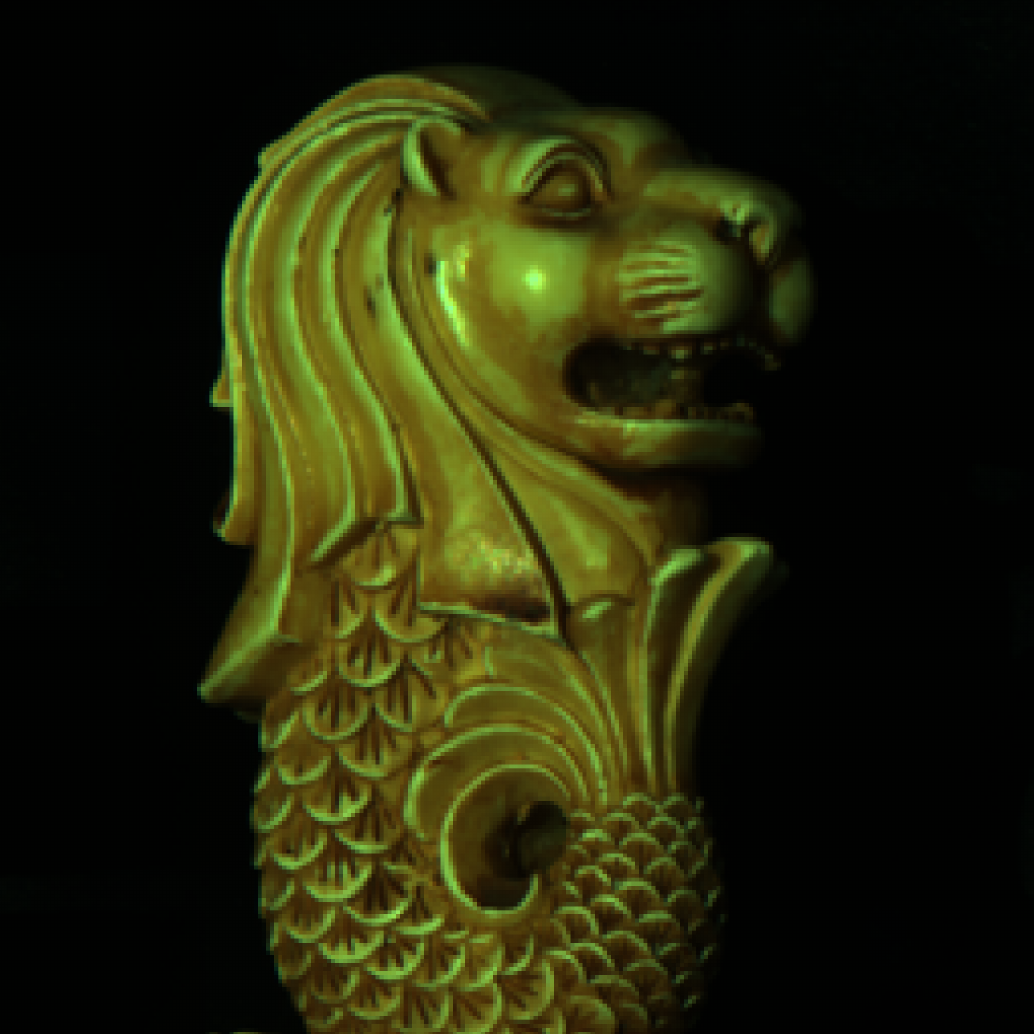}\hspace{-2.3mm}
\end{tabular}
\end{adjustbox}
\\
\begin{adjustbox}{valign=t}
\begin{tabular}{cccccccccc}
\includegraphics[width=0.08\textwidth]{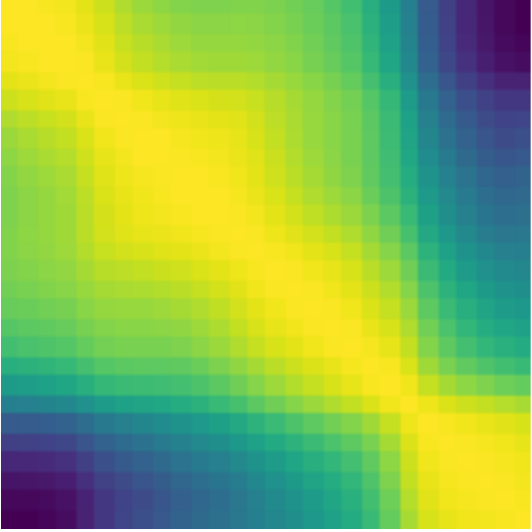}\hspace{-2.3mm}  &
\includegraphics[width=0.08\textwidth]{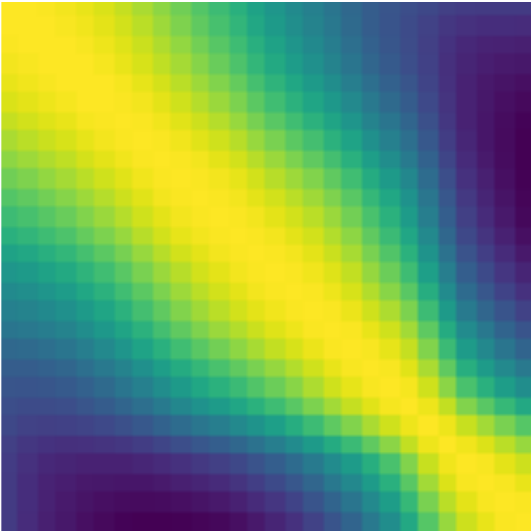}\hspace{-2.3mm}   &
\includegraphics[width=0.08\textwidth]{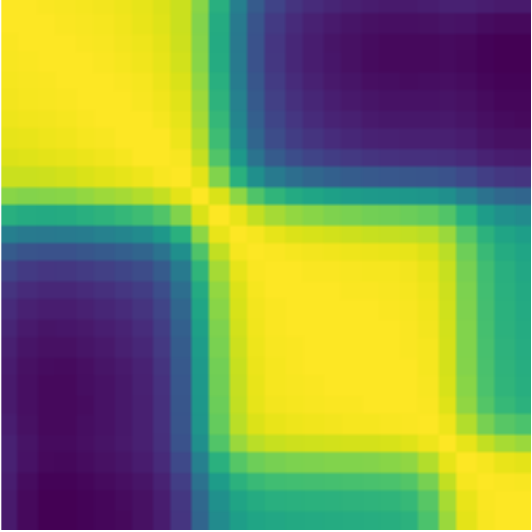}\hspace{-2.3mm}    &
\includegraphics[width=0.08\textwidth]{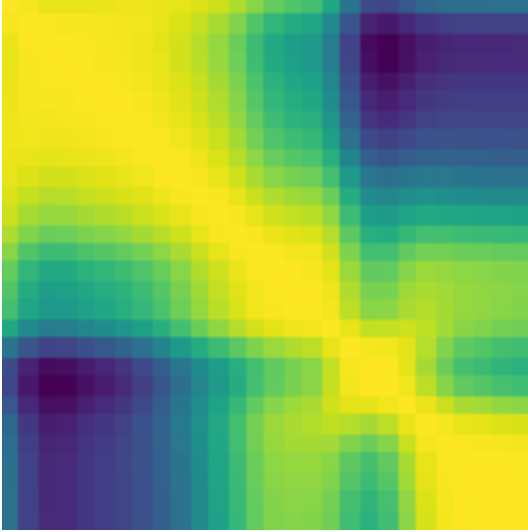}\hspace{-2.3mm} &
\includegraphics[width=0.08\textwidth]{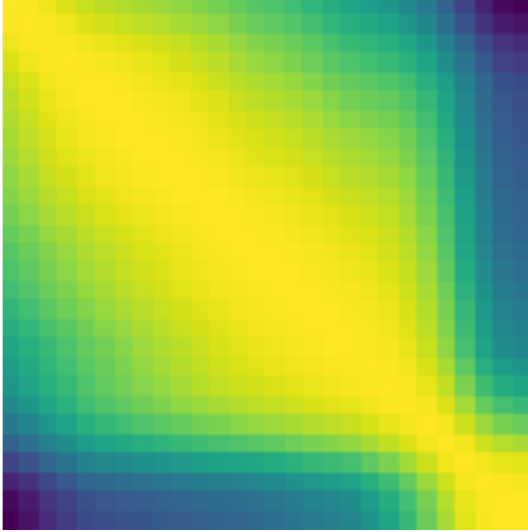}\hspace{-2.3mm}    &
\includegraphics[width=0.08\textwidth]{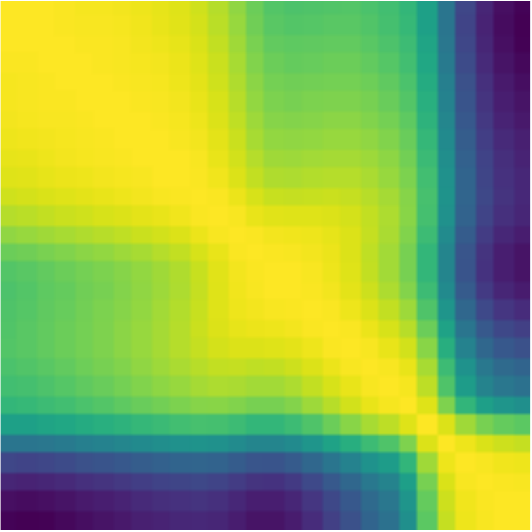}\hspace{-2.3mm}  &
\includegraphics[width=0.08\textwidth]{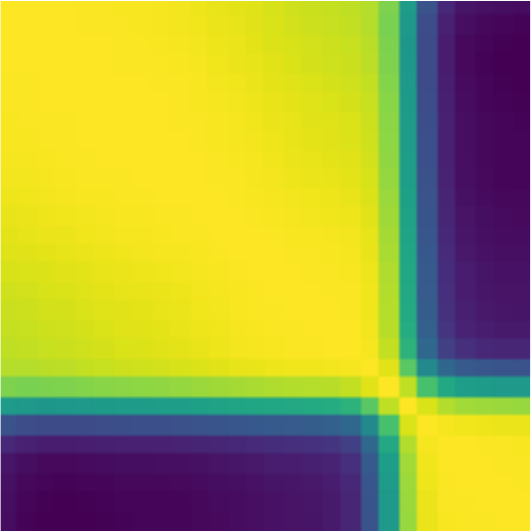}\hspace{-2.3mm} &
\includegraphics[width=0.08\textwidth]{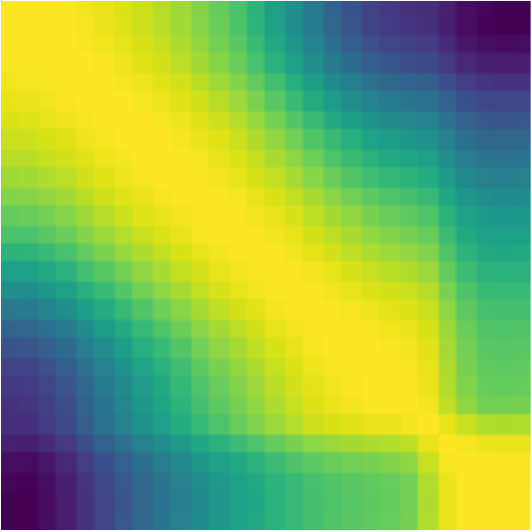}\hspace{-2.3mm}   &
\includegraphics[width=0.08\textwidth]{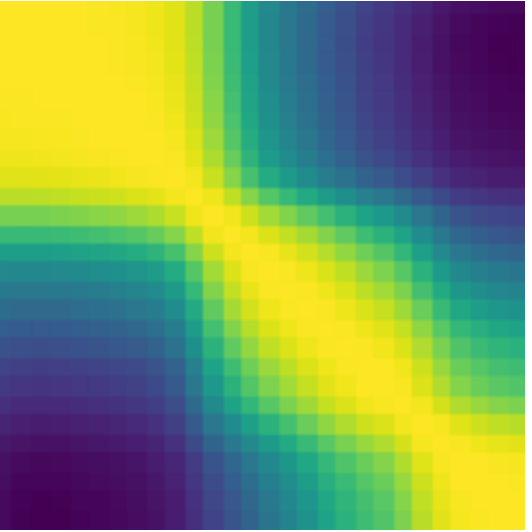}\hspace{-2.3mm}   &
\includegraphics[width=0.08\textwidth]{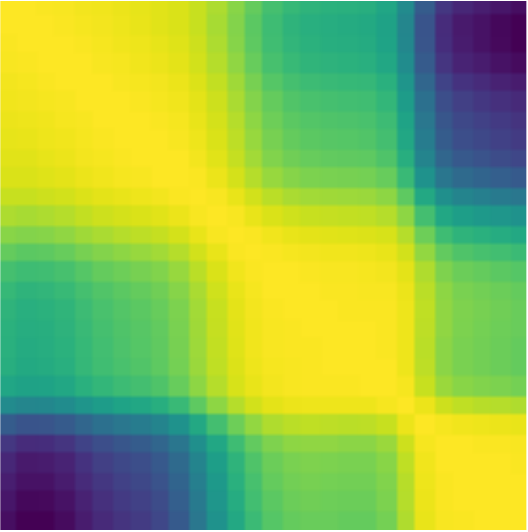}\hspace{-2.3mm}  
\end{tabular}
\end{adjustbox}
\\
\begin{adjustbox}{valign=t}
\begin{tabular}{cccccccccc}
\includegraphics[width=0.08\textwidth]{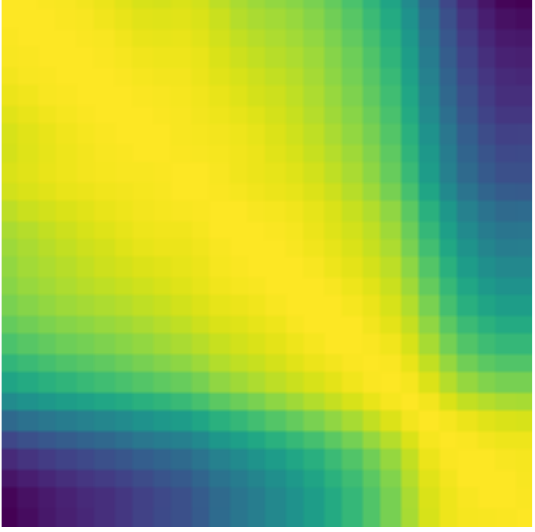}\hspace{-2.3mm}  &
\includegraphics[width=0.08\textwidth]{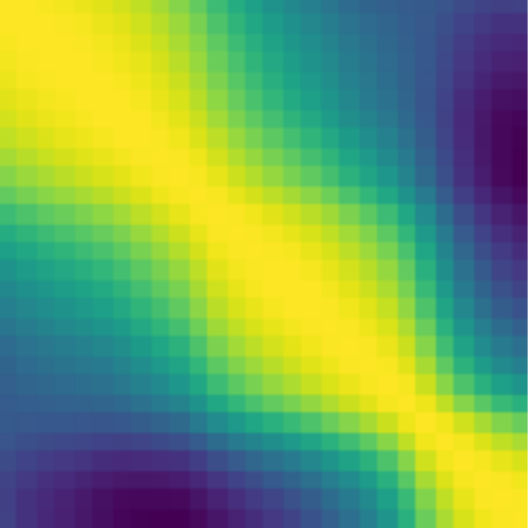}\hspace{-2.3mm}   &
\includegraphics[width=0.08\textwidth]{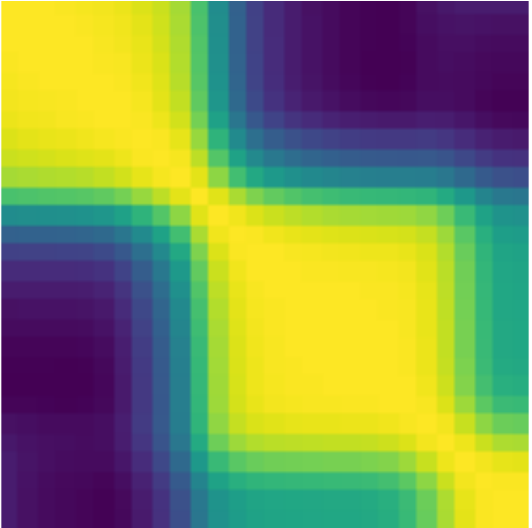}\hspace{-2.3mm}    &
\includegraphics[width=0.08\textwidth]{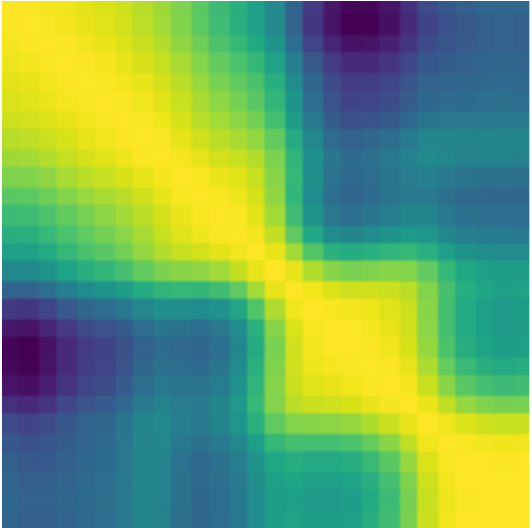}\hspace{-2.3mm} &
\includegraphics[width=0.08\textwidth]{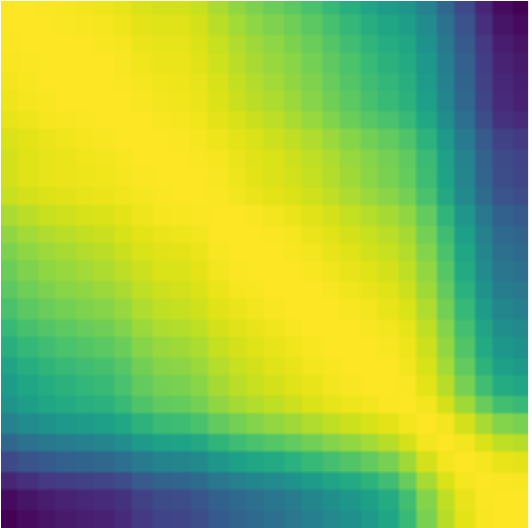}\hspace{-2.3mm}    &
\includegraphics[width=0.08\textwidth]{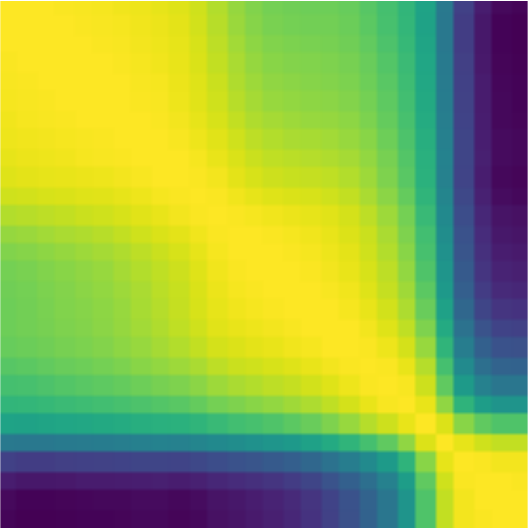}\hspace{-2.3mm}  &
\includegraphics[width=0.08\textwidth]{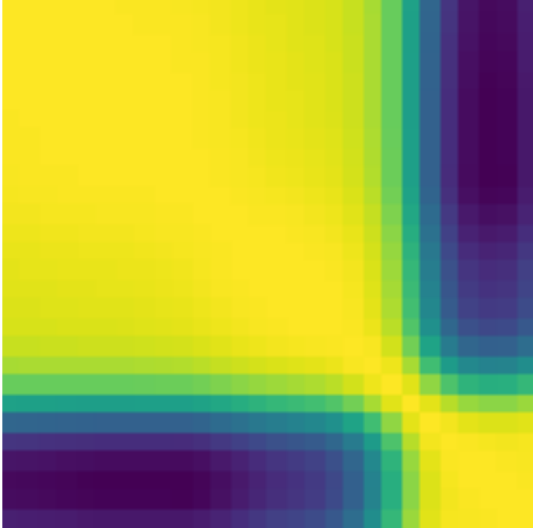}\hspace{-2.3mm} &
\includegraphics[width=0.08\textwidth]{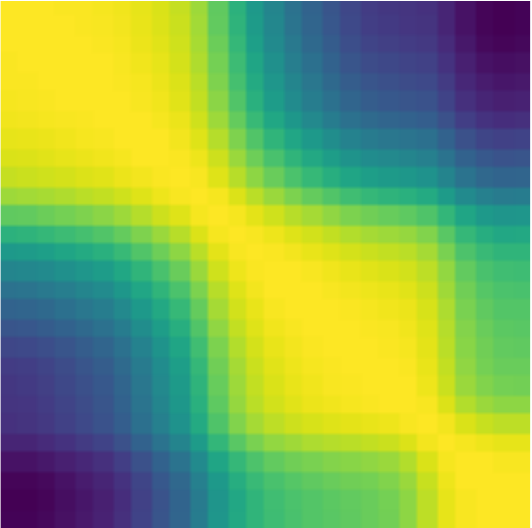}\hspace{-2.3mm}   &
\includegraphics[width=0.08\textwidth]{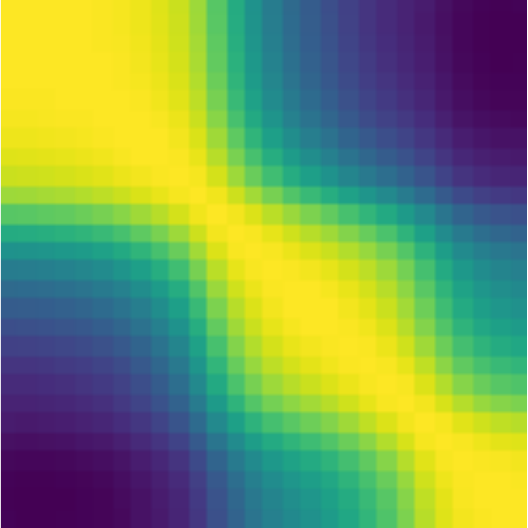}\hspace{-2.3mm}   &
\includegraphics[width=0.08\textwidth]{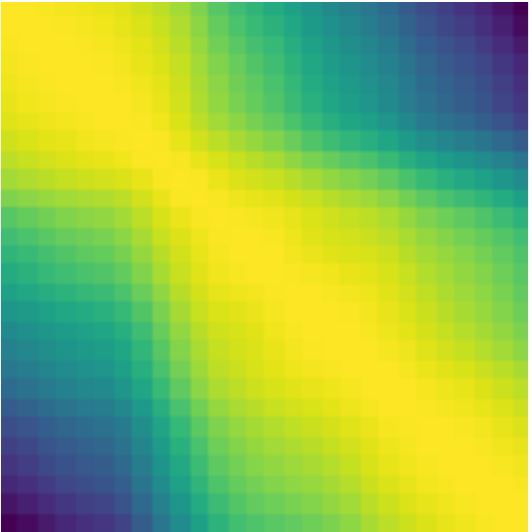}\hspace{-2.3mm}  
\end{tabular}
\end{adjustbox}
\end{tabular}
\end{adjustbox}
\hspace{-4mm}
\begin{adjustbox}{valign=t}
\begin{tabular}{c}
\includegraphics[width=0.0394\textwidth]{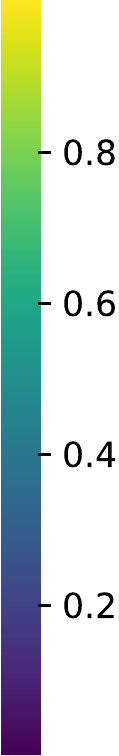}
\end{tabular}
\end{adjustbox}
\end{tabular}
\vspace{-2mm}
\caption{RGB references of the benchmark simulation test set (top line) and spectral correlation coefficient visualizations by the reference (middle line) as well as the proposed method (bottom line). Each correlation coefficient map is of the size 28$\times$28.}
\label{fig: spectral_corr}
\vspace{-2mm}
\end{figure*}

Firstly, given the prediction $\widehat{x} \in \mathrm{R}^{H \times W \times N_{\lambda}}$, we treat each spectral channel as a \texttt{R.V.} (random variable) of $HW$ dimensions and calculate the channel-wise correlations. For each hyperspectral image with $N_{\lambda}=28$, a correlation matrix of 28$\times$28 could be visualized.  We compare these matrices by the references and the predictions in Fig.~\ref{fig: spectral_corr}. The more consistent they are, the higher spectral fidelity we achieve. By observation, the correlation matrices by the predictions show highly similar visual patterns as the reference, indicating that the proposed method effectively captures long-range spectra dependencies. We also notice that there might be minor differences at the centers of the matrices between the visualizations. Rectifying this part is pretty challenging as the model need to precisely distinguish the difference between each adjacent spectral pairs. 

\begin{figure*}[t]
\scriptsize
\centering
\begin{tabular}{cc}
\begin{adjustbox}{valign=t}
\begin{tabular}{cccccc}
\includegraphics[width=0.28\textwidth]{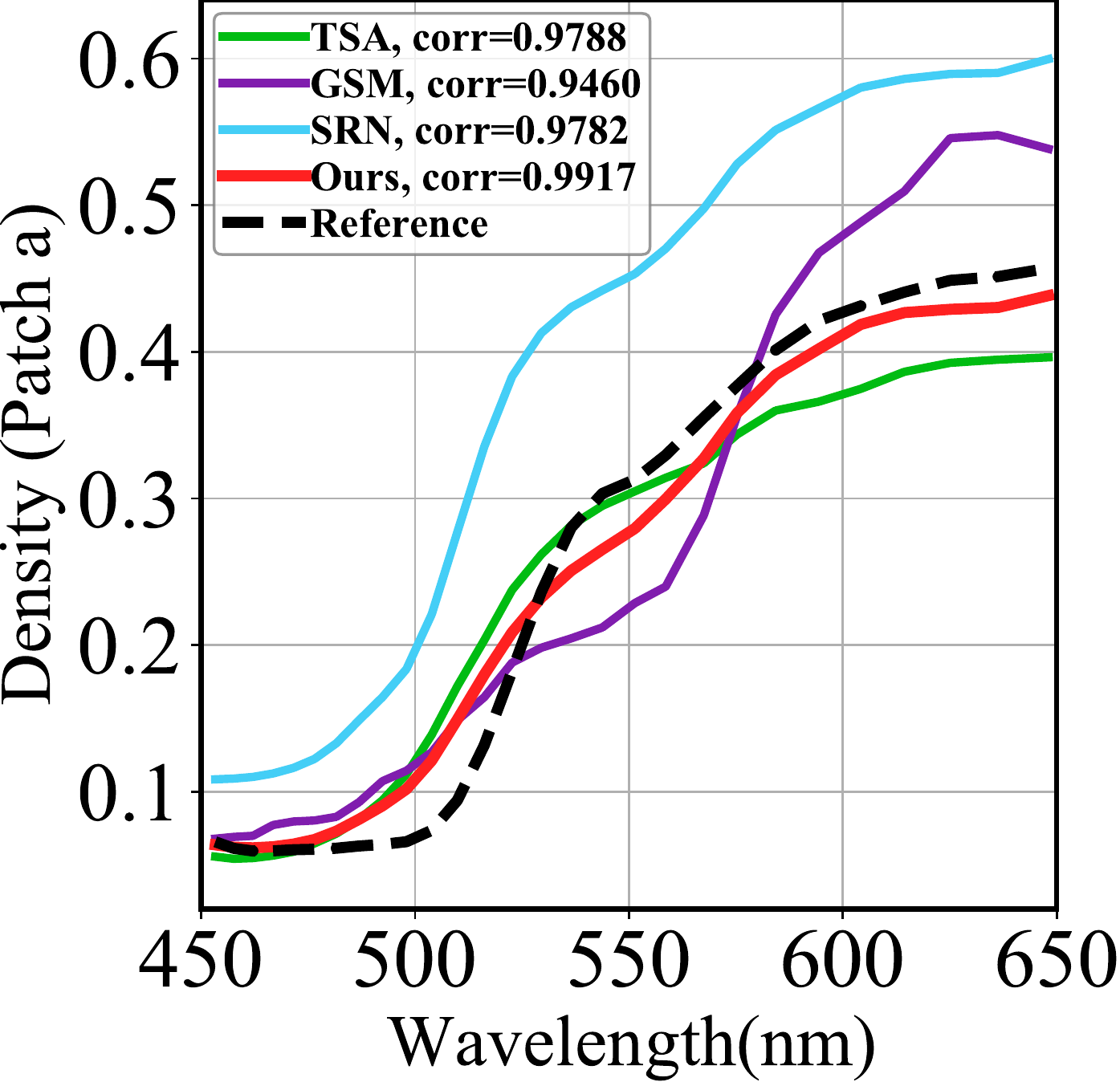}
\hspace{-6mm} 
&
\includegraphics[width=0.28\textwidth]{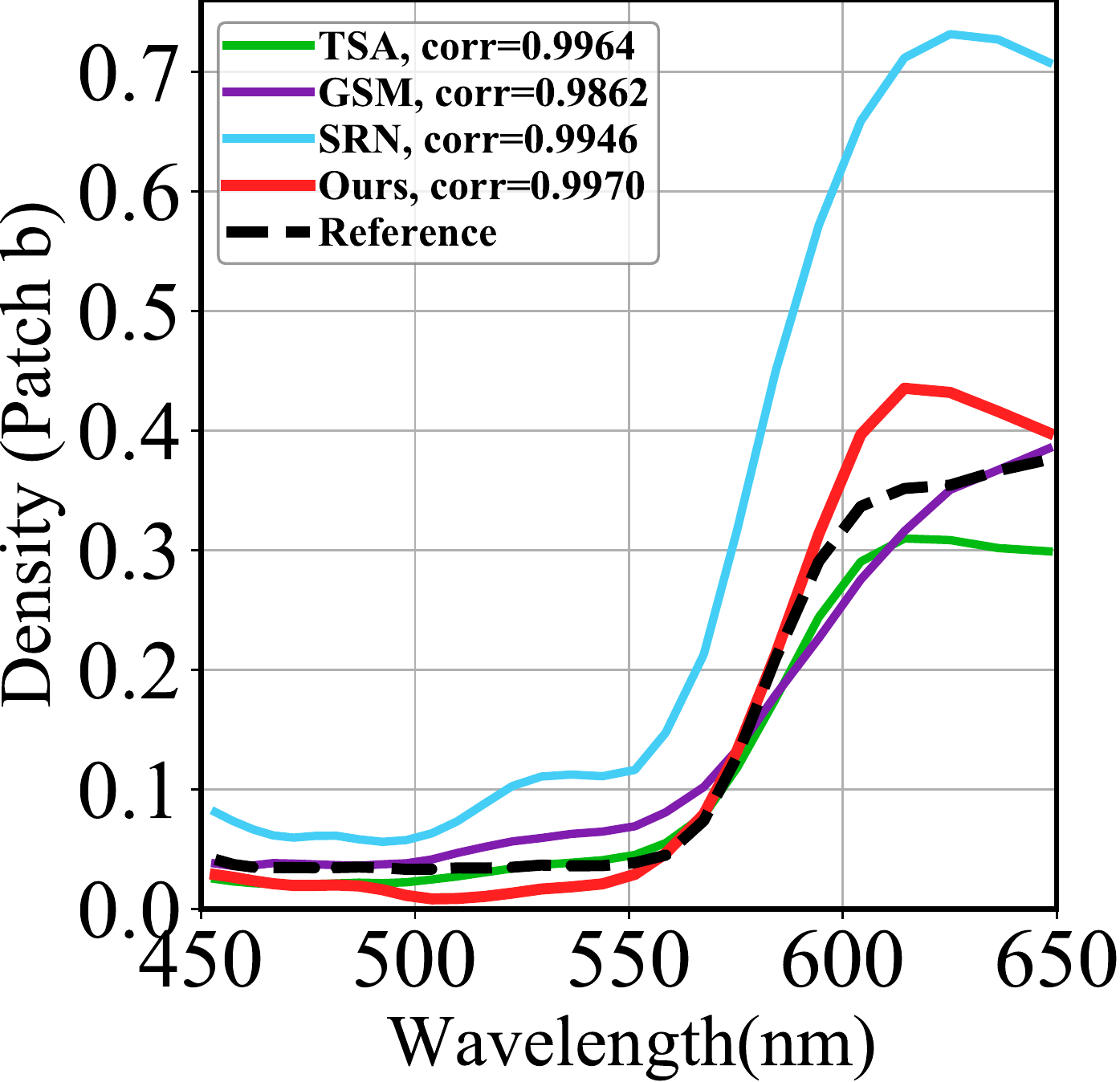}   
\hspace{-6mm} 
&
\includegraphics[width=0.28\textwidth]{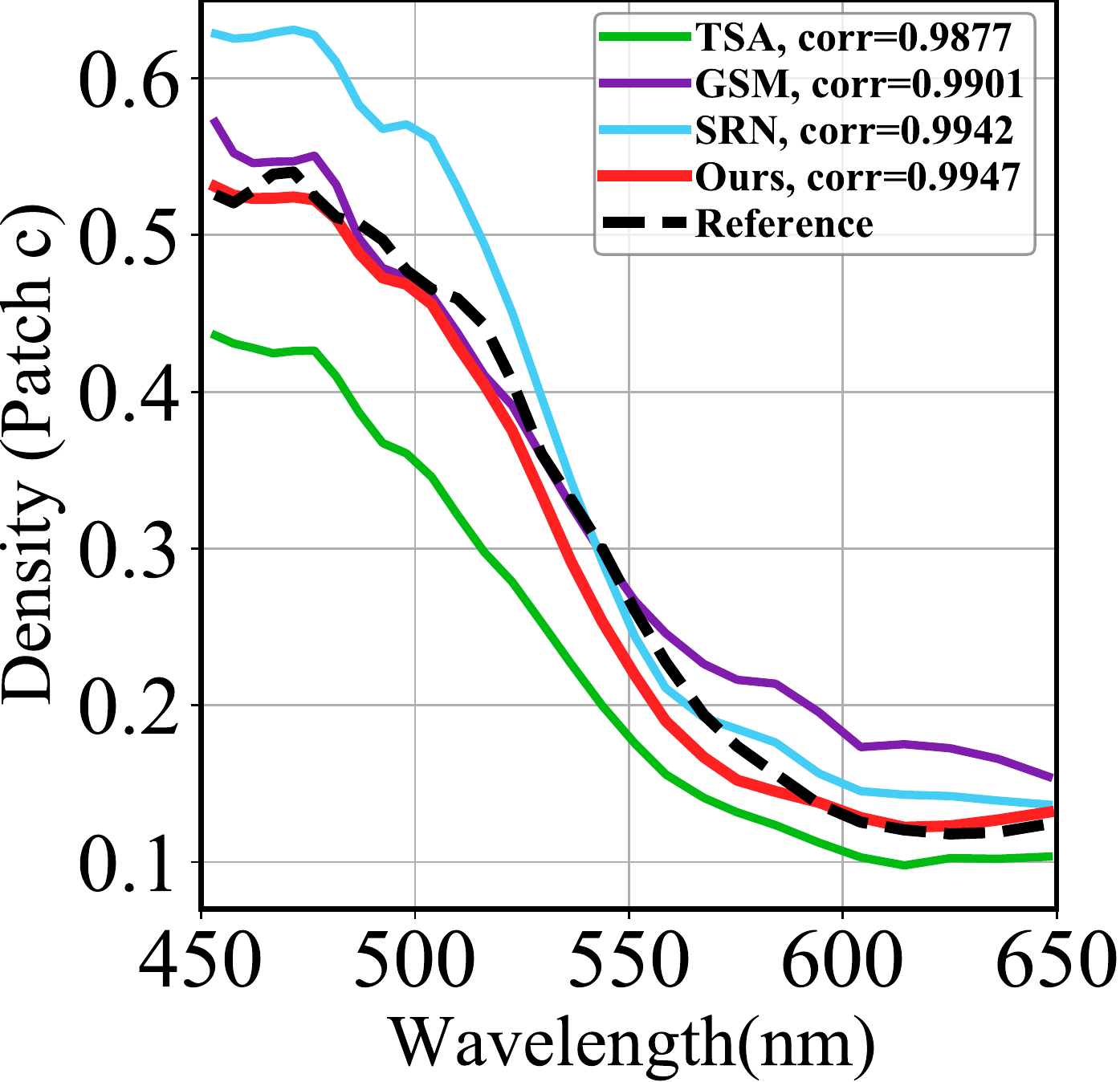}
\hspace{-6mm} 
\end{tabular}
\end{adjustbox}
\begin{adjustbox}{valign=t}
\begin{tabular}{c}
\includegraphics[width=0.086\textwidth]{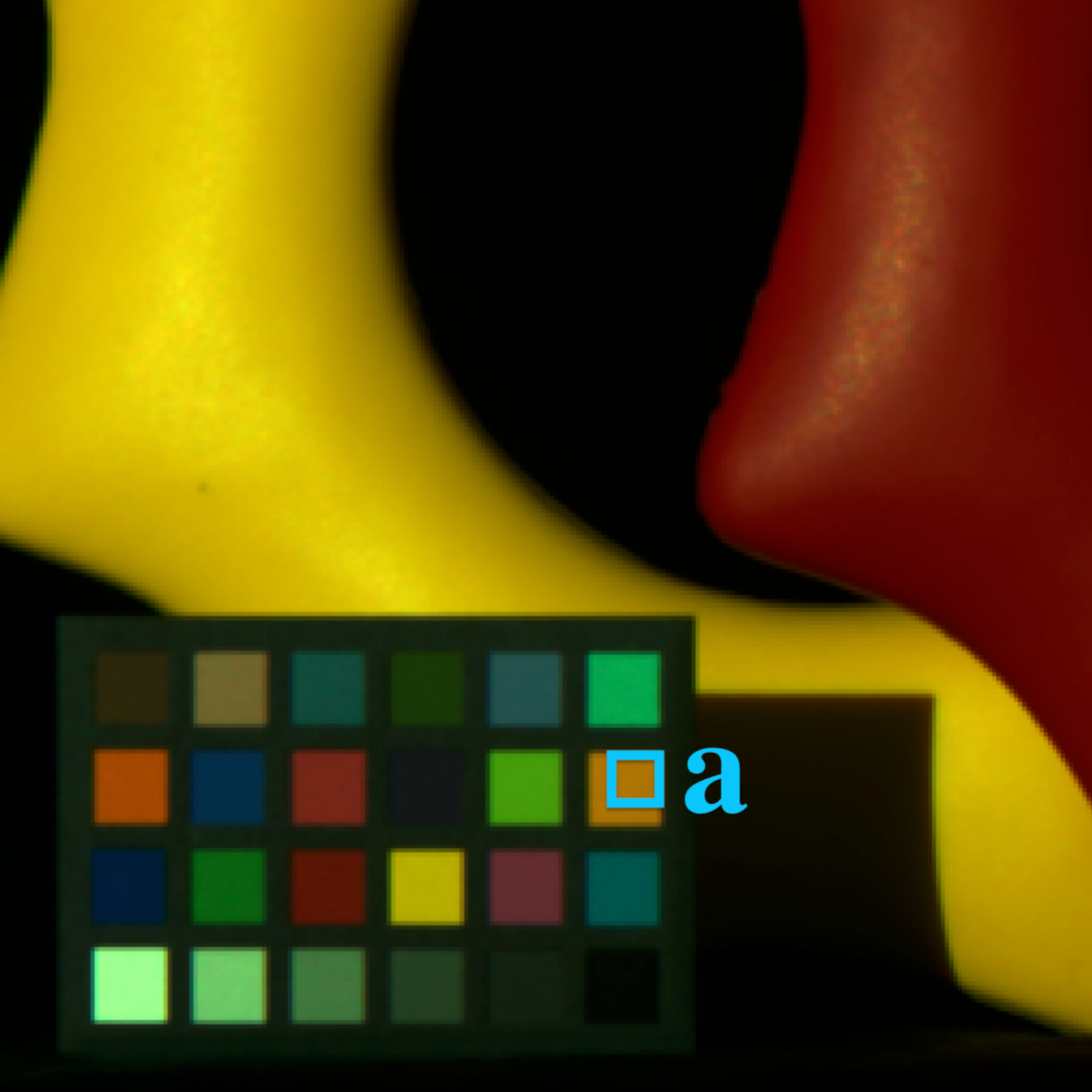} 
\\
\includegraphics[width=0.086\textwidth]{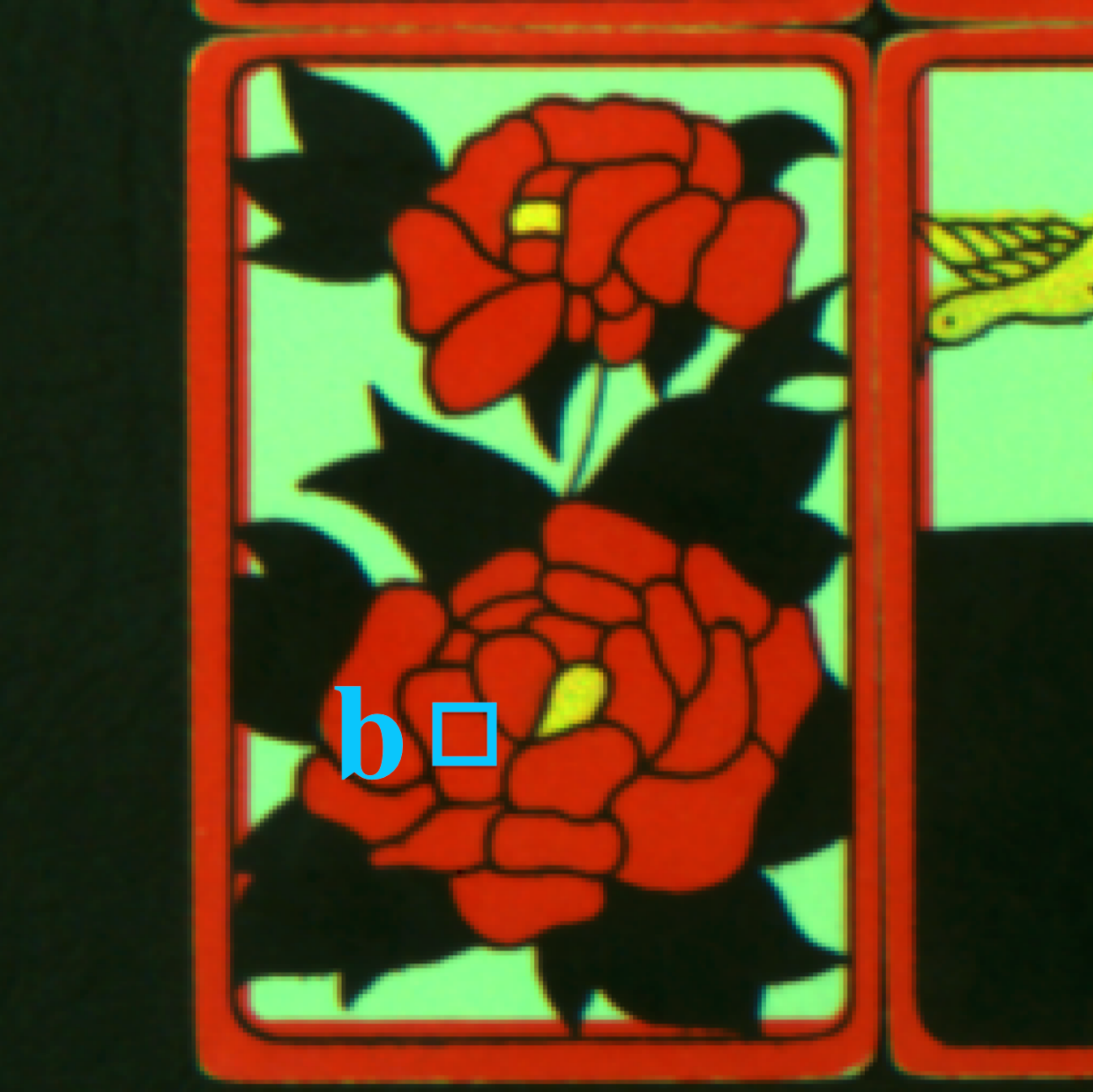}
\\
\includegraphics[width=0.086\textwidth]{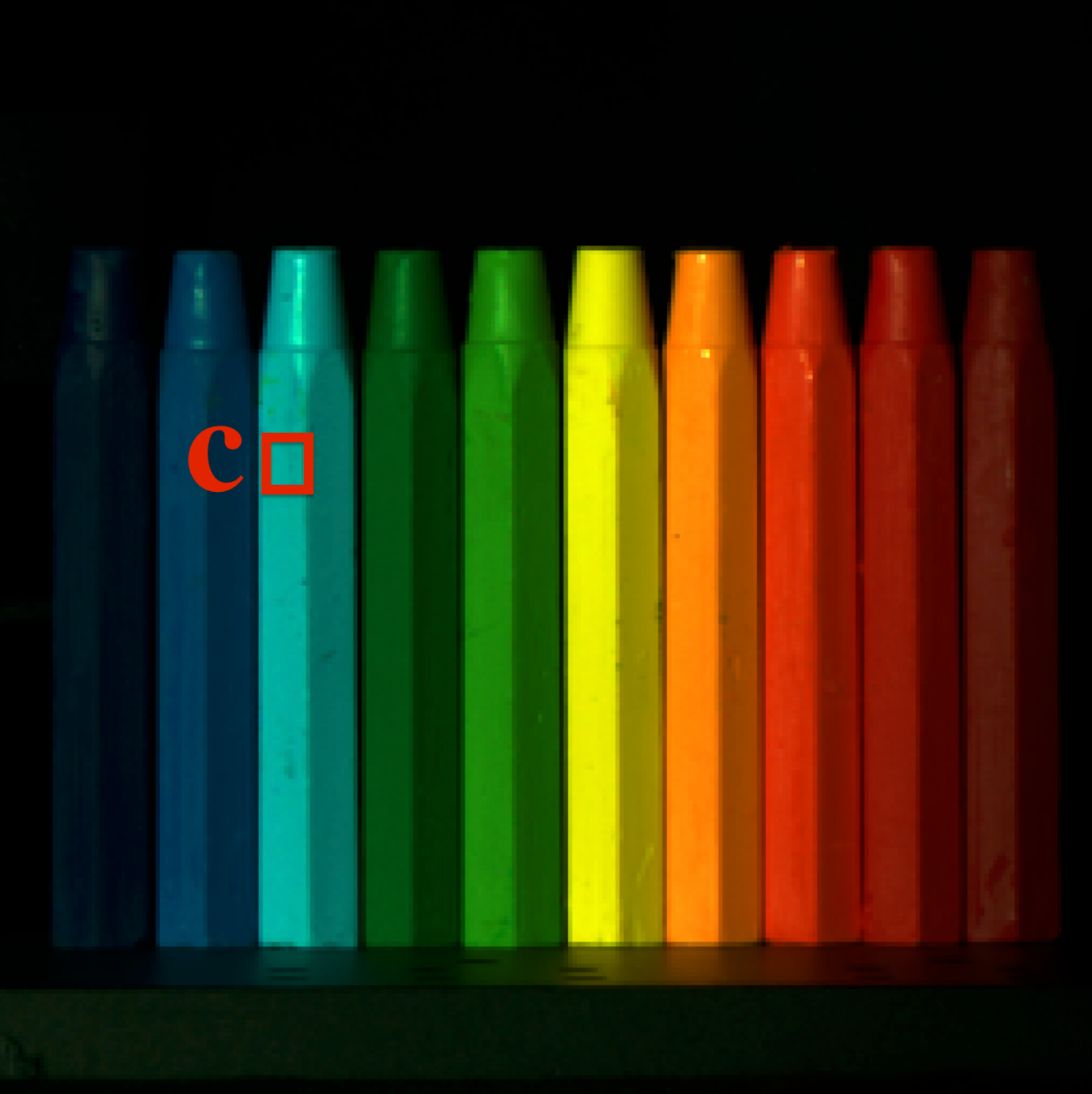}
\\
\end{tabular}
\end{adjustbox}
\end{tabular}
\vspace{-3mm}
\caption{Spectral correlation to the ground truth on exampled locations. Spatial patches (patch \texttt{a},\texttt{b},\texttt{c} as plotted in the right most column RGB references, please zoom in for better visualization) are chosen to ensure the monochromaticity. Density curves are computed upon the predictions by different methods within the chosen patch. }
\label{fig: spectral_curve_examples}
\vspace{-5mm}
\end{figure*} 

\begin{table*}[th]
\scriptsize
\caption{
Averaged spectral correlations ($\uparrow$) to the reference. For each scene, we compute the averaged correlation values upon density curves corresponding to five selected patches. Please refer to examples in Fig.~\ref{fig: spectral_curve_examples} for a detailed computational procedure.
}\label{Tab: average spectral corr}
\vspace{-2mm}
\begin{center}
\resizebox{0.98\textwidth}{!}{
\begin{tabular}{l|cccccccccc|c}
\toprule
Methods & Scene1 & Scene2 & Scene3 & Scene4 & Scene5 & Scene6 & Scene7 & Scene8 & Scene9 & Scene10 & \emph{Avg.} \\
\midrule
GSM~\cite{huang2021deep}  & 0.9903 & 0.9674 & 0.9751 & 0.9435 & 0.9891 & 0.9904 & 0.9951 & 0.9700 & 0.9870 & 0.8926 & 0.9701 \\
SRN~\cite{wang2021new} & 0.9910 & 0.9701 & 0.9753 & 0.9429 & 0.9898 & 0.9910 & 0.9959 & 0.9865 & 0.9870 & 0.9000 & 0.9730 \\
Ours & \textbf{0.9956} & \textbf{0.9910} & \textbf{0.9921} & \textbf{0.9458} & \textbf{0.9925} & \textbf{0.9982} & \textbf{0.9969} & \textbf{0.9867} & \textbf{0.9903} & \textbf{0.9004} & \textbf{0.9790} \\
\bottomrule
\end{tabular}}
\end{center}
\vspace{-2mm}
\end{table*}

\begin{figure*}[t]
\scriptsize
\centering
\begin{tabular}{cc}
\begin{adjustbox}{valign=t}
\begin{tabular}{ccccccccc}
\includegraphics[width=0.093\textwidth]{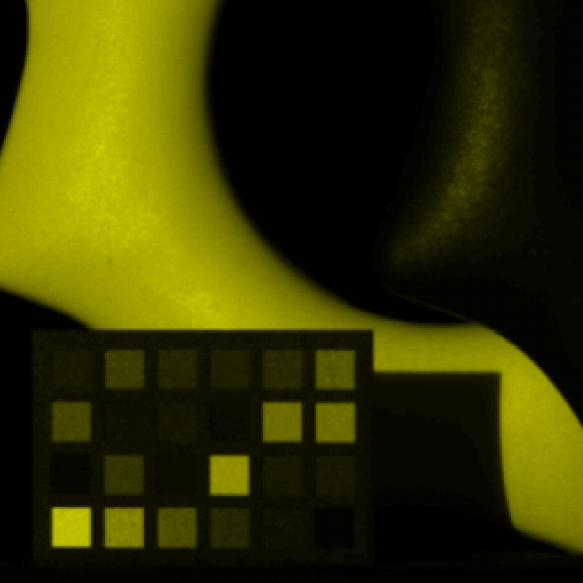}      
\hspace{-4mm} 
&
\includegraphics[width=0.093\textwidth]{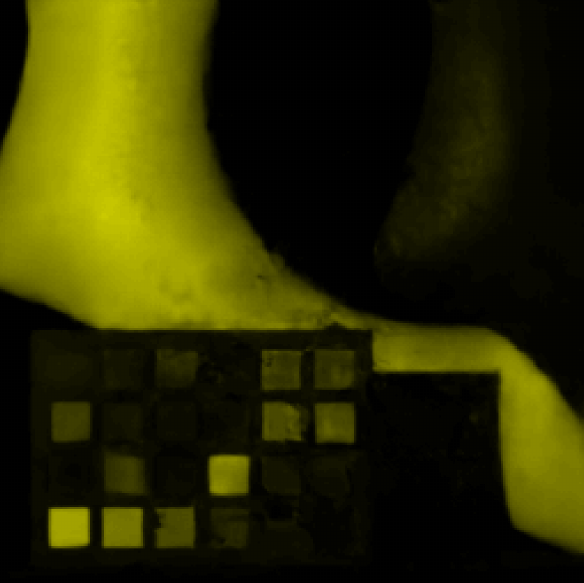}
\hspace{-4mm} 
&
\includegraphics[width=0.093\textwidth]{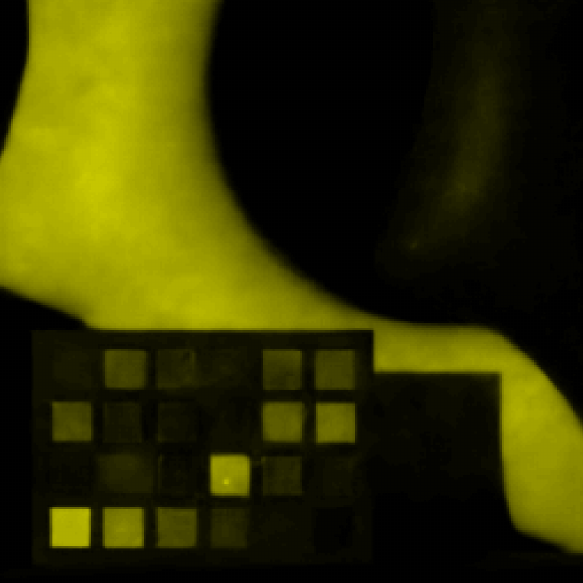}
\hspace{-4mm} 
&
\includegraphics[width=0.093\textwidth]{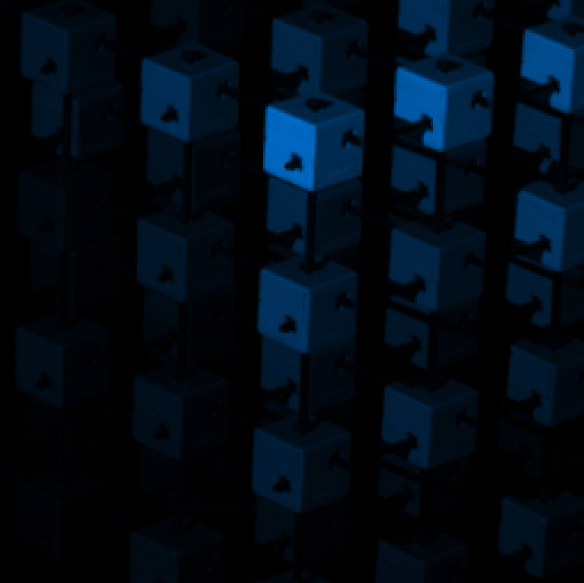}      
\hspace{-4mm} 
&
\includegraphics[width=0.093\textwidth]{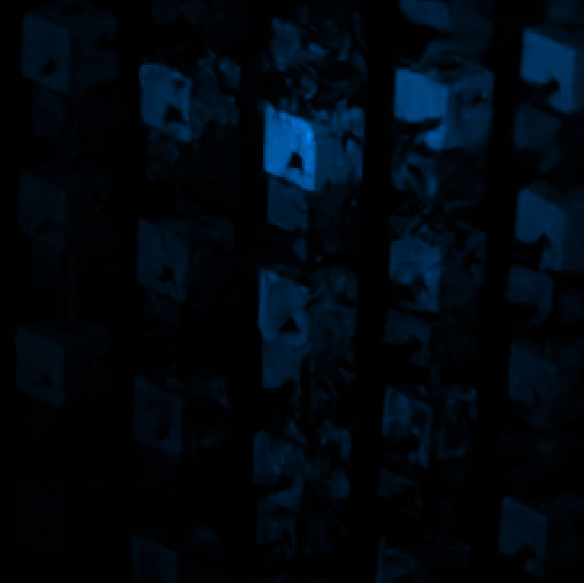}
\hspace{-4mm} 
&
\includegraphics[width=0.093\textwidth]{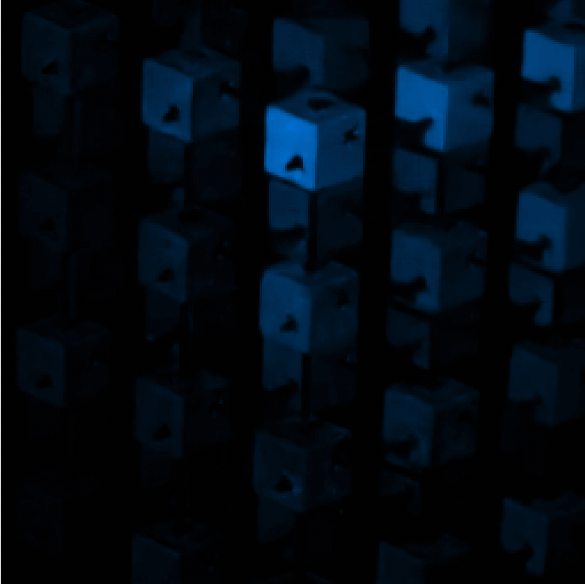}
\hspace{-4mm} 
&
\includegraphics[width=0.093\textwidth]{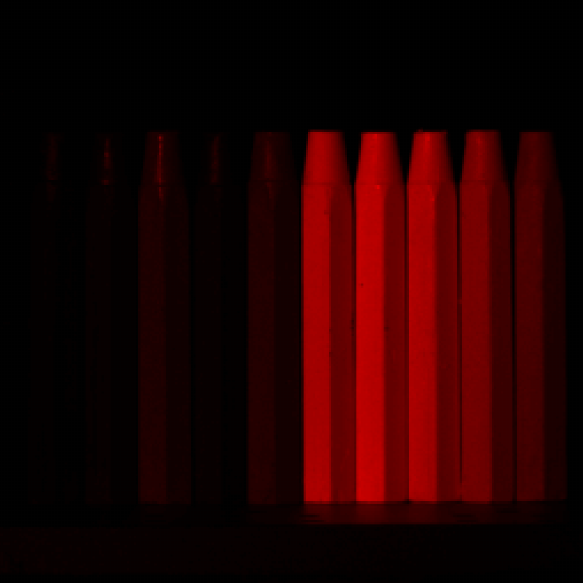}      
\hspace{-4mm} 
&
\includegraphics[width=0.093\textwidth]{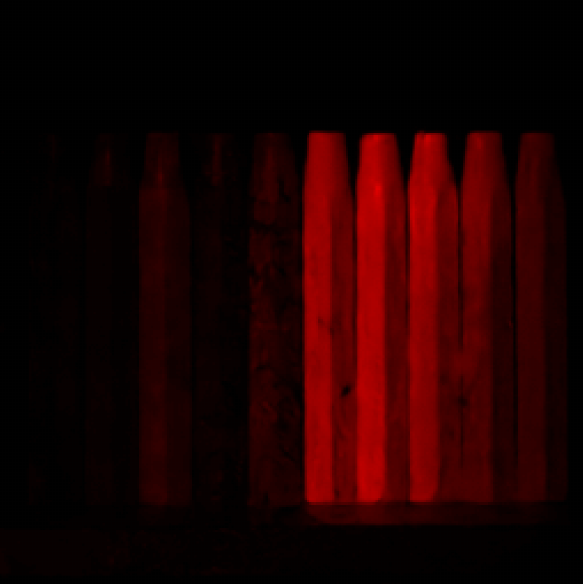}
\hspace{-4mm} 
&
\includegraphics[width=0.093\textwidth]{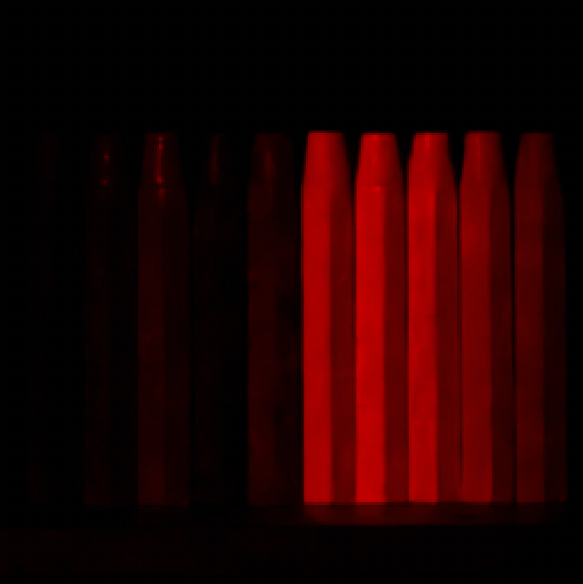}
\hspace{-4mm}
\\
\includegraphics[width=0.093\textwidth]{gt3}         
\hspace{-4mm} 
&
\includegraphics[width=0.093\textwidth]{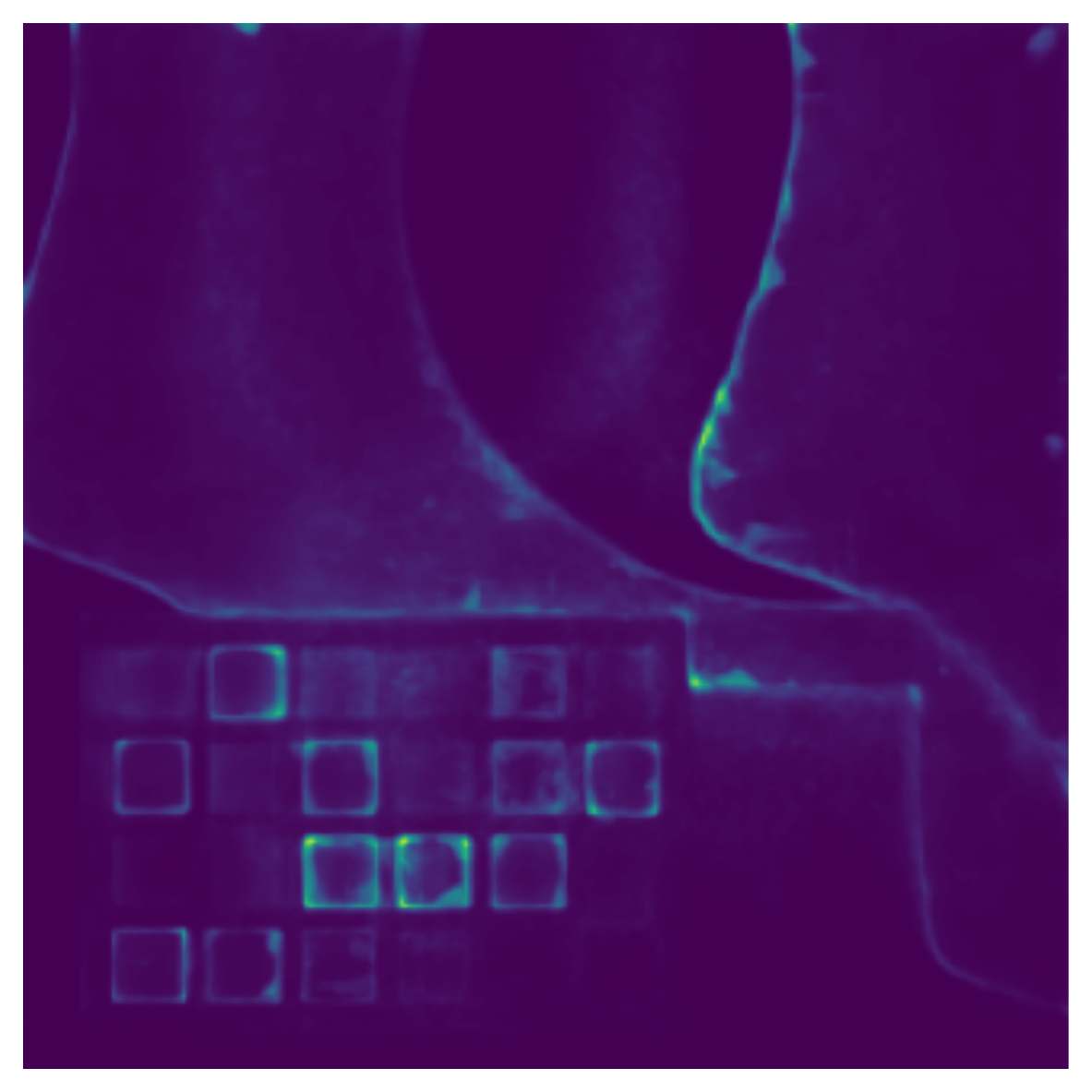}  
\hspace{-4mm} 
&
\includegraphics[width=0.093\textwidth]{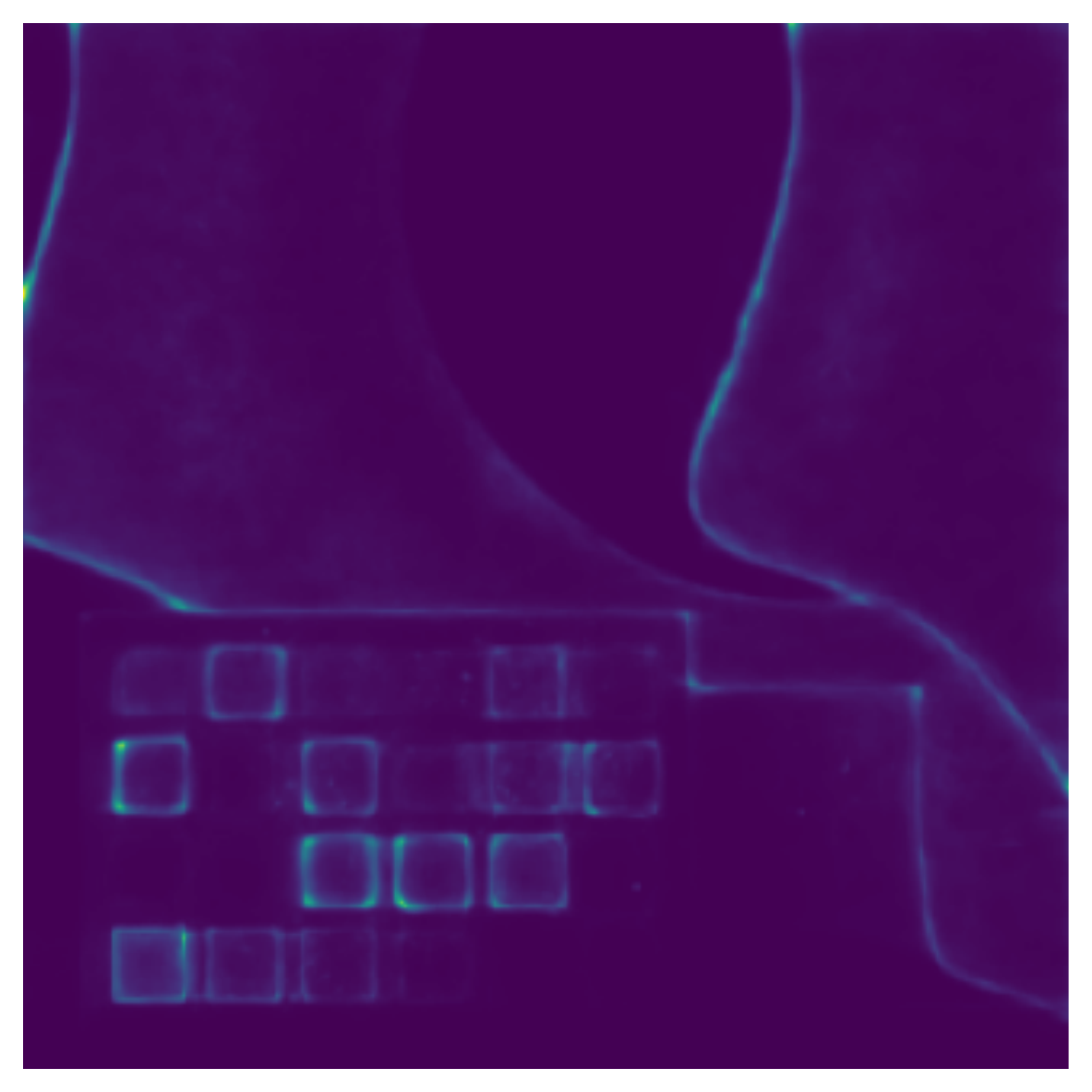} 
\hspace{-4mm} 
&
\includegraphics[width=0.093\textwidth]{gt2}                   
\hspace{-4mm} 
&
\includegraphics[width=0.093\textwidth]{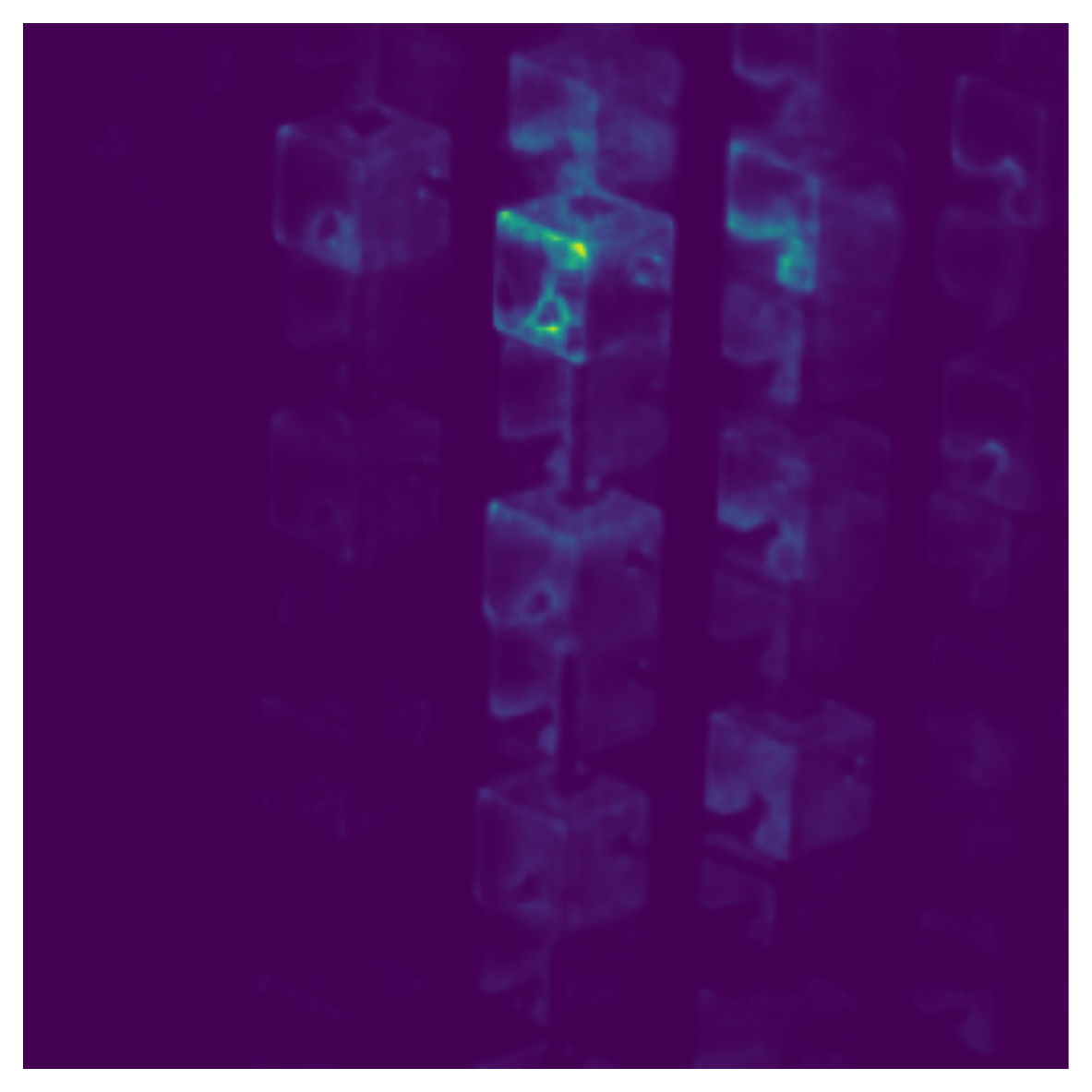}  
\hspace{-4mm} 
&
\includegraphics[width=0.093\textwidth]{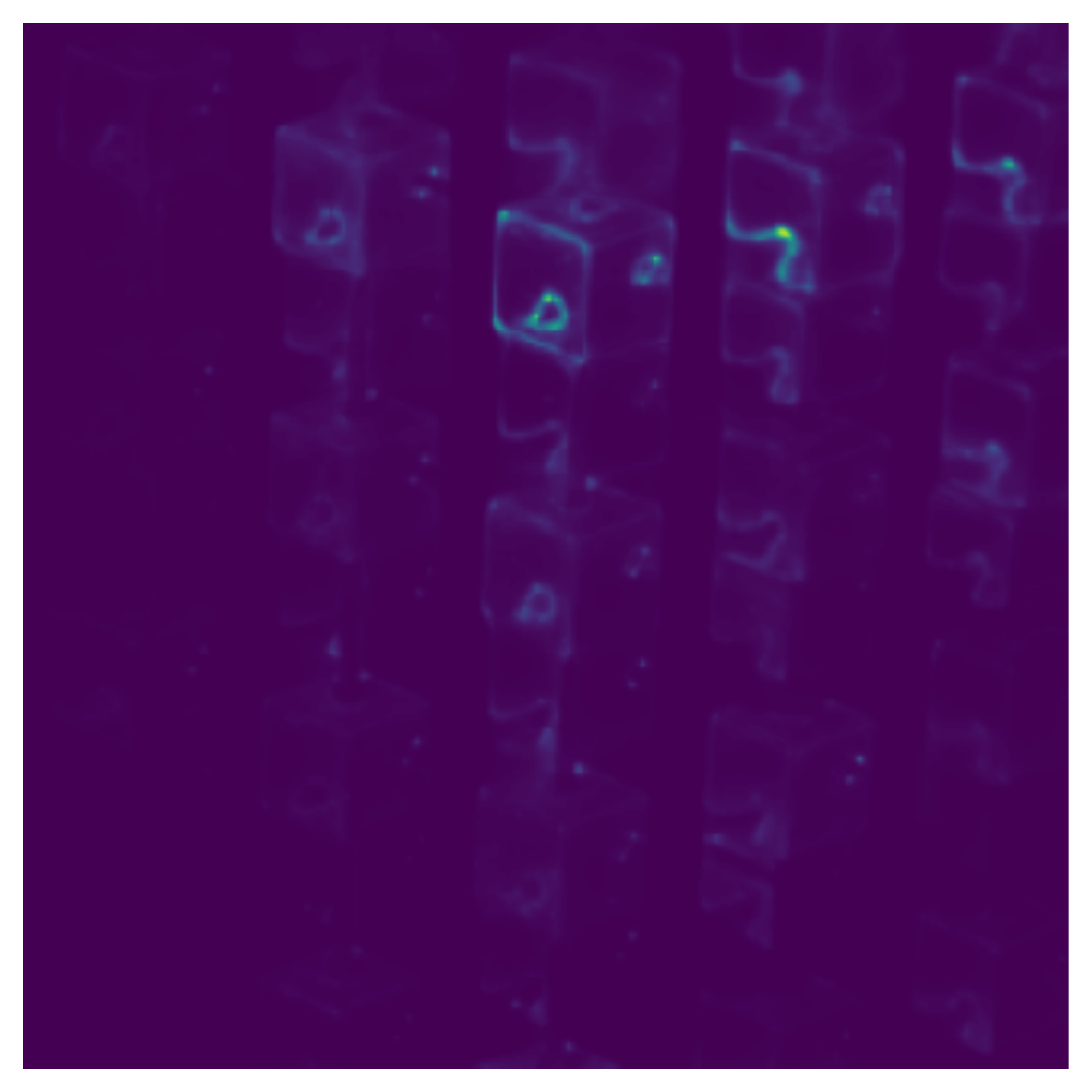} 
\hspace{-4mm} 
&
\includegraphics[width=0.093\textwidth]{gt9}                   
\hspace{-4mm} 
&
\includegraphics[width=0.093\textwidth]{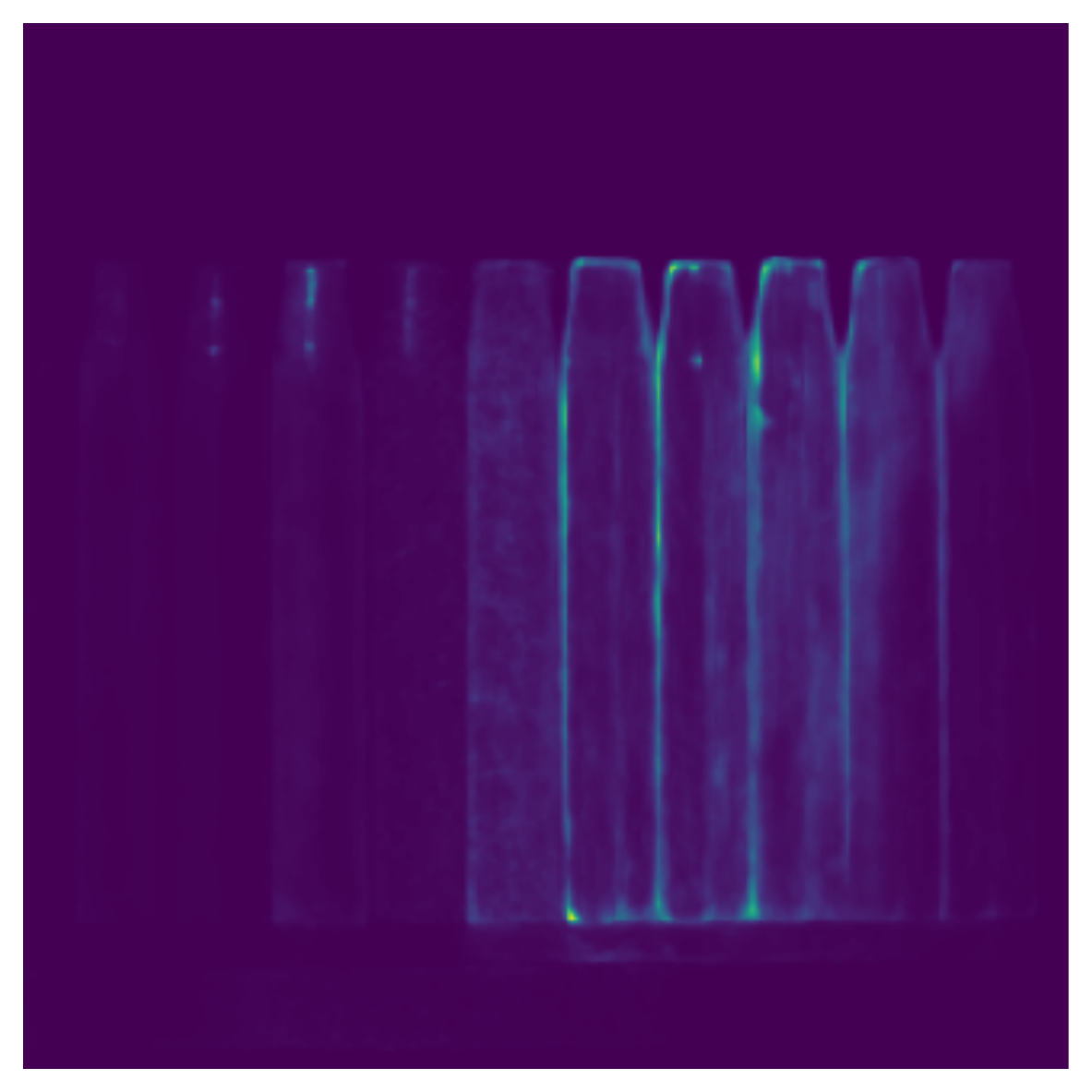}  
\hspace{-4mm} 
&
\includegraphics[width=0.093\textwidth]{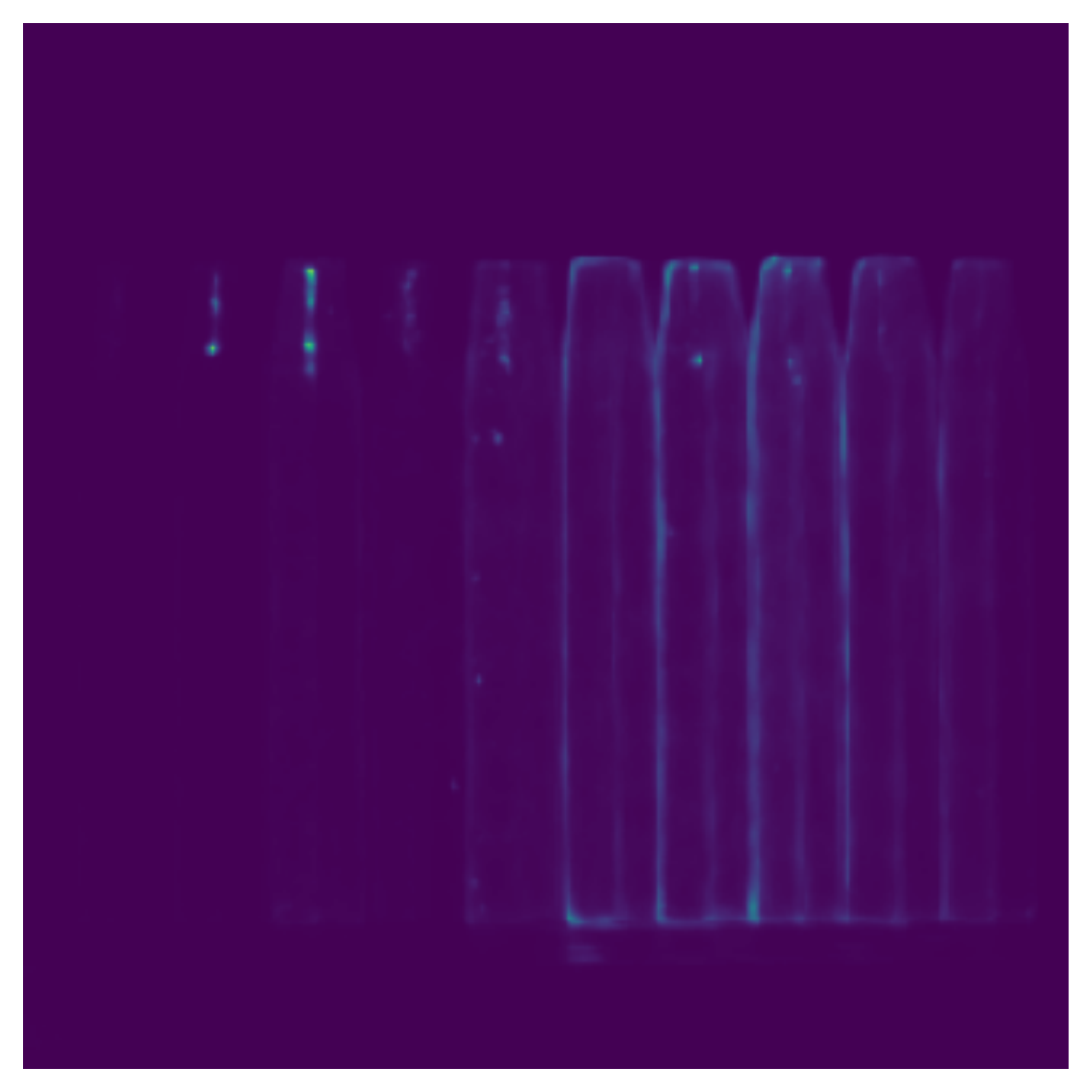} 
\hspace{-4mm}
\\
reference & GSM & Ours &  reference & GSM & Ours & reference & GSM & Ours 
\\
\end{tabular}
\end{adjustbox}
\hspace{-2mm}
\begin{adjustbox}{valign=t}
\begin{tabular}{c}
\includegraphics[width=0.0305\textwidth]{color_bar}
\end{tabular}
\end{adjustbox}
\end{tabular}
\caption{Three exampled epistemic uncertainty visualizations by GSM~\cite{huang2021deep} and the proposed method. For each example, we demonstrate averaged reconstruction results on selected wavelengths (i.e., 567.5nm, 471.6nm, and 614.4nm) in top line, and the epistemic uncertainty in the bottom line.}
\label{fig: epistemic_uncertainty}
\vspace{-2mm}
\end{figure*}

\begin{table}[t]
\caption{Ablation study of the proposed method under the traditional setting (one-to-one).}\label{Tab: ablation one-to-one}
\vspace{-2mm}
\begin{center}
\resizebox{.42\textwidth}{!}{
\begin{tabular}{lcc}
\toprule
    Settings & PSNR(dB) & SSIM\\
\midrule
    \texttt{w/o GST}   & 33.17 & 0.9288 \\

\midrule
    \texttt{w/o Bi-Opt} & 32.73 & 0.9193 \\
\midrule
    \texttt{w/o GCN} & 33.23 & 0.9286 \\
\midrule
    Ours (full model) & 33.22 & 0.9292 \\
\bottomrule
\end{tabular}}
\end{center}
\vspace{-5mm}
\end{table}

\begin{table}[t]
\caption{Ablation study of the proposed method under the setting of miscalibration (one-to-many) among 100 testing trials.}\label{Tab: ablation one-to-many}
\vspace{-2mm}
\begin{center}
\resizebox{.46\textwidth}{!}{
\begin{tabular}{lcc}
\toprule
    Settings & PSNR(dB) & SSIM\\
\midrule
    \texttt{w/o GST}  & 30.17$\pm$0.63 & 0.8865$\pm$0.0108 \\
\midrule
    \texttt{w/o Bi-Opt} & 30.30$\pm$0.06 & 0.8843$\pm$0.0011 \\
\midrule
    \texttt{w/o} GCN & 30.13$\pm$0.07 & 0.8849$\pm$0.0011 \\
\midrule
    Ours (full model) & 30.60$\pm$0.08 & 0.8881$\pm$0.0013 \\
\bottomrule
\end{tabular}}
\end{center}
\vspace{-4mm}
\end{table}

\begin{figure*}[t] 
\centering 
\includegraphics[width=0.9\textwidth]{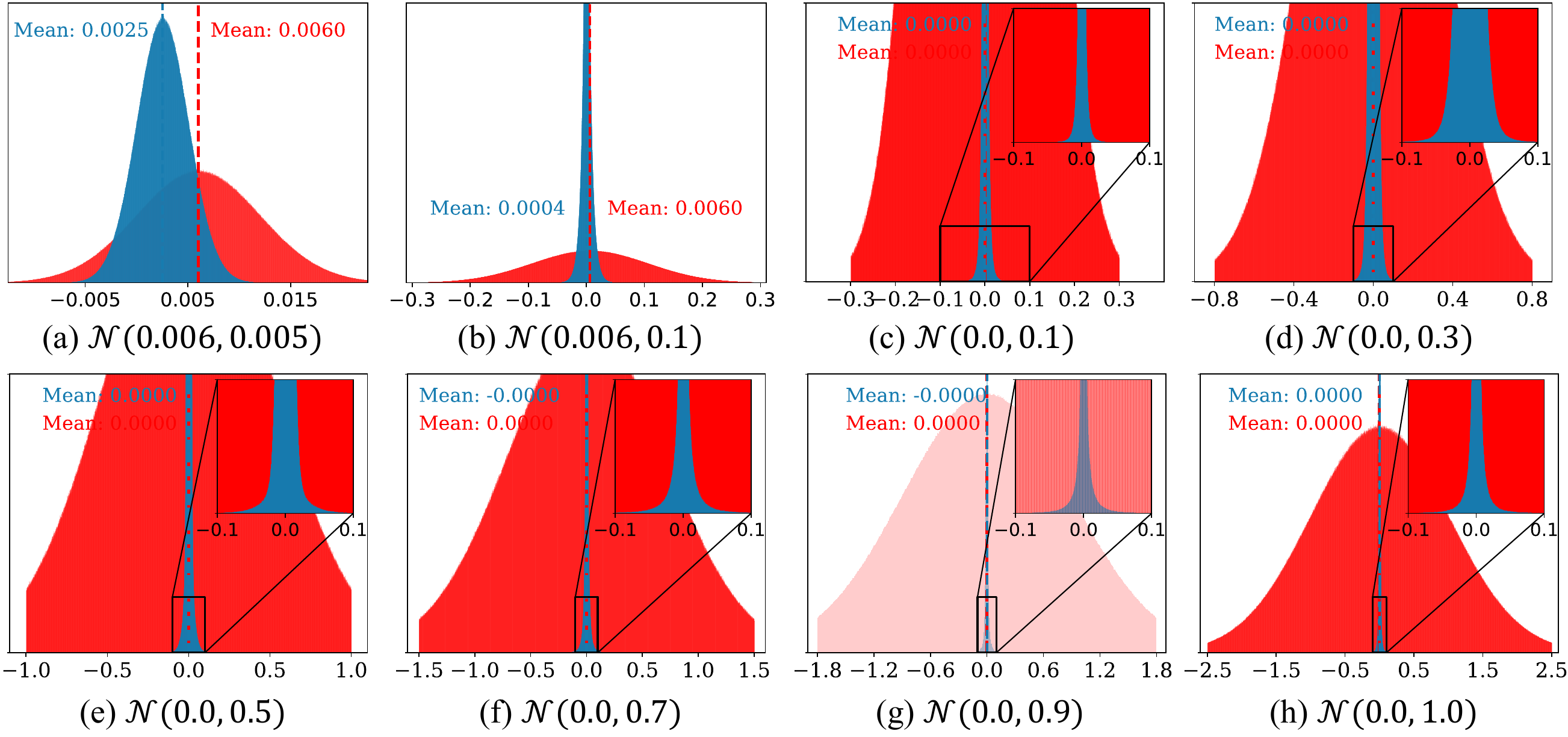}\vspace{-4mm}
\caption{Learned variational noise distribution under different priors. Eight different priors (red) are adopted in the experiment. By comparison, variational noise distributions (blue) are characterized by smaller variance. Please refer to the red curves in Fig.~\ref{fig: curves} for corresponding reconstruction performance comparison.}
\label{fig: noise distributions} 
\vspace{-2mm}
\end{figure*}

\begin{figure}[t] 
\centering 
\includegraphics[width=.65\textwidth]{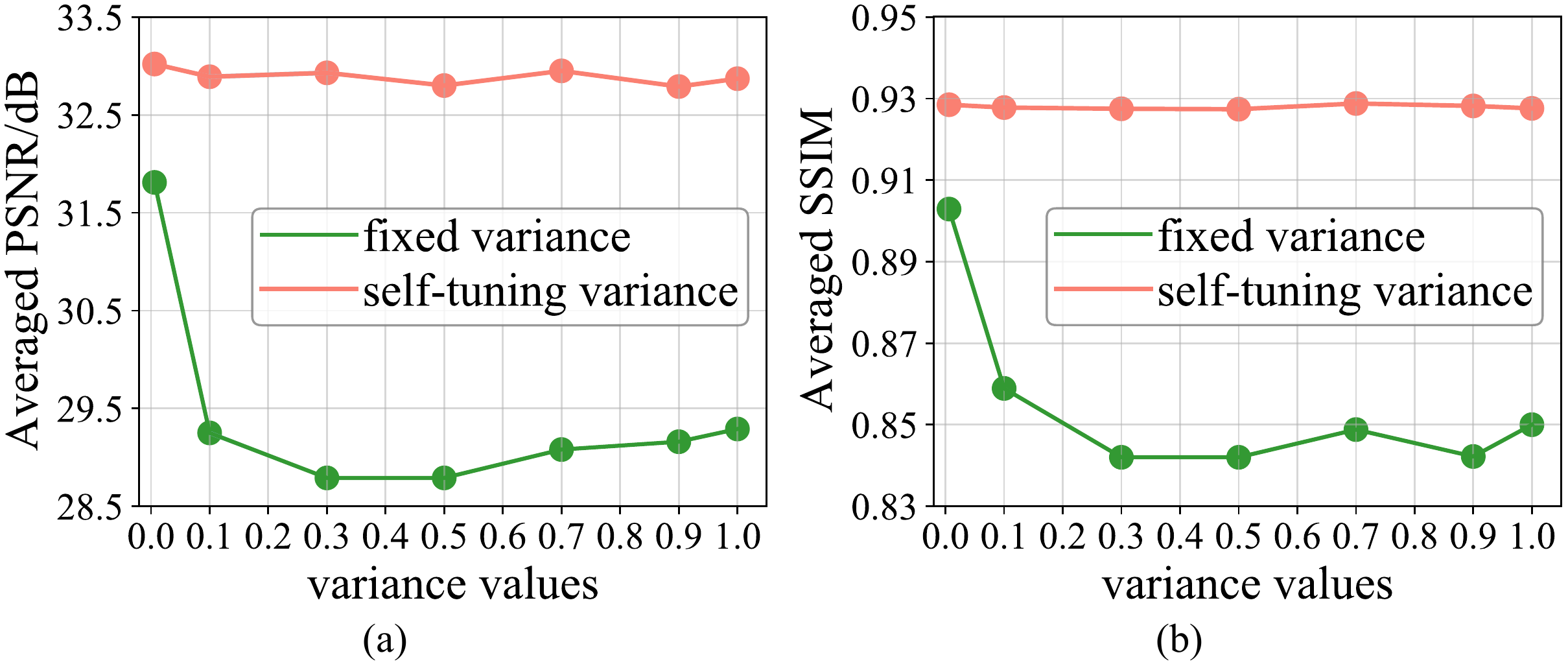}
\vspace{-4mm}
\caption{Performance comparison between fixed variance (green) and self-tuning variance (red). The PSNR is compared in (a) and SSIM is compared in (b). Reconstruction using self-tuning variance outperforms that using fixed variance with different values.} 
\label{fig: curves}
\vspace{-2mm}
\end{figure} 

\begin{figure}[t] 
\centering 
\includegraphics[width=.5\textwidth]{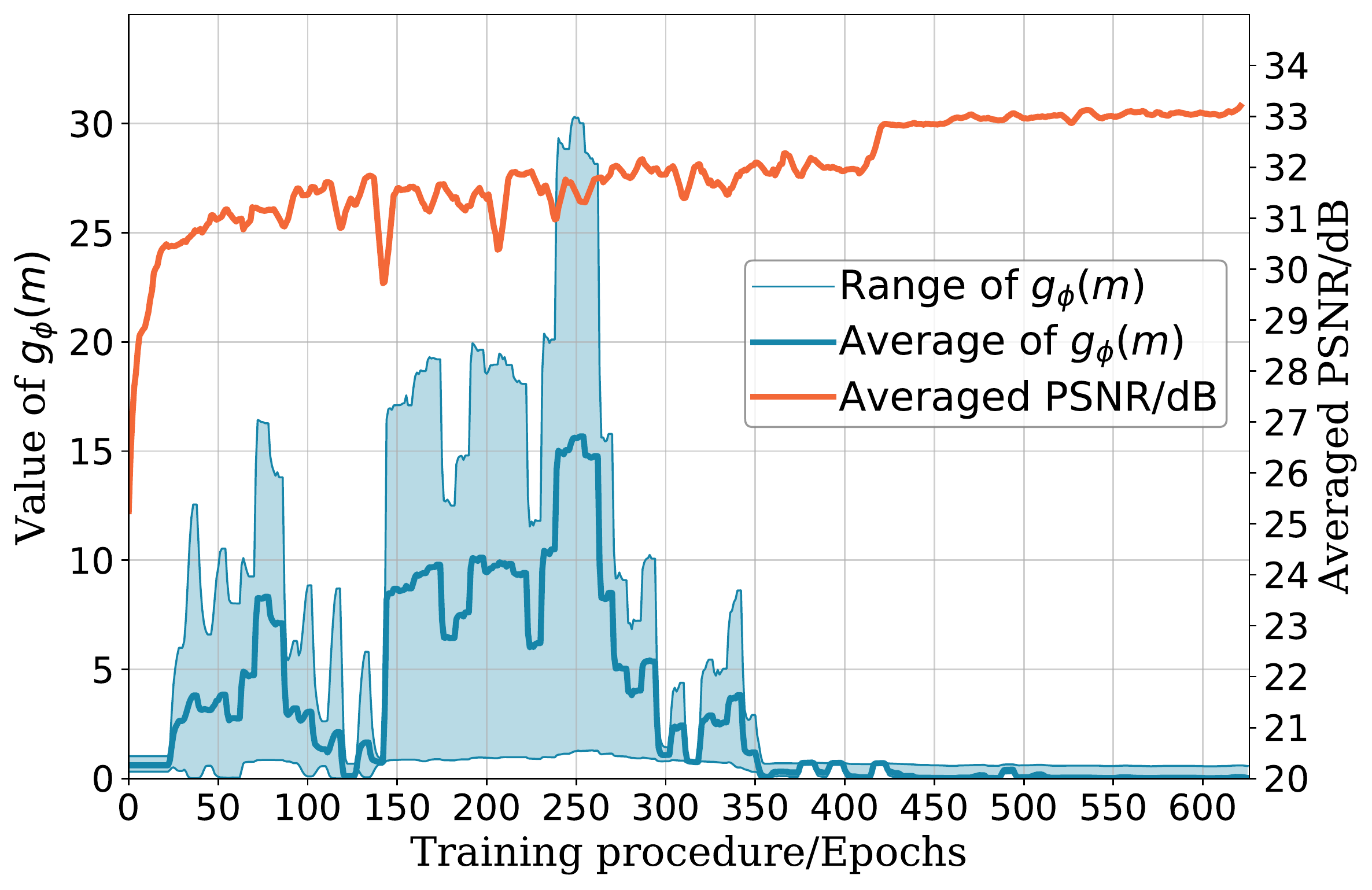}
\vspace{-3mm}
\caption{Observation on $g_{\phi}(m)$ during training. The $g_{\phi}(m)$ gradually converge to a smaller range with more epochs of training. Meanwhile, a better reconstruction performance can be observed. Both training epochs and validation epochs are jointly counted.} 
\label{fig: convergence}
\vspace{-1mm}
\end{figure} 

\begin{figure*}[t] 
\centering 
\includegraphics[width=1.0\textwidth]{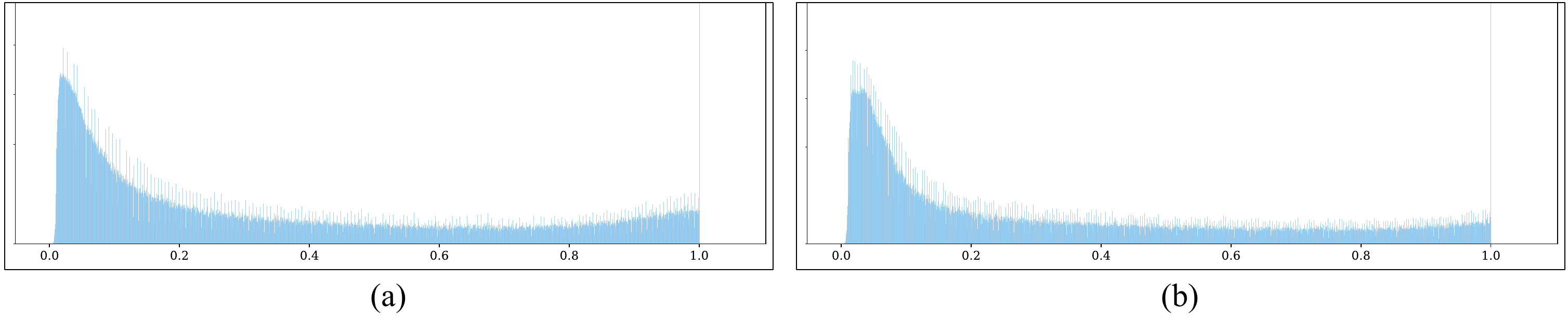}\vspace{-2mm}
\caption{Histograms of two real masks applied in this work. (a) sources from~\cite{meng2020end} and (b) sources from~\cite{meng2020snapshot}. Both masks are produced by the same fabrication process. Bin number is set to 2000 for both histograms.}
\label{fig: two real masks} 
\vspace{-3mm}
\end{figure*}

Secondly, we quantitatively compare the spectral fidelity of different methods upon density curves. As three examples demonstrated in Fig.~\ref{fig: spectral_curve_examples}, we first crop a small spatial patch from the prediction (exampled by the RGB reference on the right most column), then we draw the density curve by pixel intensities in that small patch. Finally, the correlations between the reference curve and that from the predictions are computed. Higher correlation values indicate a higher spectral fidelity for the cropped patch. The small patch is chosen to ensure the monochromaticity (wavelength). For example, if we choose the patch whose color lies in blue$\sim$cyan range (bottom-right RGB reference in Fig.~\ref{fig: spectral_curve_examples}), the energy of the density would concentrate in the 450nm$\sim$500nm.

To globally compare the spectral fidelity, we randomly choose five monochromatic patch of each scene and compute an averaged correlation value upon five density curves. We report the correlation values of ten scenes in Table~\ref{Tab: average spectral corr} and demonstrate the superiority of the proposed method, as compared to GSM~\cite{huang2021deep} and SRN~\cite{wang2021new}.

\section{Epistemic Uncertainty}\label{sec: epistemic uncertianty}
As illustrated in Section {\color{red}4.2} of the manuscript, the proposed method demonstrates low epistemic uncertainty by approximating the mask distribution. In this section, we provide more visualizations and analyses on epistemic uncertainty in Fig.~\ref{fig: epistemic_uncertainty}. Specifically, we test the well-trained models upon random real masks and repeat 100 trials. Both GSM and the proposed method are trained upon the same mask set $\mathcal{M}$ for a fair comparison. For each exampled hyperspectral images, we compare the averaged reconstruction and epistemic uncertainty on a selected spectral channel. Notably, in low-frequency regions, both methods show high confidence, while in high-frequency regions (i.e., edges), the proposed method presents a much-lower epistemic uncertainty, which would potentially benefit the down-stream applications like object detection or segmentation upon hyperspectral images.

\section{Ablation Study}\label{sec: ablation} 
In this paper, three scenarios are introduced: 1) one-to-one setting, which is the traditional setting considered by previous reconstruction methods, 2) one-to-many miscalibration, 3) many-to-many miscalibration. Notably, the third scenario enables a complete mask distribution modeling,  for which reason we put more emphasize on it and provide the ablation study accordingly in the manuscript. Following that, Table~\ref{Tab: ablation one-to-one} conducts the same ablation experiments under the traditional setting (one-to-one). For miscalibration (one-to-many), we also
do verification and report the performance in Table~\ref{Tab: ablation one-to-many}.
The ablated models include
\begin{itemize}
    \setlength\itemsep{0em}
    \item \texttt{w/o GST}: we remove the graph-based self-tuning (GST) network from the proposed method. Actually it degrades into the reconstruction backbone SRN~\cite{wang2021new} applied under corresponding scenarios. 
    \item \texttt{w/o Bi-Opt}: we simultaneously optimize all of the parameters by the lower-level loss function, i.e., Eq. ({\color{red}7}) in the manuscript, based on the original training set.
    \item \texttt{w/o GCN}: for the self-tuning network, we exchange the GCN with a convolutional layer carrying more parameters for a fair comparison. 
\end{itemize}

As shown in the Table~\ref{Tab: ablation one-to-one} and Table~\ref{Tab: ablation one-to-many}, both the GST module and bilevel optimization strategy contribute significantly for the final performance boost. While the ablated model \texttt{w/o GCN} in self-tuning network works comparably with the \texttt{Ours (full model)} under the traditional setting by PSNR, it falls behind regarding SSIM, indicating a sub-optimal reconstruction ability. 

\section{Model Discussion} \label{sec: model discussion}

\noindent \textbf{Fixed variance.} In the manuscript, we showcase that the fixed variance with distinct values only achieves sub-optimal performance compared with the self-tuning variance. In Fig.~\ref{fig: curves} (b), we also plot the SSIM curve (green) with different values of the fixed variance. Besides, the original PSNR curve (green) is shown in Fig.~\ref{fig: curves} (a). 

\noindent \textbf{Self-tuning variance under different priors.}
Due to the limitation of time and computational resource, we only discussed the self-tuning variance under three most representative noise priors in the manuscript, i.e., $\mathcal{N}(0.006, 0.005)$, $\mathcal{N}(0.006, 0.1)$ and $\mathcal{N}(0.0, 1.0)$. 
In this section, we demonstrate additional results corresponding to more noise priors. Similar to the performance curves plotted in Fig. {\color{red}7} (a) in the manuscript, we report the corresponding performances by red curves in Fig.~\ref{fig: curves}, verifying the superiority of the self-tuning variance. In Fig.~\ref{fig: noise distributions}, we visualize both the noise prior and the subsequent variational noise distributions. The similarity between all subplots -- the learned variational noise distribution gets a smaller variance than the given prior -- validates the effectiveness of mask uncertainty modeling.

In Fig.~\ref{fig: convergence}, we explore the convergence of the $g_{\phi}(m)$ during the training phase of our best model. For pre-training phase  of reconstruction network (first 20 epochs as mentioned in the manuscript), the range of $g_{\phi}(m)$ remains invariant. An interesting observation is that the fluctuation of the $g_{\phi}(m)$ value is accompanied by fluctuation of the reconstruction performance, indicating the underlying impact of the self-tuning variance. 
During the last 200 epochs (including training and validation epochs), a converged $g_{\phi}(m)$ contributes to a steady performance improvement.

\section{Dataset}\label{sec: data}
\noindent \textbf{HSI data set.} We adopt the training set provided in~\cite{meng2020end} and follow the same data augmentation operations.
Specifically, the training set contains 205 1024$\times$1024$\times$28 training samples, all of which sources from the CAVE dataset~\cite{CAVE_0293}. Our model is trained on 256$\times$256$\times$28 patches randomly cropped from these 205 samples. For a fair comparison with the other deep reconstruction networks, 
we create a validation data set by randomly splitting 40 hyperspectral images from the above 205 samples. Therefore, \textit{no new HSI data is introduced for our model training}. For the model testing, ten simulation hyperspectral images corresponding to 10 scenes shown in Fig.~\ref{fig: spectral_corr} are used for quantitative and perceptual comparison, following previous works~\cite{meng2020end,huang2021deep,wang2021new,meng2021self,meng2020gap}.

\noindent \textbf{Mask set.} Two 660$\times$660 real masks following the same fabrication process are employed in this work. Fig.~\ref{fig: two real masks} demonstrates the histograms of both masks. As mentioned in the manuscript, the training mask set $\mathcal{M}$ is built by randomly cropping 256$\times$256 patches from the first real mask~\cite{meng2020end}. For simulation data, testing masks are collected from both real masks.  Notably, \textbf{there is no overlap between training and testing mask sets.} For real HSI reconstruction, no testing mask set is available. The second 660$\times$660 real mask~\cite{meng2020snapshot} is directly applied for testing purpose, indicating the miscalibration scenario.

\bibliographystyle{plain}
\bibliography{arxiv}
\end{document}